\documentclass[a4paper,11pt,final]{article}

\usepackage{setspace,graphicx,amsmath,amsfonts,amssymb,amsthm,versionPO}
\usepackage{marginnote,enumitem,subfigure,rotating,fancyvrb}
\usepackage{float}
\usepackage[longnamesfirst]{natbib}
\usepackage[plainpages=false,breaklinks=true,colorlinks=true,urlcolor=black,citecolor=black,%
                                       linkcolor=black,bookmarks=true,bookmarksopen=true,%
                                       bookmarksopenlevel=1,pdfstartview=FitH,pdfview=FitH]{hyperref}

\usepackage{morefloats}     
\usepackage{subfigure}     
\usepackage{ctable}
\usepackage{colortbl}
\usepackage{longtable}
\usepackage{rotating}
\usepackage[english]{babel}
\usepackage{amsmath} %need for subequations
\usepackage{amssymb}           %use if color is used in text
\usepackage[all,color]{xy}
\usepackage{ragged2e}
\usepackage{rfs}
\usepackage{ifpdf}
\usepackage{color}
\usepackage{parskip}
\usepackage{marvosym}

\usepackage{titlesec}
\newcommand{\periodafter}[1]{#1.}
\titleformat{\subsubsection}[runin]
{\normalfont\bfseries}{\thesubsubsection}{1em}{\periodafter}

\excludeversion{notes}		% Include notes?
\includeversion{links}          % Turn hyperlinks on?

\iflinks{}{\hypersetup{draft=true}}
\usepackage[lmargin=1in, rmargin=1in, tmargin=1in, bmargin=1in, paperwidth=220mm, paperheight=290mm]{geometry}

\makeatletter\let\chapter\@undefined\makeatother % Undefine \chapter for todonotes

\newcommand{\expnumber}[2]{{#1}\mathrm{e}{#2}}

\def\q{\mbox{\boldmath$q$}}
\def\u{\mbox{\boldmath$u$}}
\def\U{\mbox{\boldmath$U$}}
\def\h{\mbox{\boldmath$h$}}
\def\H{\mbox{\boldmath$H$}}
\def\A{\mbox{\boldmath$A$}}
\def\C{\mbox{\boldmath$C$}}
\def\U{\mbox{\boldmath$U$}}

\def\F{\mbox{\boldmath$F$}}

\def\R{\mbox{\boldmath$R$}}
\def\I{\mbox{\boldmath$I$}}

\def\one{\mbox{\boldmath$1$}}
\def\x{\mbox{\boldmath$x$}}
\def\y{\mbox{\boldmath$y$}}
\def\b{\mbox{\boldmath$b$}}

\def\bbeta{\mbox{\boldmath$\beta$}}
\def\bPhi{\mbox{\boldmath$\Phi$}}

\def\btheta{\mbox{\boldmath$\theta$}}

\def\W{\mbox{\boldmath$W$}}

\def\bomega{\mbox{\boldmath$\omega$}}

\def\m{\mbox{\boldmath$m$}}

\def\h{\mbox{\boldmath$h$}}

\def\mA{\mathcal{A}}
\def\mD{\mathcal{D}}
\def\mH{\mathcal{H}}
\def\seq#1#2{#1{:}#2}

\usepackage{rfs}          % RFS-specific formatting of sections, etc.
\newcommand{\citeRFSNP}[1]{\citeauthor{#1}~\citeyear{#1}} % Not very elegant\ldots
\newcommand{\citeRFS}[1]{\citeauthor{#1}~(\citeyear{#1})} % Not very elegant\ldots

\begin{document}

\setlist{noitemsep}  % Reduce space between list items (itemize, enumerate, etc.)

\newcommand{\tit}{\vspace{1em}\Large Large-Scale Dynamic Predictive Regressions\vspace{2em}}

\newcommand{\abs}{\small 
We develop a novel ``decouple-recouple'' dynamic predictive strategy and contribute to the literature on forecasting and economic decision making in a data-rich environment. Under this framework, clusters of predictors generate different latent states in the form of predictive densities that are later synthesized within an implied time-varying latent factor model. As a result, the latent inter-dependencies across predictive densities and biases are sequentially learned and corrected. Unlike sparse modeling and variable selection procedures, we do not assume a priori that there is a given subset of active predictors, which characterize the predictive density of a quantity of interest. We test our procedure by investigating the predictive content of a large set of financial ratios and macroeconomic variables on both the equity premium across different industries and the inflation rate in the U.S., two contexts of topical interest in finance and macroeconomics. We find that our predictive synthesis framework generates both statistically and economically significant out-of-sample benefits while maintaining interpretability of the forecasting variables. In addition, the main empirical results highlight that our proposed framework outperforms both LASSO-type shrinkage regressions, factor based dimension reduction, sequential variable selection, and equal-weighted linear pooling methodologies.   
\vspace{0.3in}
\\
\textbf{Keywords:} Data-Rich Models, Big Data, Forecast Combination, Model Averaging, Dynamic Forecasting, Macroeconomic Forecasting, Returns Predictability. 
\\
\\
\textbf{JEL codes:} C11, C53, D83, E37, G11, G12, G17}

\title{\vspace{-2em}{\bf \tit}}%\thanks{\footnotesize We are grateful to... }}

\author{Daniele Bianchi\thanks{University of Warwick, Warwick Business School, Coventry, UK. \color{blue}\texttt{Daniele.Bianchi@wbs.ac.uk}} \and Kenichiro McAlinn\thanks{University of Chicago, Booth School of Business, Chicago, IL, USA. \color{blue}\texttt{ kenichiro.mcalinn@chicagobooth.edu}}}

\date{\vspace{2em}First draft: December 2017. \hspace{2em} This draft: \today }
%\date{\emph{Preliminary and Incomplete}}

\maketitle

\vspace*{1in}
\centerline{\bf Abstract}
\medskip
\abs
\normalsize
\setlength{\parskip}{.2cm }
\setlength{\parindent}{0.6cm}

\doublespacing

\clearpage

\section{Introduction}

The increasing availability of large datasets, both in terms of the number of variables and the number of observations, combined with the recent advancements in the field of econometrics, statistics, and machine learning, have spurred the interest in predictive models in a data-rich environment; both in finance and economics.\footnote{See, e.g., \citeRFS{Timmermann2004}, \citeRFS{de2008}, \citeRFS{monch2008}, \citeRFS{bai2010}, \citeRFS{belloni2012}, \citeRFS{billio2013}, \citeRFS{elliott2013}, \citeRFS{manzan2015}, \citeRFS{harvey2016}, \citeRFS{freyberger2017}, \citeRFS{giannone2017}, and \citeRFS{McAlinn2017}, just to name a few.} As not all predictors are necessarily relevant for a given economic decision, decision makers often pre-select the most important candidate covariates by appealing to economic theories, existing empirical literature, and their own heuristic arguments. Nevertheless, a decision maker is often still left with tens-- if not hundreds-- of sensible predictors that may possibly provide useful information. However, the out-of-sample performance of standard techniques such as ordinary least squares, maximum likelihood, or Bayesian inference with uninformative priors tends to deteriorate as the dimensionality of the data increases, which is the well known curse of dimensionality.  

Confronted with a large set of predictors, three main classes of models became popular. \emph{Sparse} modeling focus on the selection of a sub-set of variables with the highest predictive power out of a large set of predictors, and discard those with the least relevance. In the Bayesian literature, a prominent example is given by \citeRFS{george1993variable} (and more recently, \citeRFSNP{rovckova2016spike} and \citeRFSNP{rovckova2018bayesian}), which introduced variable selection through a data-augmentation approach. Similarly, \emph{regularized} models take a large number of predictors and introduces penalization to discipline the model space. LASSO-type regularization and ridge regressions are by far the most used in both research and practice. Finally, \emph{dense} modeling is based on the assumption that, a priori, all variables could bring useful information for prediction, although the impact of some of these might be small. As a result, the statistical features of a large set of predictors are assumed to be captured by a small set of common components, which could be either static or dynamic. Factor analysis is a clear example of dense statistical modeling (see, e.g., \citeRFSNP{stock2002} and \citeRFSNP{de2008} and the references therein), which is very popular in macro-econometrics. 

All these approaches entail either an implicit or explicit reduction of the model space that is intended to mitigate the curse of dimensionality. However, the question of which one of these techniques is best is still largely unresolved. For economic and financial decision making, in particular, these dimension reduction techniques always lead to a decrease in consistent interpretability, something that might be critical for policy makers, analysts, and investors. For instance, a portfolio manager interested in constructing a long-short investment strategy might not find useful to use latent factors that she cannot clearly identify as meaningful sources of risk, or similarly would not want critical, economically sound, predictors to be shrunk to zero. More importantly, \citeRFS{giannone2017} recently show, in a Bayesian setting, that the posterior distribution of the parameters of a large dimensional linear regression do not concentrate on a single sparse model, but instead spreads over different types of models depending on priors elicitation. These problems possibly undermine the usefulness of exploiting data-rich environments for economic and financial decision making.

In this paper, we propose a class of data-rich predictive synthesis techniques and contribute to the literature on predictive modeling and decision making with big data. Unlike sparse modeling, we do not assume a priori that there is sparsity in the set of predictors. For example, suppose we are interested in forecasting the one-step ahead excess returns on the stock market based on, say, a hundred viable predictors. Using standard LASSO-type shrinkage-- a typical solution-- will implicitly impose a dogmatic prior that only a small subset of those regressors is useful for predicting stock excess returns and the rest is noise, i.e., sparsity is pre-assumed. Yet, there is no guarantee that the small subset is consistent, or smooth, over time. Similarly, even with such a moderate size, the model space is about $\expnumber{1}{+30}$ possible combinations of the predictors, which prevent any reasonable convergence within the class of standard stochastic search variable selection algorithms, for example, spike-and-slab priors (see, e.g.,  \citeRFSNP{giannone2017}). 

We, in turn, retain all of the information available and \emph{decouple} a large predictive model into a set of much smaller predictive regressions, which are constructed by similarity among the set of regressors. Suppose these predictors can be classified into $J$ different subgroups, each one containing fewer regressors, according to their economic meaning. Rather than assuming a sparse structure, in our framework, we retain all of the information by estimating $J$ different predictive densities-- separately and sequentially-- one for each class of predictors, and \emph{recouple} them dynamically using the predictive synthesis approach. Note that the way the subgroups of regressors are classified in the first place is entirely independent of the decoupling-recoupling strategy. In the empirical application we classified groups of variables according to their economic meaning. However, nothing prevents to use correlation-based clustering algorithms such as, K-means, fuzzy C-means, hierarchical clustering, mixture of Gaussians, or other nearest neighbor classifications.   

Our proposed approach significantly differs from model combination of multiple small models (e.g. multiple LASSO models with different tuning parameters), such as \citeRFS{stevanovic2017}, by utilizing the theoretical foundations and recent developments of Bayesian predictive synthesis \citep[BPS:][]{West1992c,West1992d,McAlinn2017}. This makes our decouple-recouple strategy theoretically and conceptually coherent, as it regards the decoupled models as separate latent states that are learned and calibrated using the Bayes theorem in an otherwise typical dynamic linear modeling framework (see \citeRFSNP{WestHarrison1997book2}). Under this framework, the dependencies between subgroups, as well as biases within each subgroup, can be sequentially learned; information that is critical, though lost in typical model combination techniques. As a result, our framework can be thought of as an efficient model combination strategy that allows dynamic modeling and forecasting in a data-rich environment by breaking down a large dimensional problem in a sequence of smaller ones, then fully utilizing all of the information and maintaining interpretability that is key in effective decision making. 

We calibrate and implement the proposed methodology, which we call decouple-recouple synthesis (DRS), on both a macroeconomic and a finance application. More specifically, in the first application we test the performance of our decouple-recouple approach using BPS to forecast the one- and three-month ahead annual inflation rate in the U.S. over the period 1986/1 to 2015/12, a context of topical interest (see, e.g. \citeRFSNP{Cogley2005}, \citeRFSNP{Primiceri2005}, \citeRFSNP{koop2010bayesian}, and \citeRFSNP{Nakajima2010}, among others). The set of monthly macroeconomic predictors consists of an updated version of the Stock and Watson macroeconomic panel available at the Federal Reserve Bank of St.Louis. Details on the construction of the dataset can be found in \citeRFS{mccracken}. The empirical exercise involves a balanced panel of 119 monthly macroeconomic and financial variables, which are classified into eight main groups: Output and Income, Labor Market, Consumption, Orders and Inventories, Money and Credit, Interest Rate and Exchange Rates, Prices, and Stock Market.

The second application relates to forecasting monthly year-on-year total excess returns across different industries from 1970/1 to 2015/12, based on a large set of predictors, which have been chosen by previous academic studies and existing economic theory with the goal of ensuring the comparability of our results with these studies (see, e.g., \citeRFSNP{lewellen2004predicting}, \citeRFSNP{Avramov:2004}, \citeRFSNP{Goyal2008}, \citeRFSNP{Goyal2010}, and \citeRFSNP{Dangl:Halling:2012}, among others). More specifically, we collect monthly data on more than 70 pre-calculated financial ratios for all U.S. companies across eight different categories. Both returns and predictors are aggregated at the industry level by constructing value-weighted returns in excess of the risk-free rate and value-weighted aggregation of the single-firm predictors. Industry aggregation is based on the four-digit SIC codes of the existing firm at each time $t$. Those 70 ratios are classified into eight main categories: Valuation, Profitability, Capitalization, Financial Soundness, Solvency, Liquidity, Efficiency Ratios, and Other. Together with industry-specific predictors, we use additional 14 aggregate covariates obtained from existing research, which are divided in two categories; aggregate financials and macroeconomic variables (see, \citeRFSNP{Goyal2008} and \citeRFSNP{Goyal2010}).

To evaluate our approach empirically, we compare forecasts from our framework against standard Bayesian model averaging (BMA), in which the forecast densities are mixed with respect to sequentially updated model probabilities~\citep[e.g.][Sect 12.2]{PJHandCF1976,WestHarrison1997book2}, as well as against simpler, equal-weighted averages of the model-specific forecast densities using linear pools, i.e., arithmetic means of forecast densities, with some theoretical underpinnings~(e.g.~\citealt{West1984}). In addition, we compare the forecasts from our setting with a state-of-the-art LASSO-type regularization, which constraints the coefficients of least relevant variables to be null leading to sparse models ex-post, and PCA based latent factor modeling \citep{stock2002,mccracken}. While some of these strategies might seem overly simplistic, they have been shown to dominate some more complex aggregation strategies in some contexts, at least in terms of direct point forecasts in empirical studies~\citep{Genre2013}. Finally, we also compare our decouple-recouple model synthesis scheme against the marginal predictive densities computed from the group-specific set of predictors taken separately. Forecasting accuracy is primarily assessed by evaluating the out-of-sample log predictive density ratios (LPDR); at horizon $k$ and across time indices $t$. Although we mainly focus on density forecasts in this paper, we also report the root mean squared forecast error (RMSFE) over the forecast horizons of interest, which, combined with the LPDR results, paints a fuller picture of the results. 

Irrespective of the performance evaluation metrics, our decouple-recouple model synthesis scheme emerges as the best for forecasting the annual inflation rate for the U.S. economy. This holds for both one-step ahead and three-step ahead forecasts. It significantly out-performs both sequential BMA and the equal-weighted linear pooling of predictive densities. Interestingly, the LASSO performs worst among the model combination/shrinkage schemes, in terms of density forecasts. The sequential estimates of the latent inter-dependencies across classes of macroeconomic predictors show that pressure on the labor market and price levels tend to dominate other groups of predictors, with labor market being a dominant component in early 2000s, while prices tend to increase their weight in the aggregate predictive density towards the end of the test period. 

The results are possibly even more pronounced concerning the prediction of the yearly total excess returns across different industries. The differences in the LPDRs are rather stark and clearly shows a performance gap in favor of DRS. None of the alternative specifications come close to DRS when it comes to predicting one-step ahead. While the equally-weighted linear pooling turns out to be a challenging benchmark to beat, we show that LASSO-type shrinkage estimators and PCA perform poorly out-of-sample, especially when it comes to predicting the one-step ahead density of excess returns. This result is consistent with the recent evidence in \citeRFS{diebold2017}, which show the sub-optimality of LASSO estimators in out-of-sample real-time forecasting exercises. We also compare our model combination scheme against the competitors outlined above on the basis of the economic performance assuming a representative investor with power utility preferences. 

The comparison is conducted for the unconstrained as well as short-sales constrained investor at the monthly horizons, for the entire sample. We find that the economic constraints lead to higher Certainty Equivalent (CER) values at all horizons and across practically all competing specifications. Specifically, the short-sale constraint results in a higher CER (relative to the unconstrained case) of more than 100 basis points per year, on average across sectors. Consistent with the predictive accuracy results, we generally find that the DRS strategy produces higher CER improvements than the competing specifications under portfolio constraints. In addition, we show that DRS allows to reach a higher CER both in the cross-section and in the time-series, which suggests that there are economically important gains by using our methodology. 

The structure of this paper is as follows. Section 2 introduces our decouple-recouple methodology for the efficient synthesis of predictive densities. Section 3 presents the core of the paper and report the empirical results related to both the U.S. annual inflation forecasts and the total stock returns predictability across industries in the U.S. Section 4 concludes the paper with further discussion.

\section{Decouple-Recouple Strategy}

A decision maker $\mD$ is interested in predicting some quantity $y$, in order to make some informed decision based on a large set of predictors, which are all considered relevant to $\mD$, but with varying degree. In the context of macroeconomics, for example, this might be a policy maker interested in forecasting inflation using multiple macroeconomic indicators, that a policy maker can or cannot control (such as interest rates). Similar interests are also relevant in finance, with, for example, portfolio managers tasked with implementing optimal portfolio allocations on the basis of expected future returns on risky assets. 

A canonical and relevant approach is to consider a basic time series linear predictive regression (see, e.g., \citeRFSNP{Stambaugh:1999}, \citeRFSNP{pesaran2002market}, \citeRFSNP{Avramov:2004}, \citeRFSNP{lewellen2004predicting}, \citeRFSNP{Goyal2008}, and \citeRFSNP{Goyal2010}, among others); 
\begin{equation}\label{eq:ols}
	y_{t}=\bbeta'\x_{t}+\epsilon_{t}, \quad \epsilon_{t}\sim N(0,\nu),
\end{equation}
where $y_{t}$ is the quantity of interest, $\x_t$ is a $p-$dimensional vector of predictors, which could have its own dynamics, $\bbeta$ is the $p-$dimensional vector of betas, and $\epsilon_{t}$ is some observation noise (Gaussian and constant over time here to fix ideas). 

In many practically important applications, the dimension of predictors relevant to $\mD$ is large, possibly too large to directly fit something as simple as an ordinary linear regression. As a matter of fact, at least a priori, all of these predictors could provide relevant information for the decision making process of $\mD$. Under this setting, regularization or shrinkage would not be consistent with $\mD$'s decision making process, as she has no dogmatic priors on the size of the model space. Similarly, dimension reduction techniques such as principal component analysis and factor models, e.g., \citeRFS{stock2002} and \citeRFS{bernanke2005measuring}, while using all of the predictors available, reduces them to a small preset number of latent factors that are typically difficult to interpret or control, in the sense of optimal decision making.

Our decouple-recouple strategy exploits the fact that the potentially large $p-$dimensional vector of predictors can be partitioned into smaller groups $j=1{:}J$, modifying  Eq.~\eqref{eq:ols} to 
\begin{equation}\label{eq:ols2}
	y_{t}=\bbeta_1'\x_{t,1}+...+\bbeta_j'\x_{t,j}+...+\bbeta_J'\x_{t,J}+\epsilon_{t}, \quad \epsilon_{t}\sim N(0,\nu).
\end{equation}
These groups can be partitioned based on some qualitative categories (e.g. group of predictors related to the same economic phenomenon), or by some quantitative measure (e.g. clustering based on similarities, correlation, etc.), though the dimension of each partitioned group should be relatively small in order to obtain sensible estimates. The first step of our model combination strategy is to {\em decouple} Eq.~\eqref{eq:ols2} into $J$ smaller models, such as,
\begin{equation}\label{eq:ols3}
	y_{t}=\bbeta_j'\x_{t,j}+\epsilon_{t,j}, \quad \epsilon_{t,j}\sim N(0,\nu_j),
\end{equation}
for all $j=1{:}J$, producing forecast distributions $p(y_{t+k}|\mA_j)$, where $\mA_j$ denotes each subgroup, and $k\geq 1$ is the forecast horizon. Since Eq.~\eqref{eq:ols3} is a linear projection of data from each subgroup, we can consider, without loss of generality, that $p(y_{t+k}|\mA_j)$ is reflecting the information arising from that subgroup regarding the quantity of interest.

In the second step, we {\em recouple} the densities $p(y_{t+k}|\mA_j)$ for $j=1{:}J$ in order to obtain a forecast distribution $p(y_{t+k})$ reflecting and incorporating all of the information that arises from each subgroup. In the most simple setting, $p(y_{t+k}|\mA_j)$ can be recoupled via linear pooling (see, e.g., \citeRFSNP{Geweke2011} for a further discussion);
\begin{equation}\label{eq:rec}
	y_{t+k}=w_1p(y_{t+k}|\mA_1)+...+w_jp(y_{t+k}|\mA_j)+...+w_Jp(y_{t+k}|\mA_J),
\end{equation}
where weights $w_{1:J}$ are estimated by the decision maker based on past observations (e.g. using $w_{1:J}$ proportional to the marginal likelihood). The main difference between BMA and linear pooling is about the domain of $w_{1:J}$ and the estimation approach adopted.   

While this linear combination structure is conceptually and practically appealing, it does not capture the fact that we expect and understand that each $p(y_{t+k}|\mA_j)$ to be biased and dependent with each other. Arguably each subgroup $p(y_{t+k}|\mA_j)$ is always biased unless one of them is the data generating process, which is something we cannot expect in applications in economics or finance. \citeRFS{Geweke2012} formally show that even when none of the constituent models are true, linear pooling and BMA assign positive weights to several models. 

The dependence between $p(y_{t+k}|\mA_j)$ and $p(y_{t+k}|\mA_q)$, for $j\neq q$, is also a crucial aspect of model combination. As a matter of fact, optimal combination of weights should be chosen to minimize the expected loss of the \emph{combined} forecast, which, by definition, reflects both the forecasting accuracy of each sub-model and the correlation across single forecasts. For instance, it is evident that the marginal predictive power of macroeconomic variables related to the labor market is somewhat correlated with the explanatory power of output and income. In addition, correlations across predictive densities are arguably latent and dynamic. The linkages between liquidity, solvency, and aggregate macroeconomic variables changed before and after the great financial crisis of 2008/2009.  Thus, an effective recoupling step must be able to sequentially learn and recover the latent biases and inter-dependencies between the subgroups/submodels.

To address these issues, we build on the theoretical foundations and recent developments proposed in \cite{West1992c,West1992d,McAlinn2017}. Each subgroup is considered to be a latent state, whereby $p(y_{t+k}|\mA_j)$ represents a distinct prior on state $j=1,...,J$. As BPS treats the latent states within the Bayesian paradigm, the biases and inter-dependencies between the latent states can be learned and recovered via standard Bayesian updating. The difference between BPS and more general latent factor models, such as PCA, is that BPS allows to pin down each latent state, using priors $p(y_{t+k}|\mA_j)$ at each time $t$, to a group that $\mD$ specifies. In this respect, the underlying assumption of BPS is that each latent state reflects information from each subgroup/submodel, and thus retains interpretability, which is the key component of $\mD$'s decision making process. 

Before we delve into the specifics of BPS, we first outline the intuition behind why using BPS within our proposed decouple-recouple framework potentially improves predictive ability compared to shrinkage methods and factor models. To fix ideas, we reconsider the bias-variance tradeoff; a well known statistical property where an increase in model complexity increases variance and lowers bias and vice versa. 

The goal in both shrinkage methods and factor models is to arbitrarily lower model complexity to balance bias and variance, in order to potentially minimize predictive loss. In terms of LASSO-type shrinkage, increasing the tuning parameter (i.e. increasing shrinkage) leads to increased bias, so using cross-validation aims to balance the bias-variance tradeoff by balancing the tuning parameter. Similarly, in factor model the optimal number of latent factors is chosen to reduce the variance by reducing the model dimensionality at the cost of increasing the bias. Our proposed method takes a significantly different approach towards the bias-variance tradeoff by breaking a large dimensional problem into a set of small dimensional ones, while at the same time exploiting the fact that the BPS methodology can learn the biases and inter-dependencies via Bayesian learning. As this is the case, recoupling step benefits from biased models, as long as the bias has a signal that can be learned. More specifically, by decoupling the model into smaller, less complex models, we adjust for the bias-- that characterize each group-- that is sequentially learned and corrected, while maintaining the low variance from each model. This flips the bias-variance tradeoff around, exploiting the weakness of low complexity models to an advantage in the recoupling step, potentially improving predictive performance. 

\subsection{Bayesian Predictive Synthesis}

In the general framework of BPS, the decision maker $\mD$ is interested in predicting some quantity $y$ and aims to incorporate information from $J$ individual models labeled $\mA_j$, $(j=\seq1J).$ 
$\mD$ has some prior information $p(y)$ about the quantity of interest, and each $\mA_j$ provides their own prior distribution about what they believe the outcome of the quantity is in the form of a predictive distribution $h_j(x_j)=p(y|\mA_j)$; the collection of which defines the information set $\mH = \{ h_1(\cdot), \ldots, h_J(\cdot) \} $.
The question BPS tackles is this: how should a Bayesian decision maker consolidate these prior distributions ($\mD$'s own and of  $\mA_{1{:}J}$) and learn, update, and calibrate in order to improve forecasts?

A formal prior-posterior updating scheme posits that, for a given prior $p(y)$, and (prior) information set $\mH$ provided by $\mA_{1{:}J}$, we can update using the Bayes theorem to obtain a posterior $p(y|\mH)$. 
Due to the complexity of $\mH$-- a set of $J$ density functions with cross-sectional time-varying dependencies as well as individual biases--  $p(y,\mH)=p(y)p(\mH|y)$ is impractical since $p(\mH|y)$ is difficult to define.
The works of \citeRFS{West1992c} and \citeRFS{West1992d} extend the basic theorem of \cite{GenestSchervish1985}, in the context of incorporating multiple prior information provided by experts, to show that, under a specific consistency condition, $\mD$'s posterior density takes the form 
\begin{equation}\label{eq:theorem1}
	p(y|\mH)=\int \alpha(y|\x)h(\x)d\x\qquad\text{where}\qquad h(\x)=\prod_{j=1}^{J}h_j\left(x_j\right).
\end{equation}
Here, $\x=x_{\seq1J} = (x_1,\ldots,x_J)'$ is a $J-$dimensional latent vector of states with priors provided by each $\mA_{j}$, and $\alpha(y|\x)$ is a conditional density function, which reflects how the decision maker believes these latent states to be synthesized. 
The only requirement of  Eq.~\eqref{eq:theorem1}, so that it is a coherent Bayesian posterior, is that it must be consistent with $\mD$'s prior, i.e.,
\begin{equation}\label{eq:mx}
	p(y)=\int \alpha(y|\x) m(\x) d\x\qquad\textrm{where}\qquad m(\x) = E\left[h(\x) \right],
\end{equation}
the expectation in the last formula being over $\mD$'s belief of what $p(\mH)$ should be. 
Critically, the representation of Eq.~\eqref{eq:theorem1} does not require a full specification of $p(y,\mH)$, but only the conditional density $\alpha(y|\x)$ and the marginal expectation function $m(\x)$. 
These two functions alone allows to incorporate any prior knowledge in the form of models' predictions in terms of biases, predictive accuracy, and more importantly, inter-dependencies among each other.
It is important to note that the theory does not specify the form of $\alpha(y|\x)$.
In fact, \cite{McAlinn2017} show that many forecast combination methods, from linear combinations (including Bayesian model averaging) to more recently developed density pooling methods \citep[e.g.][]{Aastveit2014,Fawcett2014,Pettenuzzo2015}, are special cases of BPS.

Now, suppose $\mD$ is interested in the more critical and relevant task of one-step ahead forecasting. 
$\mD$ wants to predict $y_t$ and receives current forecast densities $\mH_{t} = \{ h_{t1}(x_{t1}),\ldots, h_{tJ}(x_{tJ}) \}$ from the set of models. 
The full information set used by $\mD$ is thus $\{ y_{1:t-1}, \mH_{1:t} \}$, the past data of $y$ and historical information of predictive distributions coming from $\mA_{1{:}J}$. 
Extending  Eq.~\eqref{eq:theorem1} to a dynamic context \citep[as done in][]{McAlinn2017}, $\mD$ has a dynamic posterior distribution of the forecast of $y_{t}$ at time $t-1$ of the form
\begin{equation}\label{eq:theorem}
p(y_{t}|\bPhi_{t},\y_{\seq1{t-1}},\mH_{\seq1t}) \equiv p(y_{t}|\bPhi_{t},\mH_{t})=\int \alpha_{t}(y_{t}|\x_{t},\bPhi_{t})\prod_{j=\seq1J}h_{tj}(x_{tj})dx_{tj}
\end{equation}
where $\x_{t}=x_{t,\seq1J}$ is a $J-$dimensional latent agent state vector at time $t$, $\alpha_{t}(y_{t}|\x_{t},\bPhi_{t})$ is $\mD$'s conditional synthesis function for $y_{t}$ given the latent states $\x_{t},$ and $\bPhi_{t}$ represents some time-varying parameters learned and calibrated over $1{:}t$.

This general framework implies that $\x_t$ is the realization of the inherent dynamic latent factors at time $t$ and a synthesis is achieved by recoupling these separate latent predictive densities to the time series $y_t$ through the time-varying conditional distribution $\alpha_{t}(y_{t}|\x_{t},\bPhi_{t})$. 
Though the theory does not specify $\alpha_{t}(y_{t}|\x_{t},\bPhi_{t})$, a  natural choice-- as with \cite{McAlinn2017}--  is to impose linear dynamics, such that,
\begin{equation} \label{eq:BPSnormalalphadynamic}
\alpha_t(y_t|\x_t,\bPhi_t) = N(y_t|\F_t'\btheta_t,v_t) \quad\textrm{with}\quad \F_t=(1, \x_t')' \quad\textrm{and}\quad\btheta_t=(\theta_{t0},\theta_{t1},...,\theta_{tJ})',
\end{equation} 
where $\btheta_t$ represents a $(J+1)-$vector of time-varying synthesis coefficients.
 Observation noise is reflected in the innovation variance term $v_t$, and the general time-varying parameters $\bPhi_t$ is defined as $\bPhi_t=\left(\btheta_t,v_t\right)$. 
 The evolution of these parameters is needed to complete the model specification. 
 We follow existing literature in dynamic linear models and assume that both $\btheta_t$ and $v_t$ evolve as a random walk to allow for stochastic changes over time as is traditional in the Bayesian time series literature (see \citealt{WestHarrison1997book2,Prado2010}). 
 Thus, we consider
\begin{subequations}
\label{DLM}
\begin{align}
	y_t&=\F_t'\btheta_t+\nu_t, \quad \nu_t\sim N(0,v_t), \label{eq:DLMa} \\
	\btheta_t&=\btheta_{t-1}+\bomega_t, \quad \bomega_t\sim N(0, v_t\W_t),\label{eq:DLMb}
\end{align}
\end{subequations}
where $v_t\W_t$ represents the innovations covariance for the dynamics of $\btheta_t$ and $v_t$ the residuals variance in predicting $y_t$, which is based on past information and the set of models' predictive densities. 
The residuals $\nu_t$ and the evolution {\em innovations} $\bomega_s$ are independent over time and mutually independent for all $t,s$. 
The dynamics of $\W_t$ is imposed by a standard, single discount factor specification as in \cite{WestHarrison1997book2} (Ch.6.3) and \cite{Prado2010} (Ch.4.3). 
The residual variance $v_t$ follows a beta-gamma random-walk volatility model such that $v_t=v_{t-1}\delta/\gamma_t$, where $\delta\in\left(0,1\right]$ is a discount parameter, and $\gamma_t\sim Beta\left(\delta n_{t-1}/2,\left(1-\delta\right)n_{t-1}/2\right)$ are innovations independent over time and independent of $v_s,\bomega_r$ for all $t,s,r$, with $n_{t}=\delta n_{t-1} + 1$, the degrees of freedom parameter. 

With the $\x_t$ vectors in each $\F_t$  treated as latent variables, a dynamic latent factor model is defined through Eqs.~\eqref{DLM}. When forecasting each $t$, the latent states are conceived as arising as single draws from the set of models' predictive densities $h_{tj}(\cdot)$, the latter becoming available at time $t-1$ for forecasting $y_t$. 
Note that $x_{tj}$ are drawn {\em independently} (for $t$) from
\begin{equation}\label{eq:dfmh}
 p(\x_t| \bPhi_t,\y_{\seq1{t-1}},\mH_{\seq1t}) \equiv p(\x_t|\mH_t) = \prod_{j=\seq1J} h_{tj}(x_{tj}) 
\end{equation}
 with $\x_t,\x_s$ conditionally independent for all $t\ne s$. 
Importantly, the independence of the $x_{tj}$, conditional on $h_{tj}$, must not be confused with the question of modeling and estimation of the dependencies among predictive densities.
$\mD$'s modeling and estimation of the biases and inter-dependencies among these models are effectively mapped on and  reflected through the time-varying parameters $\bPhi_t=\left(\btheta_t,v_t\right)$. 

Further discussion on the dynamic synthesis function is in order. While we choose a simple and flexible dynamic form for the synthesis function, $\alpha_t(y_t|\x_t,\bPhi_t)$, in theory we do not need to imply any certain structure for the synthesis of model-specific predictive densities.  For instance, one can set cross-sectional correlations to be high if different models are known to give identical predictions; similarly, if we believe there are clear regime changes that favor certain models at given periods of time, a regime switching approach or an indicator in the state equation might be suitable. We also note that most methods in the forecast combination literature focus on weights that are restricted to the unit simplex, as well as the weights summing to one. 
For weights summing to one, we can apply the technique used in \cite{Irie2016}, where the sum of weights are always restricted to the same value. For weights restricted to the unit simplex but not summing to one, it is significantly more complicated, as we now have a non-linear state space model. Although the benefit of having weights restricted to the unit simplex is interpretability, there is no real gain in terms of forecasting accuracy in such restriction \citep{diebold1991}, just as portfolios allowed to hold short positions can improve on long only portfolios.  In the dynamic setting in Eqs.~\eqref{DLM}, restricting the weights possibly leads to an under-performance compared to the unrestricted case.  For example, consider the case where all models overestimate the quantity of interest by some positive value. Under the restrictive case, there is no combination of weights that can achieve that quantity, while the unrestrictive case can by imposing some negative coefficients. 
For these reasons, we utilize the unrestricted dynamic weighting scheme implied by Eqs.~\eqref{DLM} instead of the conventional restricted variations.

\subsection{Estimation Strategy}

Estimation for the decouple step is straightforward, depending on the model assumptions for each subgroup.
For (dynamic) linear regressions, we can sample each $h_j(x_j)=p(y|\mA_j)$ using conjugate updates.
As for the recouple step using BPS, some discussion is needed.
In particular, the joint posterior distribution of the latent states and the structural parameters is not available in closed form. 
In our framework, the latent states are represented by the predictive densities of the models, $\mA_j,j=1,...,J$, and the synthesis parameters, $\bPhi_t$. 
We implement a Markov Chain Monte Carlo (MCMC) approach using an efficient Gibbs sampling scheme, which is detailed in Appendix \ref{sec:AppendixA}. 
Marginal posterior distributions of quantities of interest are computed as mixtures of the model-dependent marginal predictive densities weighted by the synthesis implied by $\alpha_t(y_t|\x_t,\bPhi_t)$. 
Integration over the models space is performed using our MCMC scheme, which provides consistent estimates of the latent states and parameters. 

Posterior estimates of the latent states $\x_t$ provide insights into the nature of the conditional dependencies across the subgroups, as well as subgroup characteristics.
The MCMC algorithm involves a sequence of standard steps in a customized two-component block Gibbs sampler: the first component simulates from the conditional posterior distribution of the latent states given the data, past forecasts from the subgroups, and the synthesis parameters. 
This is the ``learning'' step, whereby we learn the biases and inter-dependencies of the latent states. 
The second step samples the predictive synthesis parameters, that is, we ``synthesize'' the models' predictions by effectively mapping the biases and inter-dependencies learned in the first step onto parameters in a dynamic manner.
 The second step involves the FFBS algorithm central to MCMC in all conditionally normal DLMs (\citealt{Schnatter1994}; \citealt[][Sect 15.2]{WestHarrison1997book2}; \citealt[][Sect 4.5]{Prado2010}). 
 At each iteration of the sampler we sequentially cycle through the above steps.
In our sequential learning and forecasting context, the full MCMC analysis is redone at each time point as time evolves and new data are observed. 
Standing at time $T$, the historical information $\{ y_{\seq1T}, \mH_{\seq1T}\}$ is available and initial prior $\btheta_0\sim N(\m_0, \C_0 v_0/s_0 )$ and $1/v_0\sim G(n_0/2, n_0s_0/2),$ and discount factors $(\beta,\delta)$ are specified. 

\subsection{Forecasting}

In terms of forecasting, at time $t$, we generate predictive distributions of the object of interest as follows: (i) For each sampled $\bPhi_t$ from the posterior MCMC above, draw $v_{t+1}$ from its stochastic dynamics, and then  $\btheta_{t+1}$ conditional on $\btheta_t,v_{t+1}$ from Eq.~\eqref{DLMb}-- this  gives a draw $\bPhi_{t+1} = \{ \btheta_{t+1}, v_{t+1} \}$ from $p(\bPhi_{t+1} |y_{\seq1t}, \mH_{\seq1t} )$; (ii) draw $\x_{t+1}$ via independent sampling from $h_{t+1,j}(x_{t+1,j}),$  $(j=\seq1J);$ (iii) conditional on the parameters and latent states draw $y_{t+1}$ from Eq.~\eqref{DLMa}. Repeating, this generates a random sample from the 1-step ahead forecast distribution for time $t+1$. 

Forecasting over multiple horizons is often of equal or greater importance than 1-step ahead forecasting. 
However, forecasting over longer horizons is typically more difficult than over shorter horizons, since predictors that are effective in the short term might not be effective in the long term. 
The BPS modeling framework provides a natural and flexible procedure to recouple subgroups over multiple horizons. 

In the BPS framework, there are two ways to forecast over multiple horizons, through traditional DLM updating or through customized synthesis.
The former, direct approach follows traditional DLM updating and forecasting via simulation as for 1-step ahead, where the synthesis parameters are simulated forward from time $t$ to $t+k$.
The latter, customized synthesis involves a trivial modification, in which the model at time $t-1$ for predicting $y_t$ is modified so that the $k$-step ahead forecast densities made at time $t-k,$  i.e., $h_{t-k,j}(x_{tj})$ replace $h_{tj}(x_{tj})$. 
While the former is theoretically correct, it does not address how effective predictors (and therefore subgroups) can drastically change over time as it relies wholly on the model as fitted, even though one might be mainly interested in forecasting several steps ahead. 
\cite{McAlinn2017} find that, compared to the direct approach, the customized synthesis approach significantly improves multi-step ahead forecasts, since the dynamic model parameters, $\{ \btheta_t, v_t\}$, are now explicitly geared to the $k$-step horizon.

\section{Empirical Study}

To shed light on the predictive ability of our decouple-recouple model synthesis strategy, we calibrate and test the models in two different scenarios: (1) a macroeconomic application, which relates to the monthly forecast on the U.S. annual inflation using a large set of macroeconomic and financial variables, and (2) a finance application concerning the forecasting of the one-month ahead stock returns in excess of the risk-free rate across different industries. For both applications, for the decouple step we use a dynamic linear model \citep[DLM:][]{WestHarrison1997book2,Prado2010}, for each subgroup, $j=1{:}J$,
\begin{align}\label{eq:dlmde}
	y_{t}&=\bbeta_{tj}'\x_{tj}+\epsilon_{tj}, \quad \epsilon_{tj}\sim N(0,\nu_{tj}),\\
	\bbeta_{tj} &=\bbeta_{t-1,j}+\u_{tj}, \quad \u_{tj}\sim N(0,\nu_{tj}\U_{tj}), \nonumber
\end{align}
where the coefficients follow a random walk and the observation variance evolves with discount stochastic volatility, see, e.g., \citeRFS{Dangl:Halling:2012}, \citeRFS{koop2013}, \citeRFS{GruberWest2016BA}, \citeRFS{GruberWest2017ECOSTA} and \citeRFS{ZhaoXieWest2016ASMBI}. Priors for each decoupled predictive regression are assumed rather uninformative, such as $\bbeta_{0j}|v_{0j}\sim N(\m_{0j}, (v_{0j}/s_{0j})\I)$ with $\m_{0j}={\bf 0}'$  and $1/v_{0j}\sim G(n_{0j}/2,n_{0j}s_{0j}/2)$  with $n_{0j}=10,s_0=0.01$. The discount factors for the conditional volatilities in Eq.~\eqref{eq:dlmde} are set to $(\beta,\delta)=(0.95,0.99)$. For the recouple step, we follow the synthesis function in Eq.~\eqref{eq:BPSnormalalphadynamic}, with the following marginal priors: $\btheta_0|v_0\sim N(\m_0, (v_0/s_0)\I)$ with $\m_0=(0,\one'/J)'$  and $1/v_0\sim G(n_0/2,n_0s_0/2)$ with $n_0=10,s_0=0.01$. The discount factors are the same as in the decouple step.

For both studies, we compare forecasts from our framework with the predictive densities from each subgroup regressions, a LASSO shrinkage regression estimated in a expanding window fashion with leave-one-out cross validation, latent factor modeling (PCA), linear pooling with equal weights, and standard Bayesian model averaging, in which the forecast densities are mixed with respect to sequentially updated model probabilities~\citep[e.g.][Sect 12.2]{PJHandCF1976,WestHarrison1997book2}.\footnote{The subgroup-specific predictive density can be interpreted as a model combination scheme, whereby the weights are restricted to be inside the unit circle and the $jth$ submodel is restricted to have weight equal to one.} In the finance application we also compare DRS against the prediction from the historical average, along the line of \citeRFS{campbell2007} and \citeRFS{Goyal2008}. 

We compute and compare the DRS point forecasts based on the RMSFE over the forecasting horizon of interest. In comparing density forecasts with DRS, we also evaluate log predictive density ratios (LPDR); at horizon $k$ and across time indices $t$, that is,
\begin{align}
	\mathrm{LPDR}_t(k)=\sum_{i=1}^t\mathrm{log}\{p(y_{i+k}|y_{1{:}i}, \mathcal{M}_s)/p(y_{i+k}|y_{1{:}i},\mathcal{M}_0)\},
	\label{eq:lpdr}
\end{align}
where $p(y_{t+k}|y_{1{:}t},\mathcal{M}_s)$ is the predictive density computed at time $t$ for the horizon $t+k$ under the model or model combination aggregation strategy indexed by $\mathcal{M}_s$, compared against our forecasting framework labeled by $\mathcal{M}_0$. As used by several authors recently~\cite[e.g.][]{Nakajima2010,Aastveit2015}, LPDR measures provide a direct statistical assessment of relative accuracy at multiple horizons that extend traditional 1-step focused Bayes' factors. They weigh and compare dispersion of forecast densities along with location, and elaborate on raw RMSFE measures; comparing both measurements, i.e., point and density forecasts, gives a broader understanding of the predictive abilities of the different strategies.

\subsection{Macroeconomic application: Forecasting Inflation}

The first application concerns monthly forecasting of annual inflation in the U.S., a context of topical interest \citep{Cogley2005,Primiceri2005,Koop2009,Nakajima2010}. We consider a balanced panel of $N=128$ monthly macroeconomic and financial variables over the period 1986/1 to 2015/12. A detailed description of how variables are collected and constructed is provided in \citeRFS{mccracken}. These variables are classified into eight main categories depending on their economic meaning: Output and Income, Labor Market, Consumption and Orders, Orders and Inventories, Money and Credit, Interest Rate and Exchange Rates, Prices, and Stock Market.
%\paragraph{Decouple-recouple specifications.} 
%\paragraph{Analysis.} 
The empirical application is conducted as follows; first, the decoupled models are analyzed in parallel over 1986/1-1993/6 as a training period, simply estimating the DLM in Eq.~\eqref{eq:dlmde} forward filtering to the end of that period to calibrate the forecasts from each subgroup. This continues over 1993/7-2015/12, but with the calibration of recouple strategies which, at each quarter $t$ during this period, is run with the MCMC-based DRS analysis using data from 1993/7 up to time $t$. We discard the forecast results from 1993/7-2000/12 as training data and compare predictive performance from 2001/1-2015/12. The time frame includes key periods that tests the robustness of the framework, such as the inflating and bursting of the dot.com bubble, the building up of the Iraq war, the 9/11 terrorist attacks, the sub-prime mortgage crisis and the subsequent great recession of 2008-2009. These periods exhibit sharp shocks to the U.S. economy in general, and possibly provide shifts in relevant predictors and their inter-dependencies. We consider both 1- and 3-step ahead forecasts, in order to reflect interests and demand in practice.

Panel A of Table~\ref{tab:Table003} shows that our decouple-recouple strategy using BPS improves the one-step ahead out-of-sample forecasting accuracy relative to the group-specific models, LASSO, PCA, equal-weight averaging, and BMA. The RMSE of DRS is about half of the one obtained by LASSO-type shrinkage, a quarter compared to that of PCA, and significantly lower than equal-weight linear pooling and  BMA. In general, our decouple-recouple strategy exhibits improvements of 4\% up to over 250\% in comparison to the models and strategies considered. For each group-specific model, we note that both the Labor Market and Prices achieve similarly good point forecasts, which suggests that the labor market and price levels might be intertwined and dominate the aggregate predictive density. 
\[
\left[\text{Insert Table \ref{tab:Table003} about here}\right]
\]

\begin{figure}[p]
%\captionsetup[subfigure]{position=top}
\caption{US inflation rate forecasting: Out-of-sample log predictive density ratio}
\centering
  \vspace{0.1in}
\begin{justify} 
\footnotesize{This figure shows the dynamics of the out-of-sample Log Predictive Density Ratio (LPDR) as in Eq.~\eqref{eq:lpdr} obtained for each of the group-specific predictors, by taking the results from a set of competing model combination/shrinkage schemes, e.g., Equal Weight, and Bayesian Model Averaging (BMA). LASSO not included due to scaling. The sample period is 01:2001-12:2015, monthly. The objective function is the one-step ahead density forecast of annual inflation. }
\end{justify}
%\vspace*{-2em}
\includegraphics[width=.8\textwidth]{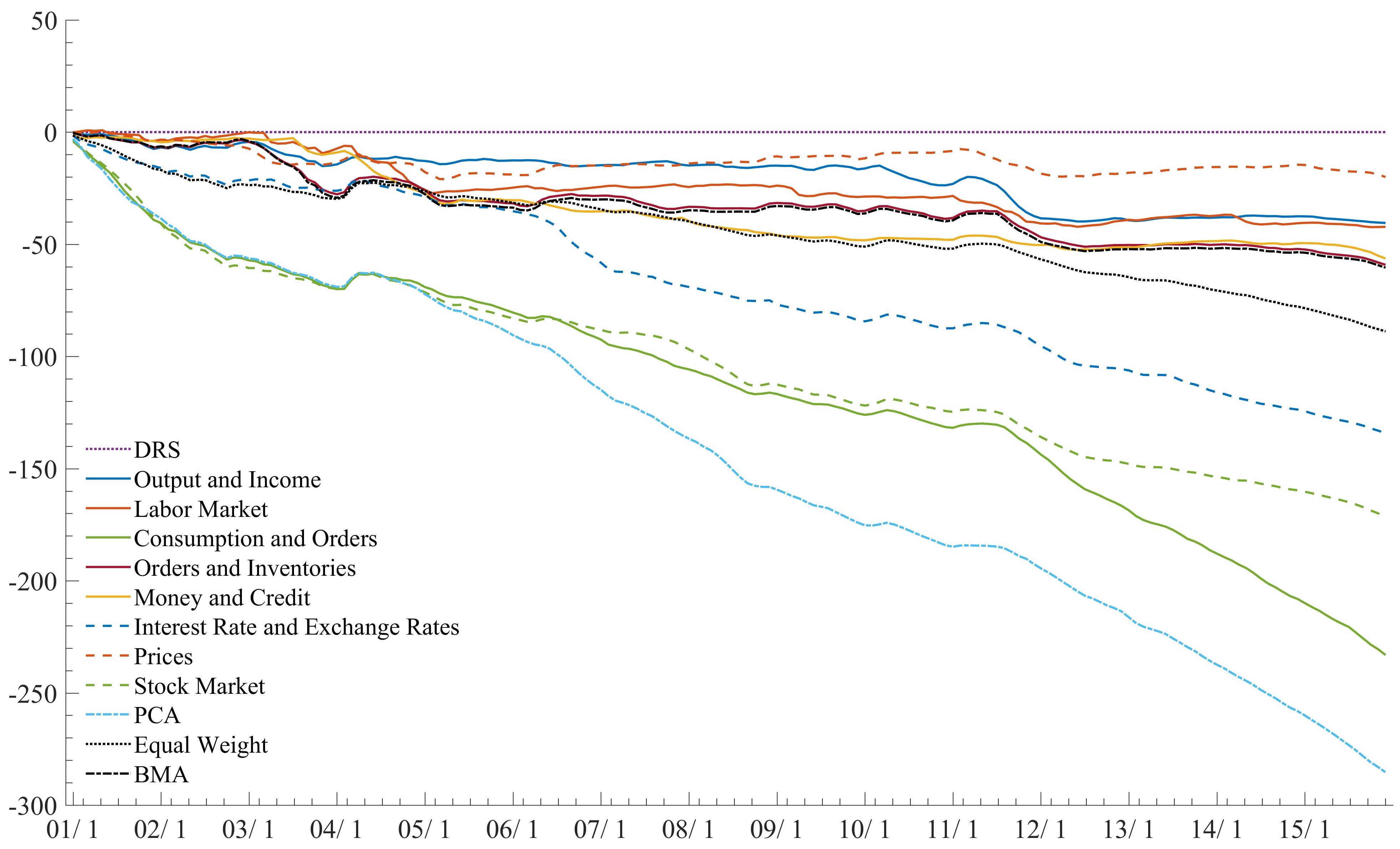} 
\label{fig:LPDR_Inflation}
 \end{figure}

Delving further into the dynamics of the LDPR, Figure~\ref{fig:LPDR_Inflation} shows that the out-performance of DRS, with respect to the benchmarking model combination/shrinkage schemes tend to increase. The out-of-sample performance of the LASSO sensibly deteriorates when it comes to predict the overall one-step ahead distribution of future inflation. Similarly, both the equal weight and BMA show a significant -50\% in terms of density forecast accuracy. Interestingly, both Labor Market and Prices, on their own, outperforms the competing combination/shrinkage schemes, except for DRS. Output and Income, Orders and Inventories, and Money and Credit, also perform well, with Output and Income outperforming Labor Market in terms of density forecasts. 
\[
\left[\text{Insert Figure \ref{fig:LPDR_Inflation} about here}\right]
\]

On the other hand, we note that Consumption, Interest Rate and Exchange Rates, and the Stock Market, perform the worst compared to the rest by a large margin. LASSO fails poorly in this exercise due to the persistence of the data, and erratic, inconsistent regularization the LASSO estimator imposes. In terms of equal weight and BMA, we observe that BMA does outperform equal weight, though this is because the BMA weights degenerated quickly to Orders and Inventories, which highlights the problematic nature of BMA, as it acts more as a model selection device rather than a recoupling procedure.

\begin{figure}[p]
%\captionsetup[subfigure]{position=top}
\caption{US inflation forecasting: On-line posterior means of predictive synthesis coefficients}
\centering
  \vspace{0.1in}
\begin{justify} 
\footnotesize{This figure shows the one-step ahead latent interdependencies across groups of predictive densities-- measured through the predictive coefficients-- used in the recoupling step. These latent components are sequentially computed at each of the $t=\seq1{180}$ months. Top panel shows the results for the one-step ahead forecasting exercise, while bottom panel shows the same results but now for a three-period ahead forecast objective function.}
\end{justify}
%\vspace*{-2em}
\subfigure[1-step ahead]{\includegraphics[width=.8\textwidth]{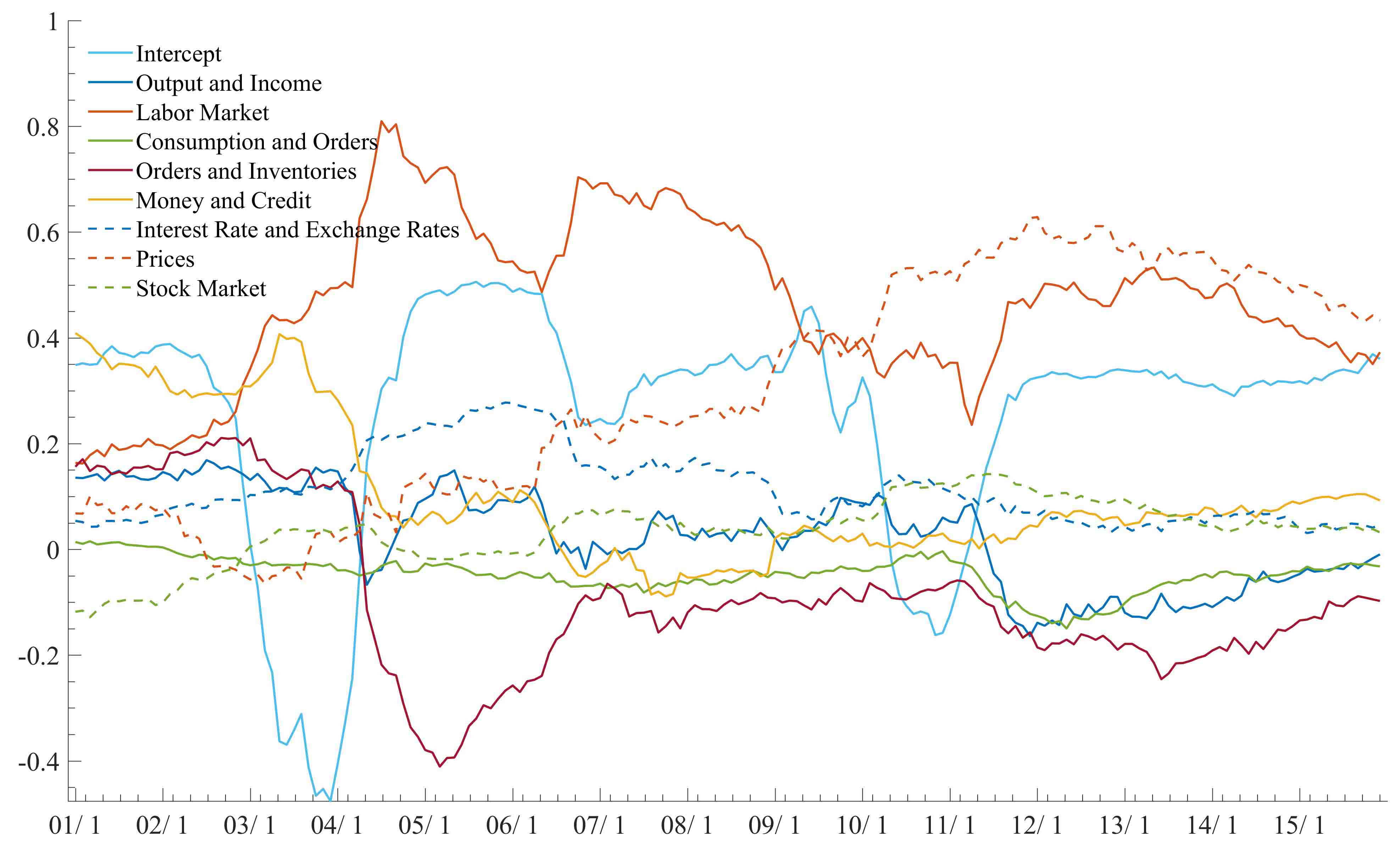}}\\
\subfigure[3-step ahead]{\includegraphics[width=.8\textwidth]{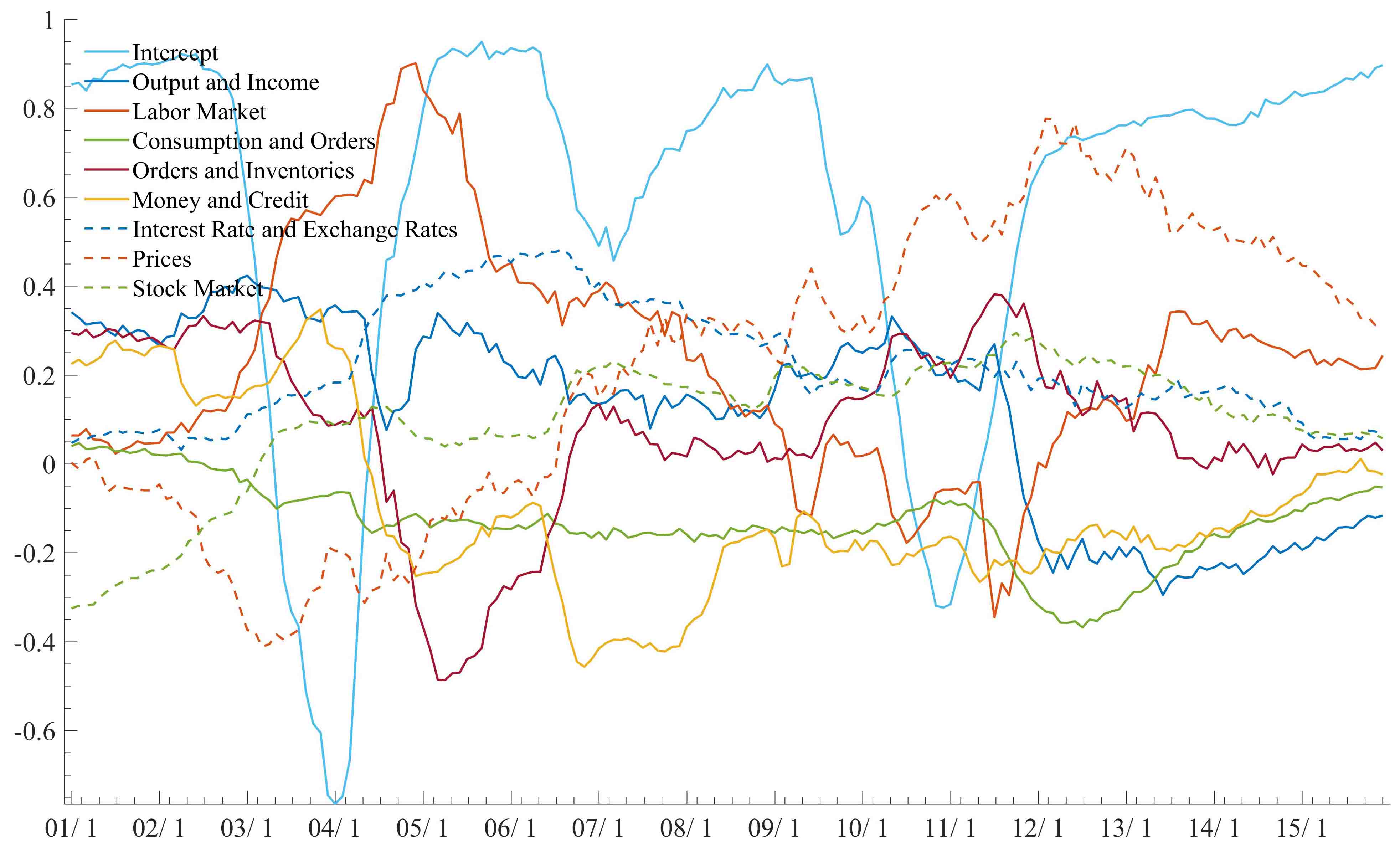}}  
\label{fig:BPS_Inflation}
 \end{figure}

Top panel of Figure~\ref{fig:BPS_Inflation} highlights a critical component of using BPS in the recouple step, namely learning the biases and inter-dependencies among and between the subgroups in order to maintain economic interpretability. Looking at the overall bias, i.e., the conditional intercept, they clearly switch sign in the aftermath of the short recession in the early 2000s and the financial crisis of 2008/2009. Since the parameters of the BPS are considered to be latent states, the conditional intercept can be considered as a free-roaming component, which is not directly pinned down by any group of predictors. In this respect, and for this application, the time variation in the conditional intercept of BPS can be thought of as a reflection of unanticipated economic shocks, which then affect inflation forecasts with some lag. 
\[
\left[\text{Insert Figure \ref{fig:BPS_Inflation} about here}\right]
\]

We next note that prior to the dot.com bubble, Money and Credit have the largest weight, though after the crisis, it essentially goes to zero throughout the rest of the testing period. One large trend in coefficients is with Labor Market, Prices, and Orders and Inventories. After the dot.com crash, we see a large increase in weight assigned to Labor Market, making it the group with the highest impact on the predictive density for most of the period. A similar pattern also emerges with Orders and Inventories, though this is in the negative. Yet, Labor Market does not always represent the group with the largest weight towards the end of the sample. In the aftermath of the the dot-com crash the marginal weight of Prices trends significantly upwards, crossing Labor Market around the sub-prime mortgage crisis, making it the highest weighted group and the end of the test period.

Panel B of Table \ref{tab:Table003} shows that DRS for the 3-step ahead forecasts reflect a critical benefit of using BPS for the recoupling step for multi-step ahead evaluation (bottom panel of Figure~\ref{fig:BPS_Inflation}). As a whole, the results are relatively similar to that of the 1-step ahead forecasts, with DRS outperforming all other methods, though the order of performance is changing. Compared to the results from the 1-step ahead forecasts (Figure~\ref{fig:BPS_Inflation}), we note some specific differences that are key to understand long term dynamics. For one, the conditional intercept is clearly amplified, compared to the 1-step ahead conditional intercept. This is natural, as we expect forecast performance to deteriorate as the forecast horizon moves further away, and thus more reliant on the free-roaming component of the latent states. Further, we note a significant decrease in importance of Labor Market before and after the great recession, and an increase of Interest Rate and Exchange Rates after the dot.com bubble. This is a stark contrast to the results of the 1-step ahead forecasts and reflects an interesting dynamic shift in importance of each subgroup that highlights the flexible specification of BPS for multi-step ahead modeling.

\begin{figure}[p]
%\captionsetup[subfigure]{position=top}
\caption{US inflation rate forecasting: Retrospective latent dependencies}
\centering
  \vspace{0.1in}
\begin{justify} 
\footnotesize{This figure shows the retrospective latent interdependencies across groups of predictive densities used in the recoupling step. The latent dependencies are measured using the MC-empirical R$^2$, i.e., variation explained of one model given the other models. These latent components are sequentially computed at each of the $t=\seq1{180}$ months.}
\end{justify}
%\vspace*{-2em}
\includegraphics[width=1\textwidth]{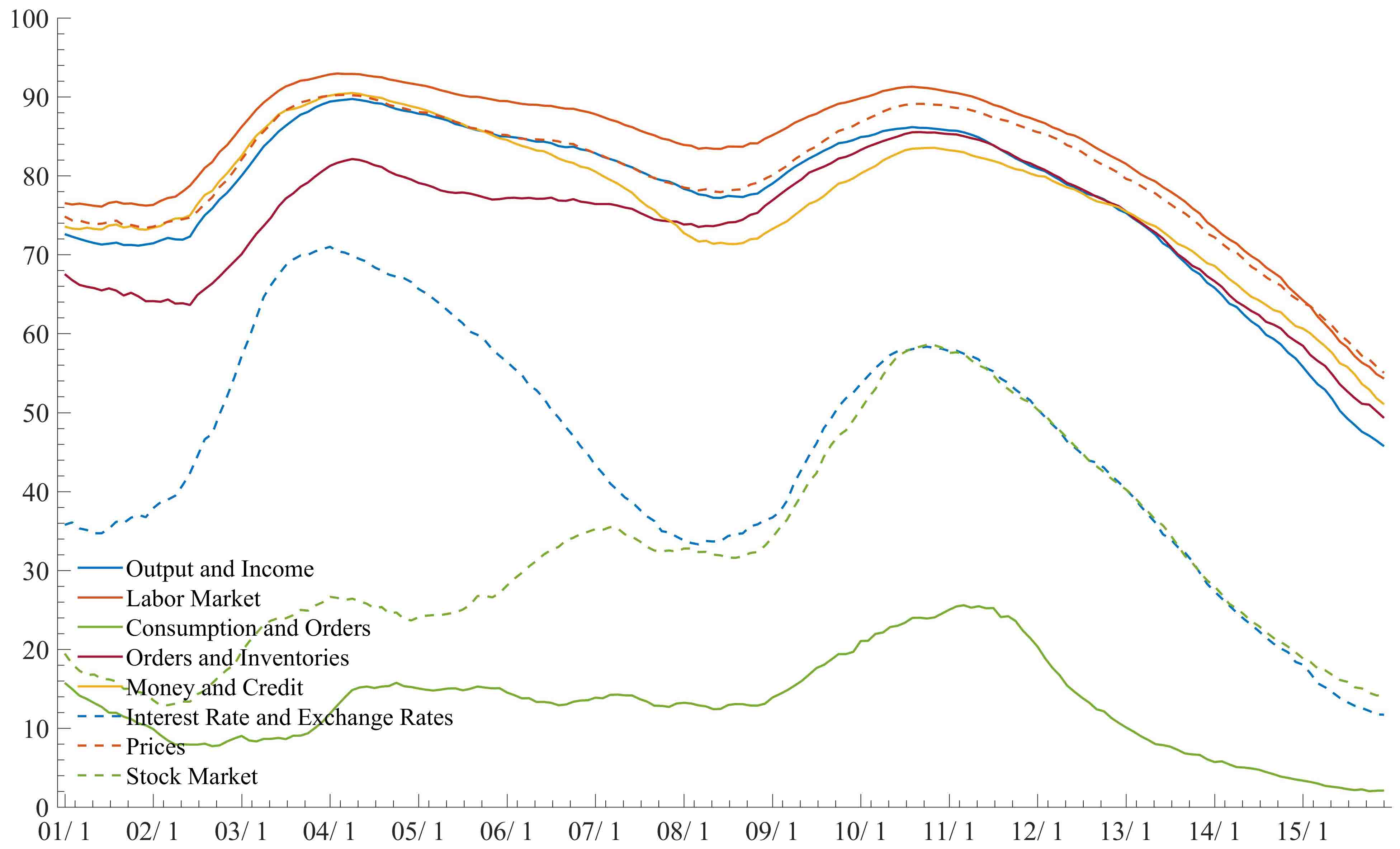} 
\label{fig:1r2_Inflation}
 \end{figure}

Finally, we explore the retrospective dependencies of the latent states for the one-step ahead inflation forecasting exercise. For this, we measure the MC-empirical R$^2$, which is the variation of one of the retrospective posterior latent states explained by the other latent states. Retrospective, here, means that these measures are computed using all of the data in the testing period, rather than the one-step ahead coefficients of Figure~\ref{fig:BPS_Inflation}. Figure~\ref{fig:1r2_Inflation} shows the MC-empirical R$^2$ for one of the latent states, given all of the other latent states; e.g., variation of Output and Income given Labor Market, Consumption and Orders, etc. There are some clear patters that emerge. Most latent states are highly dependent with each other, with Output and Income, Labor Market, Orders and Inventories, Money and Credit, and Prices grouping up over the whole period, with increased dependencies measure after the crisis of 2008/2009.
\[
\left[\text{Insert Figure \ref{fig:1r2_Inflation} about here}\right]
\]

We also note that there are clear trends in terms of decrease in dependencies before the crisis and sharp increase after. This is indicative of the closeness of these groups, as well as how they shift through different economic paradigms. Most interesting is how Interest Rate and Exchange Rates increase during the dot.com bubble, almost to the level of the other highly dependent states, and drops down, and then syncs almost perfectly with Stock Market after 2008. We can infer from this that the dependency characteristics of Interest Rate and Exchange Rates and Stock Market have changed dramatically over the testing period, with the Stock Market being significantly less dependent to the broader macroeconomy, including Interest Rate and Exchange Rates, the crisis of 2008/2009 shifting the two characteristics to be similar, and finally tapering off at the end again to be less dependent to the other latent states (though we note this is a general trend in all of the latent states).

\begin{figure}[p]
%\captionsetup[subfigure]{position=top}
\caption{US inflation rate forecasting: Retrospective latent dependencies (paired)}
\centering
  \vspace{0.1in}
\begin{justify} 
\footnotesize{This figure shows the retrospective paired latent interdependencies across groups of predictive densities used in the recoupling step. The latent dependencies are measured using the paired MC-empirical R$^2$, i.e., variation explained of one model given another model, for Labor Market (top) and Prices (bottom). These latent components are sequentially computed at each of the $t=\seq1{180}$ months.}
\end{justify}
%\vspace*{-2em}
\includegraphics[width=1\textwidth]{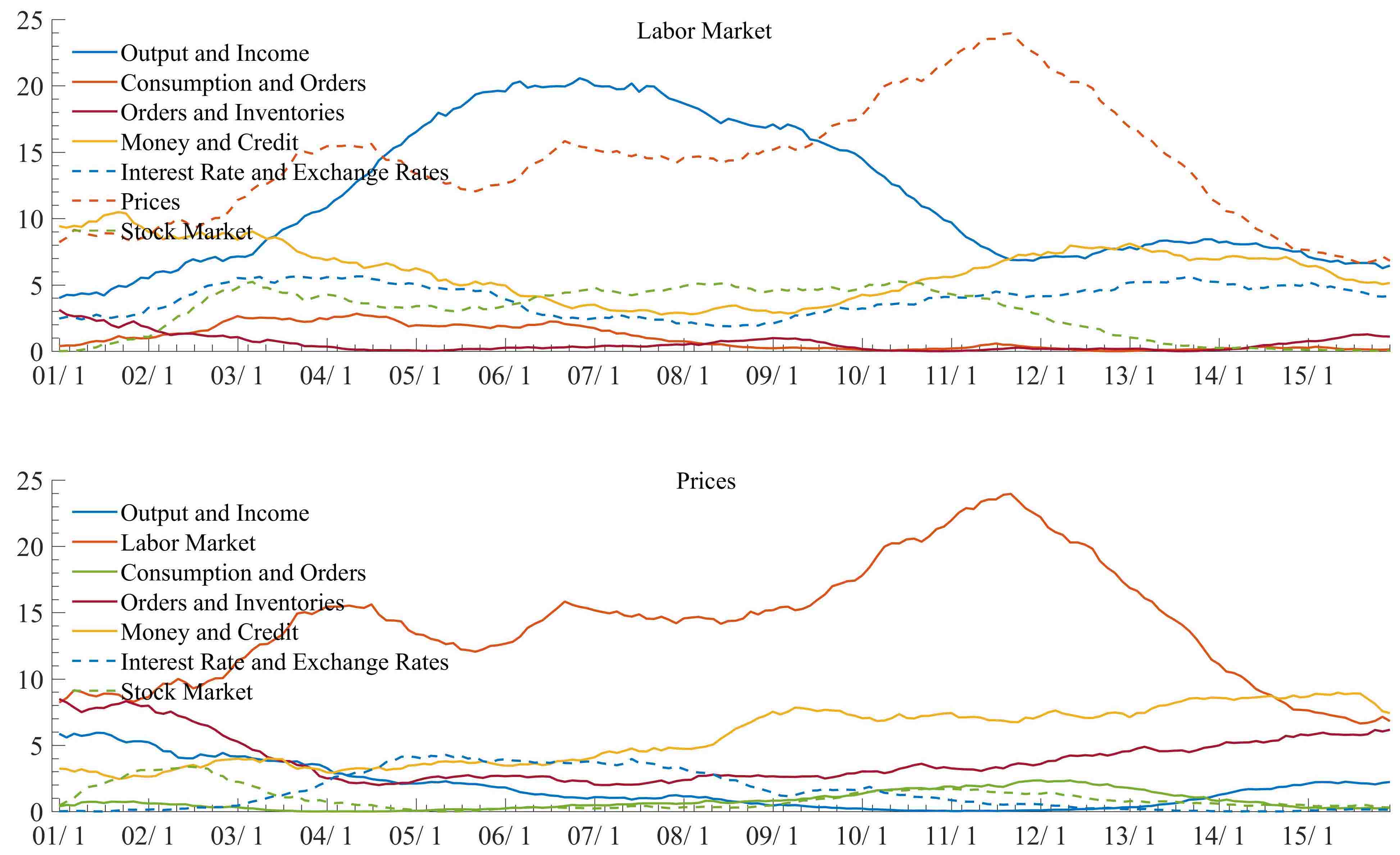} 
\label{fig:1r2per_Inflation}
 \end{figure}

Figure \ref{fig:1r2per_Inflation} further explores the retrospective dependencies showing the pairwise MC-empirical R$^2$, which measures the variation explained of one state given another, but now focusing solely on the pair of states. Based on the results in Table \ref{tab:Table003} we focus on two of the most prominent states: Labor Market (top panel) and Prices (bottom panel). Notice that, due to the symmetry in the dependence structure of the latent predictive densities, the relationship between Labor Market vs Prices and Prices vs Labor Market are the same. The rest have relatively low dependence, with some notable exceptions.
\[
\left[\text{Insert Figure \ref{fig:1r2per_Inflation} about here}\right]
\]

For one, we find that Labor Market and Output and Income to be highly dependent around the build up of the sub-prime mortgage bubble and the consequent great financial crisis of 2008/2009. Money and Credit almost has an inverse relationship, with it decreasing during that period and increasing otherwise. On the other hand, we find that, in terms of Prices, there is a gradual increase of Money and Credit and Orders and Inventories. These changes in coefficients, as well as the retrospective dependencies, are indicative of the structural changes in the economy brought on by crises and shocks, showing that recoupling using BPS successfully learns these trends and is able to provide economic interpretability to the analysis, compared to, for example, BMA, which degenerated to one of the groups, or LASSO, which dogmatically shrinks certain factors to zero.

%\newpage
%
%
%The other groups also provide insight to the economy, though not as strong as the others mentioned above.
%For example, we note that Interest Rate and Exchange Rate sees a jump after the dot-com crash and gradually goes down as time goes on.
%Stock Market also sees a jump around the subprime mortgage crisis as well, though much smaller compared to the other subgroups.

\subsection{Finance application: Forecasting Industry Stock Returns}

We consider a large set of predictors to forecast monthly total excess returns across different industries from 1970/1 to 2015/12. The choice of the predictors is guided by previous academic studies and existing economic theory with the goal of ensuring the comparability of our results with these studies (see, e.g., \citeRFSNP{lewellen2004predicting}, \citeRFSNP{Avramov:2004}, \citeRFSNP{Goyal2008}, \citeRFSNP{Goyal2010}, and \citeRFSNP{Dangl:Halling:2012}, among others). We collect monthly data on more than 70 pre-calculated financial ratios for all U.S. companies across eight different categories. Both returns and predictors are aggregated at the industry level by constructing value-weighted returns in excess of the risk-free rate and value-weighted aggregation of the single-firm predictors. Industry aggregation is based on the four-digit SIC codes of the existing firm at each time $t$. We use the ten industry classification codes obtained from Kenneth French's website. Those 70 ratios are classified in eight main categories: Valuation, Profitability, Capitalization, Financial Soundness, Solvency, Liquidity, Efficiency Ratios, and Other. 

Together with industry-specific predictors, we use additional 14 aggregate covariates obtained from existing research, which are divided in two categories; aggregate financials and macroeconomic variables. In particular, following \citeRFS{Goyal2008} and \citeRFS{Goyal2010}, the market-level, aggregate, financial predictors consist of the square root of the sum of daily squared (de-meaned) returns on the value-weighted market portfolio (svar), the ratio of 12-month moving sums of net issues divided by the total end-of-year market capitalization (ntis), the default yield spread (dfy) calculated as the difference between BAA and AAA-rated corporate bond yields, and the term spread (tms) calculated as the difference between the long term yield on government bonds and the Treasury-bill. Additionally, we consider the traded liquidity factor (liq) of \citeRFS{pastor2003}, and the year-on-year growth rate of the amount of loans and leases in Bank credit for all commercial banks. 

As far as the aggregate macroeconomic predictors are concerned, we utilize the inflation rate (infl), measured as the monthly growth rate of the CPI All Urban Consumers index, the real interest rate (rit) measured as the return on the treasury bill minus inflation rate, the year-on-year growth rate of the initial claims for unemployment (icu), the year-on-year growth rate of the new private housing units authorized by building permits (house), the year-on-year growth of aggregate industrial production (ip), the year-on-year growth of the manufacturers' new orders (mno), the M2 monetary aggregate growth (M2), and the year-on-year growth of the consumer confidence index (conf) based on a survey of 5,000 US households. 

%\paragraph{Decouple-recouple specifications.} 
%For the decouple step, we use a static linear regression, for each subgroup, $j=1{:}10$, and each industry, $n=1{:}10$,
%\begin{align}\label{eq:dlmde2}
	%y_{tn}&=\bbeta_{jn}'\x_{tjn}+\epsilon_{tjn}, \quad \epsilon_{tjn}\sim N(0,\nu_{jn}),
%\end{align}
%where $y_{tn}$ is the stock returns in excess of the risk-free rate for the industry $n=1,...,10$, $\x_{tjn}$ the $p$-dimensional vector of predictors clustered in group $j$ for industry $n$, $\bbeta_{jn}$ the corresponding slope parameters, and $\epsilon_{tjn}$ the residual at time $t$ for industry $n$ and conditional on the $j$th group of covariates. For priors on the coefficients and observation variance, we use standard normal-inverse-gamma conjugate priors using the non-informative Jeffreys prior. 

One comment is in order. The DLM specification in Eq.\eqref{eq:dlmde} is attractive due to its parsimony, ease to compute, and the smoothness it induces to the parameters. However, one might argue that an alternative Markov-switching dynamics on the model parameters would be preferable given the equity premium can vary abruptly according to different market phases. However, such dynamics would imply that conditional $\beta$'s are fixed/constant within regimes even though they can differ between regimes. This might appear to be somewhat at odds with the empirical evidence in some areas of financial econometrics and empirical asset pricing more generally, in which time-variation in the parameters is characterized by a series of small changes rather than by a few discrete breaks (see, e.g., \citealt{Jostova:Philipov:2005}, \citealt{Nardari:Scruggs:2007}, \citealt{Adrian:Franzoni:2009}, \citealt{Pastor:Stambaugh:09}, \citealt{binsbergen2010predictive}, \citealt{Pastor:Stambaugh:12}, and \citealt{bianchi2017}, among others). For the recouple step, we follow the synthesis function in Eq.~\eqref{eq:BPSnormalalphadynamic}, with the following priors: $\btheta_{0n}|v_{0n}\sim N(\m_{0n}, (v_{0n}/s_{0n})\I)$ with $\m_{0n}={\bf 0}'$  and $1/v_{0n}\sim G(n_{0n}/2,n_{0n}s_{0n}/2)$ with $n_{0n}=12,s_{0n}=0.01$. The discount factors are $(\beta,\delta)=(0.99,0.95)$.

%\paragraph{Analysis.} 

The empirical application is designed similarly to the macroeconomic study. We used, as training period for the decoupled models, the sample 1970/1-1992/9, fitting the liner regression in a expanding window manner for each industry. Over the period 1992/10-2000/6, we continue the calibration of the recouple strategies, and finally the evaluation period is from 2000/7-2015/12, where we compare the predictive results. We discard the forecast results from 1993/7-2000/12 as training data and compare predictive performance from 2001/1-2000/12. The time frame includes key periods, such as the early 2000s-- marked by the passing of the Gramm-Leach-Bliley act, the inflating and bursting of
the dot.com bubble, the ensuing financial scandals such as Enron and Worldcom and the 9/11 attacks-- and the great financial crisis of 2008/2009, which has been previously led by the burst of the sub-prime mortgage crisis (see, e.g., \citeRFSNP{bianchi2017dissecting}). Arguably, these periods exhibit sharp changes in financial markets, and more generally might lead to in both biases and the dynamics of the latent inter-dependencies among relevant predictors. 
 
Panel A of Table~\ref{tab:Table001} shows that our decouple-recouple strategy using BPS improves the out-of-sample forecasting accuracy relative to the group-specific models, LASSO, PCA, equal-weight averaging, and BMA. Consistent with previous literature, the recursively computed equal-weighted linear-pooling is a challenging benchmark to beat by a large margin. The performance gap between Equal Weight and DRS is not as significant compared to others across industries. The out-of-sample performance of the LASSO and PCA are worse than other competing model combination schemes as well as the HA. These results hold for all the ten industries under investigation. 
\[
\left[\text{Insert Table \ref{tab:Table001} about here}\right]
\]

Similar to the macroeconomic study, the performance gap in favor of DRS is quite luminous related to the log predictive density ratios. In fact, as seen in Panel B of Table~\ref{tab:Table001}, none of the alternative specifications come close to DRS when it comes to predicting one-step ahead. With the only partial exception of the Energy sector, DRS strongly outperforms both the competing model combination/shrinkage schemes and the group-specific predictive densities. Two comments are in order. First, while both the equal-weight linear pooling and the sequential BMA tend to outperform the group-specific predictive regressions, the LASSO strongly underperforms when it comes to predicting the density of future excess returns. This result is consistent with the recent evidence in \citeRFS{diebold2017}. They show that simple average combination schemes are highly competitive with respect to standard LASSO shrinkage algorithm. In particular, they show that good out-of-sample performances are hard to achieve in real-time forecasting exercise, due to the intrinsic difficulty of small-sample real-time cross validation of the LASSO tuning parameter. %In fact, only the historical mean can out-perform a simple equal-weight linear pooling of group-specific predictive densities. 

\begin{figure}[p]
%\captionsetup[subfigure]{position=top}
\caption{US equity return forecasting: Out-of-sample log predictive density ratio}
\centering
  \vspace{0.1in}
\begin{justify} 
\footnotesize{This figure shows the dynamics of the out-of-sample Log Predictive Density Ratio (LPDR) as in Eq.~\eqref{eq:lpdr} obtained for each of the group-specific predictors, by taking the historical average of the stock returns (HA), and the results from a set of competing model combination/shrinkage schemes, e.g., LASSO, Equal Weight, and Bayesian Model Averaging (BMA). For the ease of exposition we report the results for four representative industries, namely, Consumer Durables, Consumer Non-Durables, Telecomm, Health, Shops, and Other. Industry aggregation is based on the four-digit SIC codes of the existing firm at each time $t$ following the industry classification from Kenneth French's website. The sample period is 01:1970-12:2015, monthly.}
\end{justify}
\hspace*{-3em} 
\subfigure[Consumer Durable]{\includegraphics[width=0.5\textwidth]{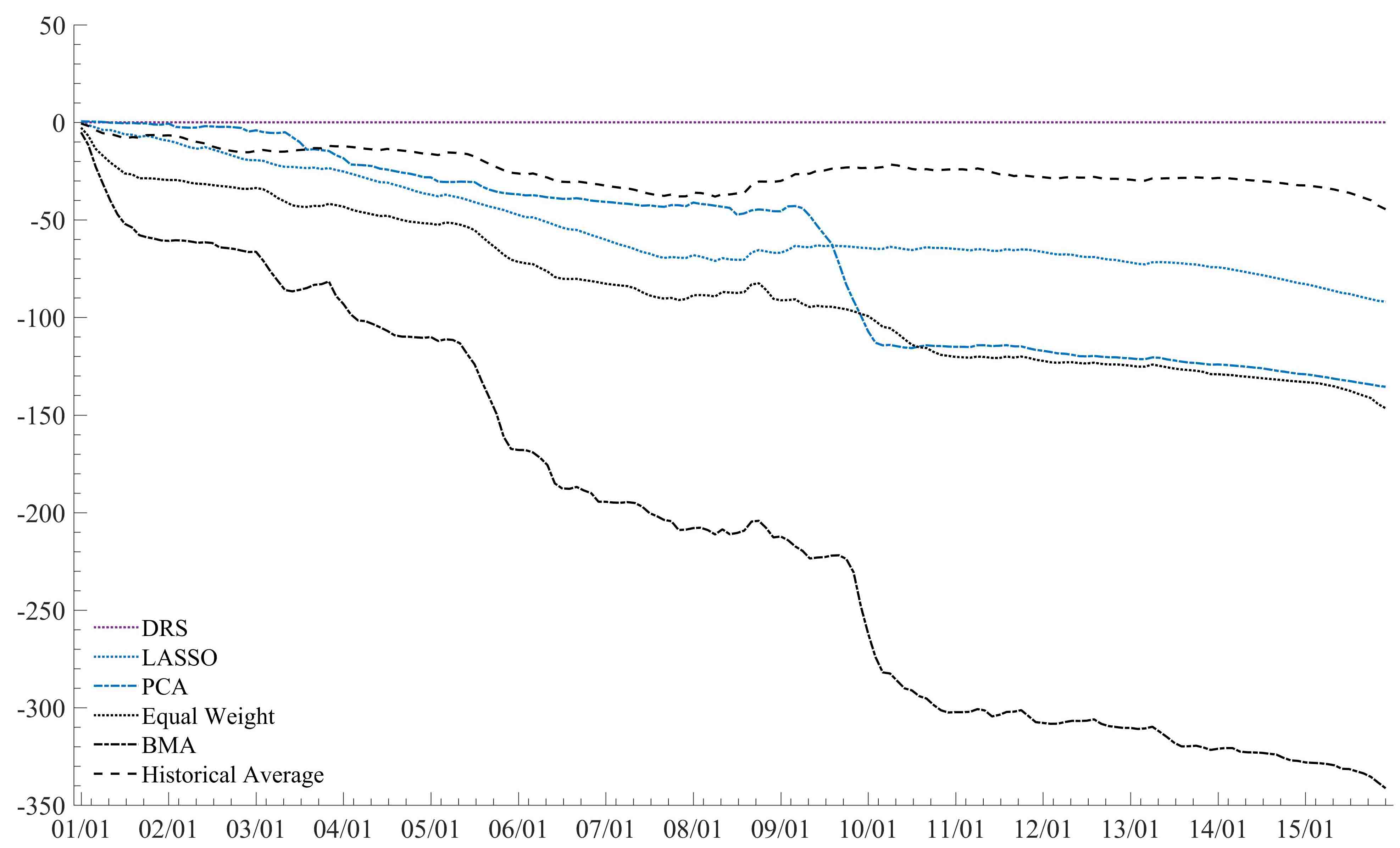}}
\subfigure[Cons. Non-Durable]{\includegraphics[width=0.5\textwidth]{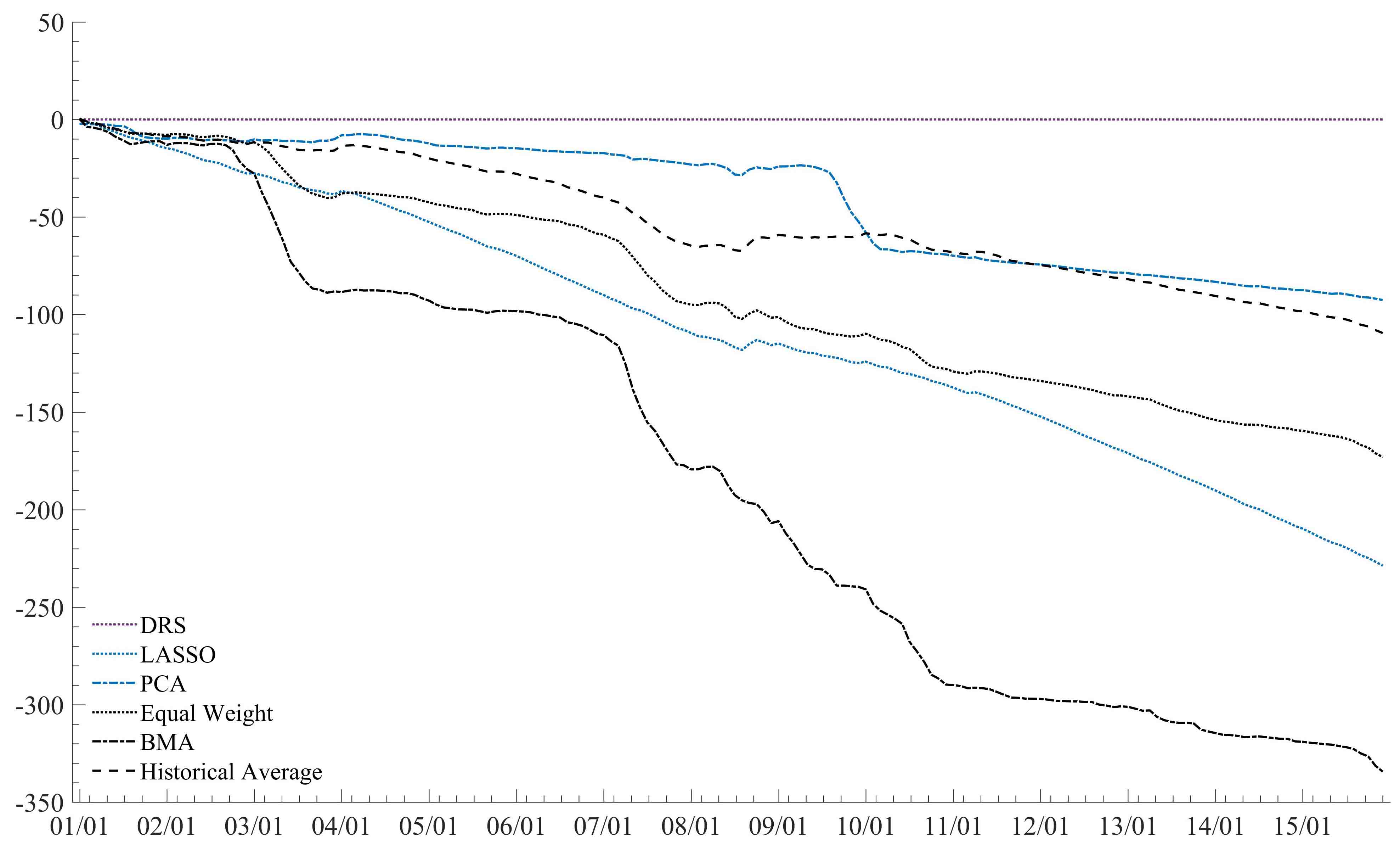}\hspace*{-2.5em}} \\
\hspace*{-3em} 
\subfigure[Telecomm]{\includegraphics[width=0.5\textwidth]{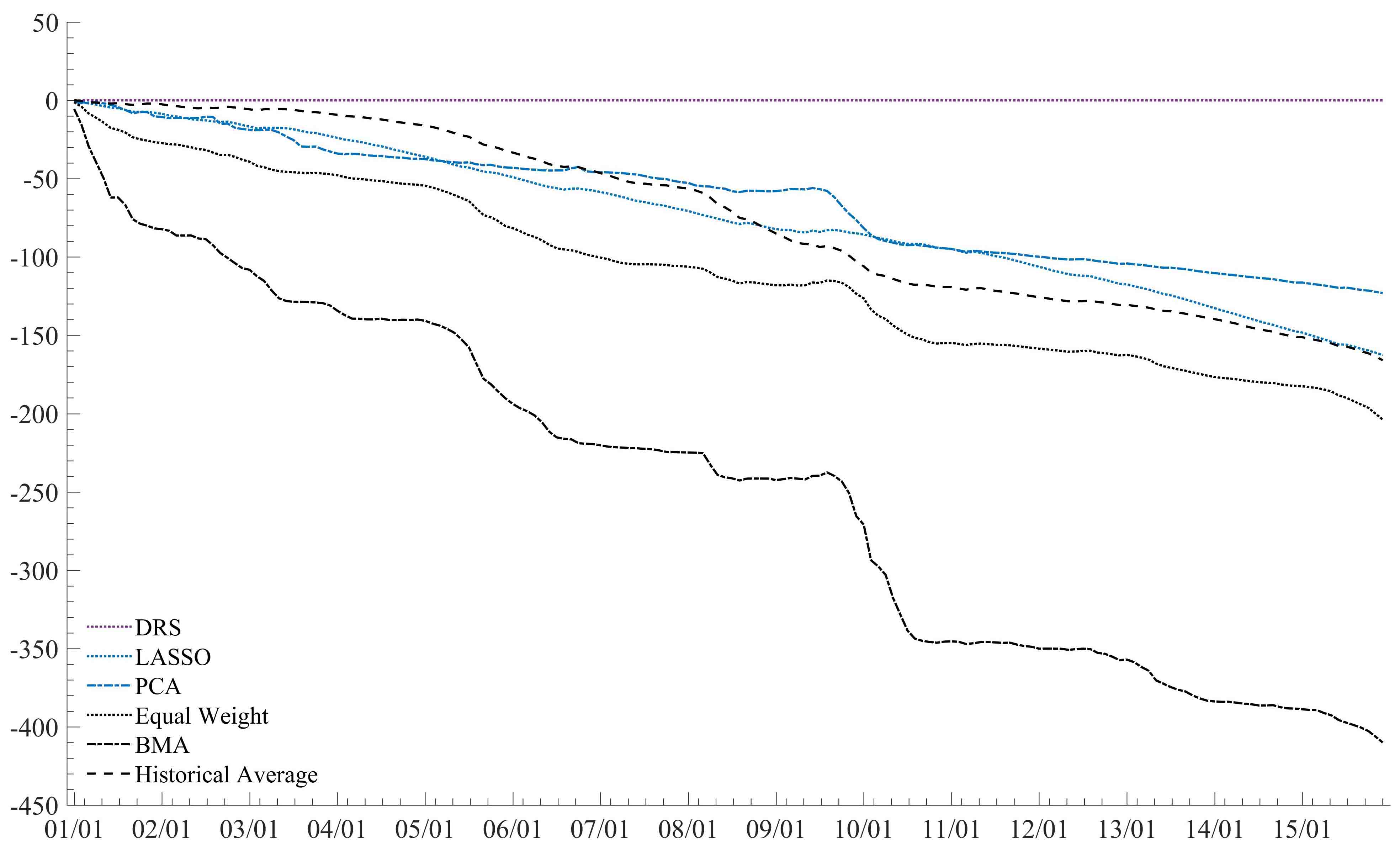}}
\subfigure[Other]{\includegraphics[width=0.5\textwidth]{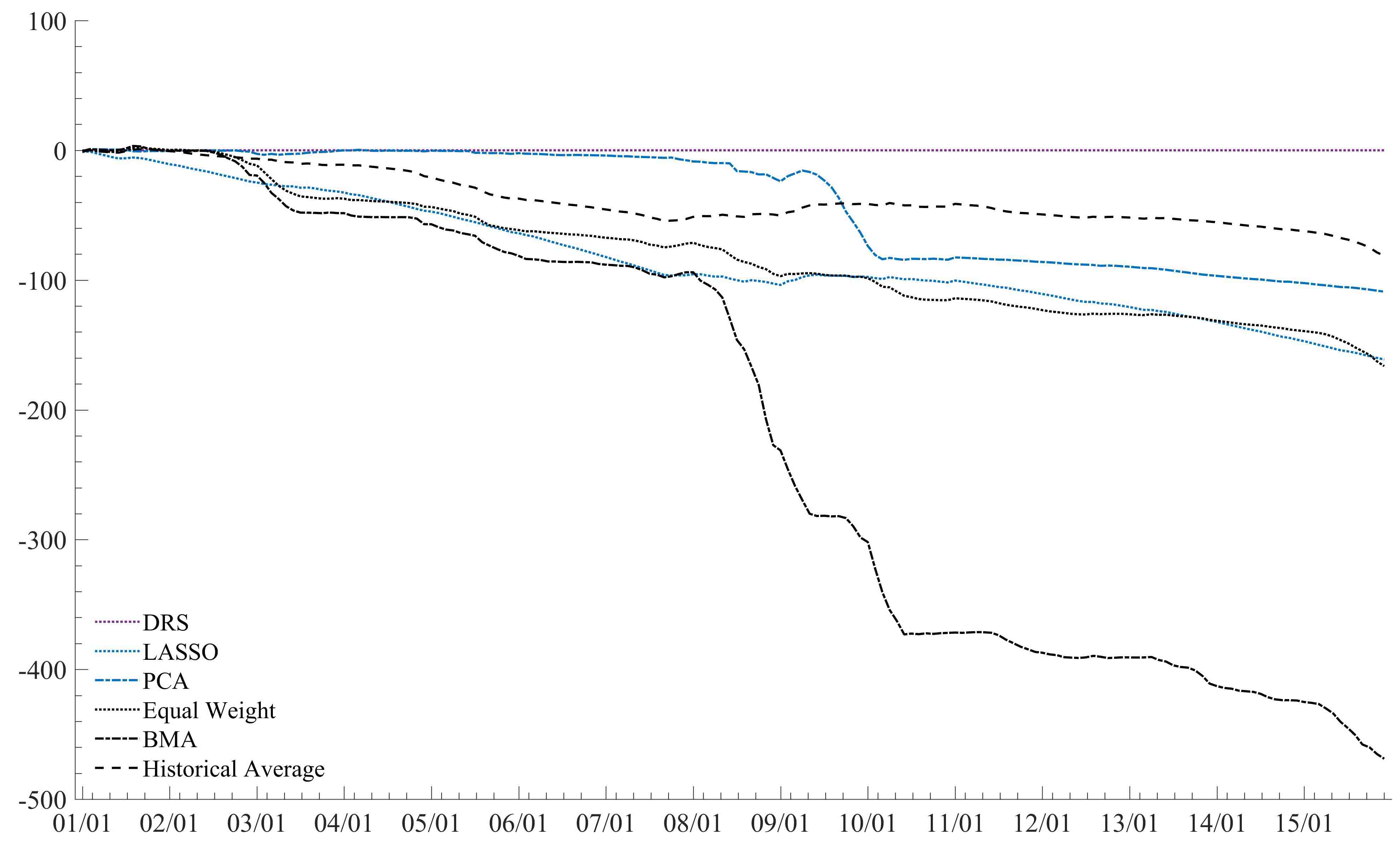}\hspace*{-2.5em}}\\  
\hspace*{-3em} 
\subfigure[Health]{\includegraphics[width=0.5\textwidth]{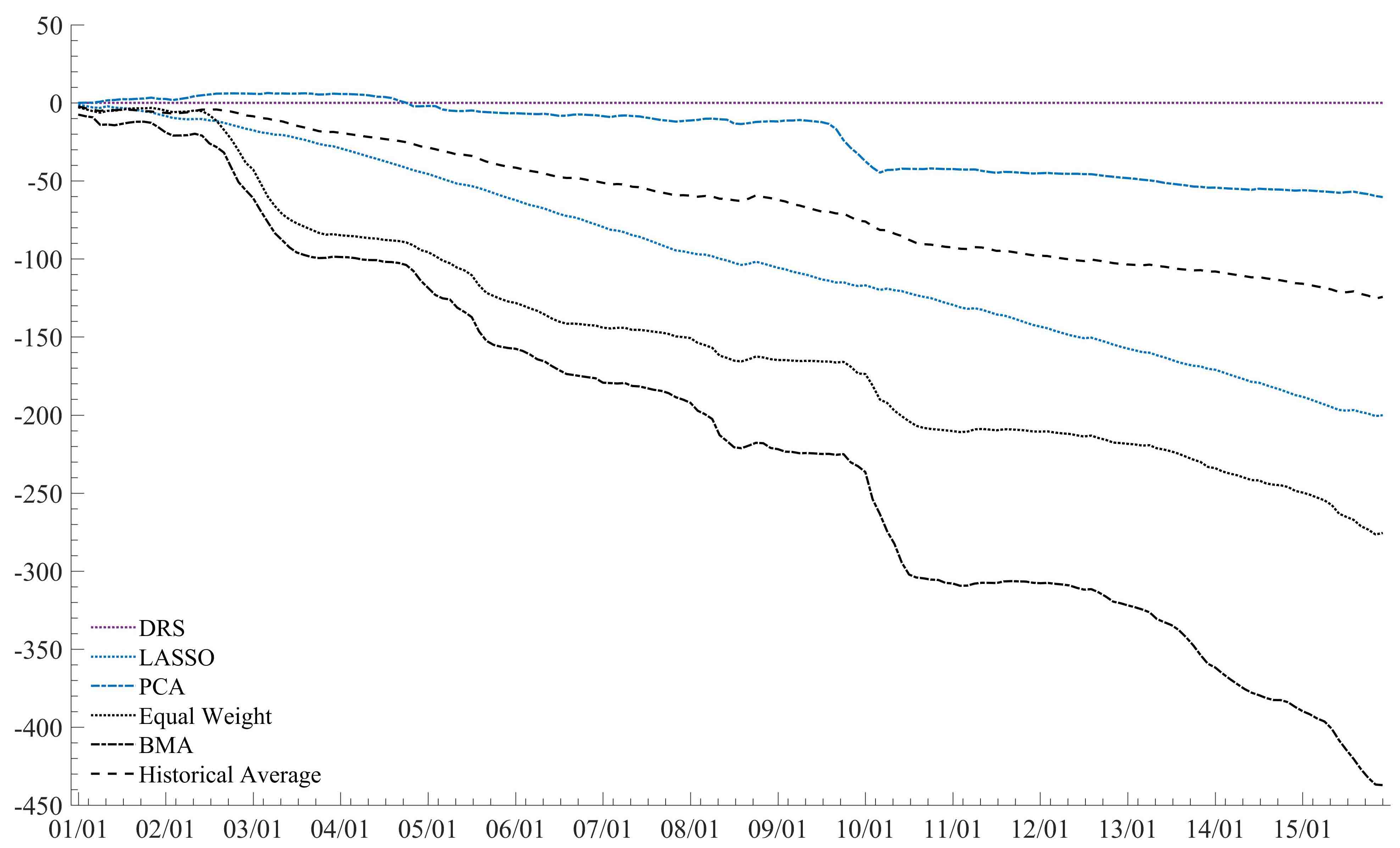}}
\subfigure[Shops]{\includegraphics[width=0.5\textwidth]{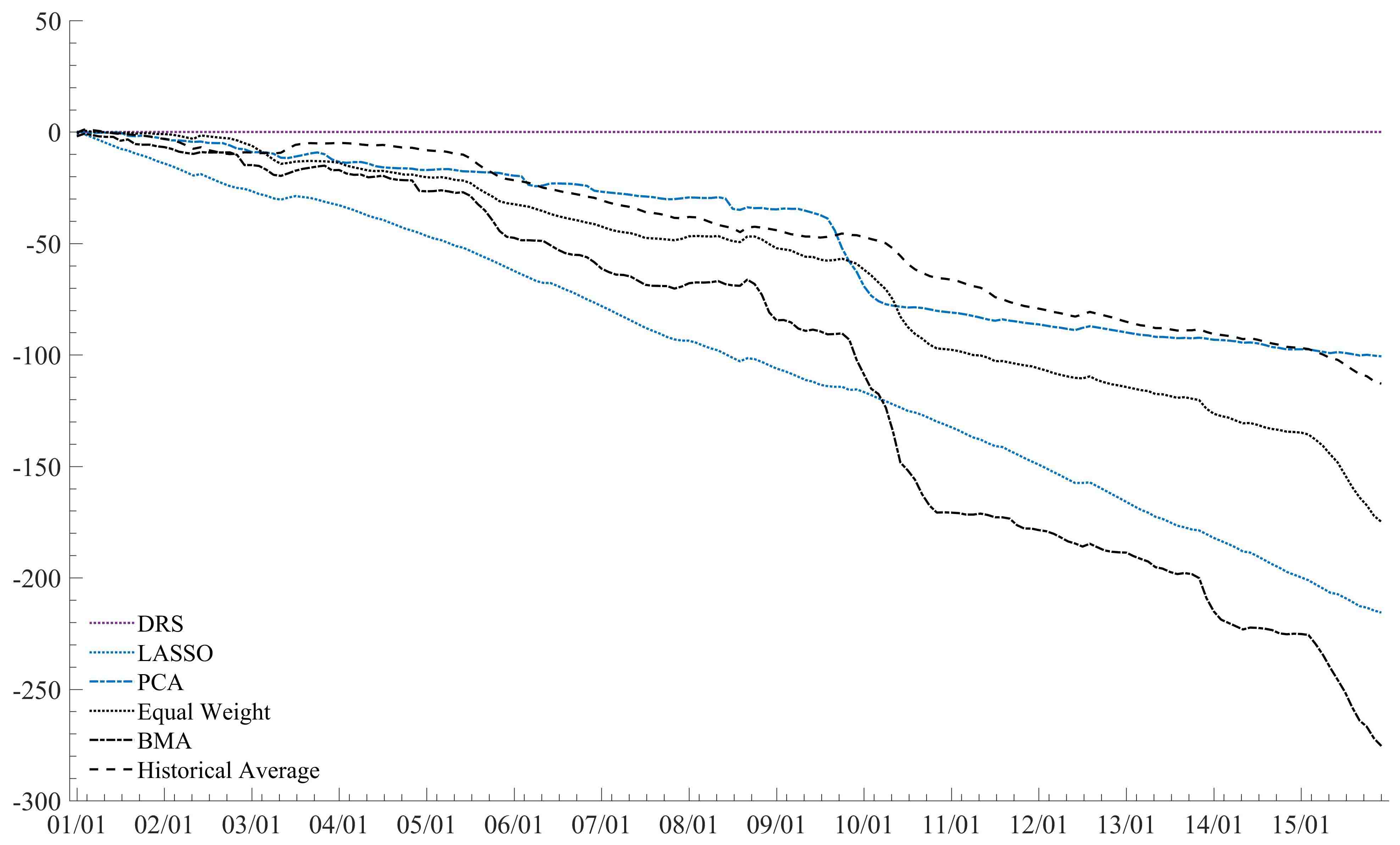}\hspace*{-2.5em}}  
  \label{fig:LPDR_Finance}
 \end{figure}

Delving further into the dynamics of the LPDR, Figure~\ref{fig:LPDR_Finance} shows the whole out-of-sample path of density forecasting accuracy across modeling specifications. For the ease of exposition we report the results for Consumer Durable, Consumer Non-Durable, Manufacturing, Telecomm, HiTech, and Other industries. The results for the remaining industries are quantitatively similar and available upon request. Top-left panel shows the out-of-sample path for the Consumer Durable sector. The DRS compares favorably against alternative predictive regressions. Similar results are also evident in other sectors. As a whole, Figure \ref{fig:LPDR_Finance} shows clear evidence of how the competing model combination/shrinkage schemes possibly fails to rapidly adapt to structural changes. Although the performance, pre-crisis, is good, it is notable that there is a large loss in predictive performance after the great recession in 2008/2009. DRS consistently shows a performance robust to shifts and shocks and stays in the best group of forecasts throughout the testing sample.
\[
\left[\text{Insert Figure \ref{fig:LPDR_Finance} about here}\right]
\]

The out-of-sample performance of the LASSO sensibly deteriorates when it comes to predicting the overall one-step ahead distribution of excess returns. The equal-weight linear-pooling turns out to out-perform the competing combination schemes but DRS, as well as the group-specific predictive regressions. Arguably, the strong outperformance of DRS is due to its ability to quickly adjust to different market phases and structural changes in the latent inter-dependencies across groups of predictors, as highlighted by the DLM-type of dynamics in Eqs.~\eqref{DLM}. 

\begin{figure}[p]
%\captionsetup[subfigure]{position=top}
\caption{US equity return forecasting: On-line posterior means of synthesis coefficients}
\centering
  \vspace{0.1in}
\begin{justify} 
\footnotesize{This figure shows the one-step ahead latent interdependencies across groups of predictive densities-- measured through the predictive coefficients-- used in the recoupling step. For the ease of exposition we report the results for four representative industries, namely, Consumer Durables, Consumer non-Durables, Manufacturing, Shops, Utils and Other. Industry aggregation is based on the four-digit SIC codes of the existing firm at each time $t$ following the industry classification from Kenneth French's website. The sample period is 01:1970-12:2015, monthly.}
\end{justify}
%\vspace*{-2em}
\hspace*{-4em} 
\subfigure[Consumer Durable]{\includegraphics[width=0.5\textwidth]{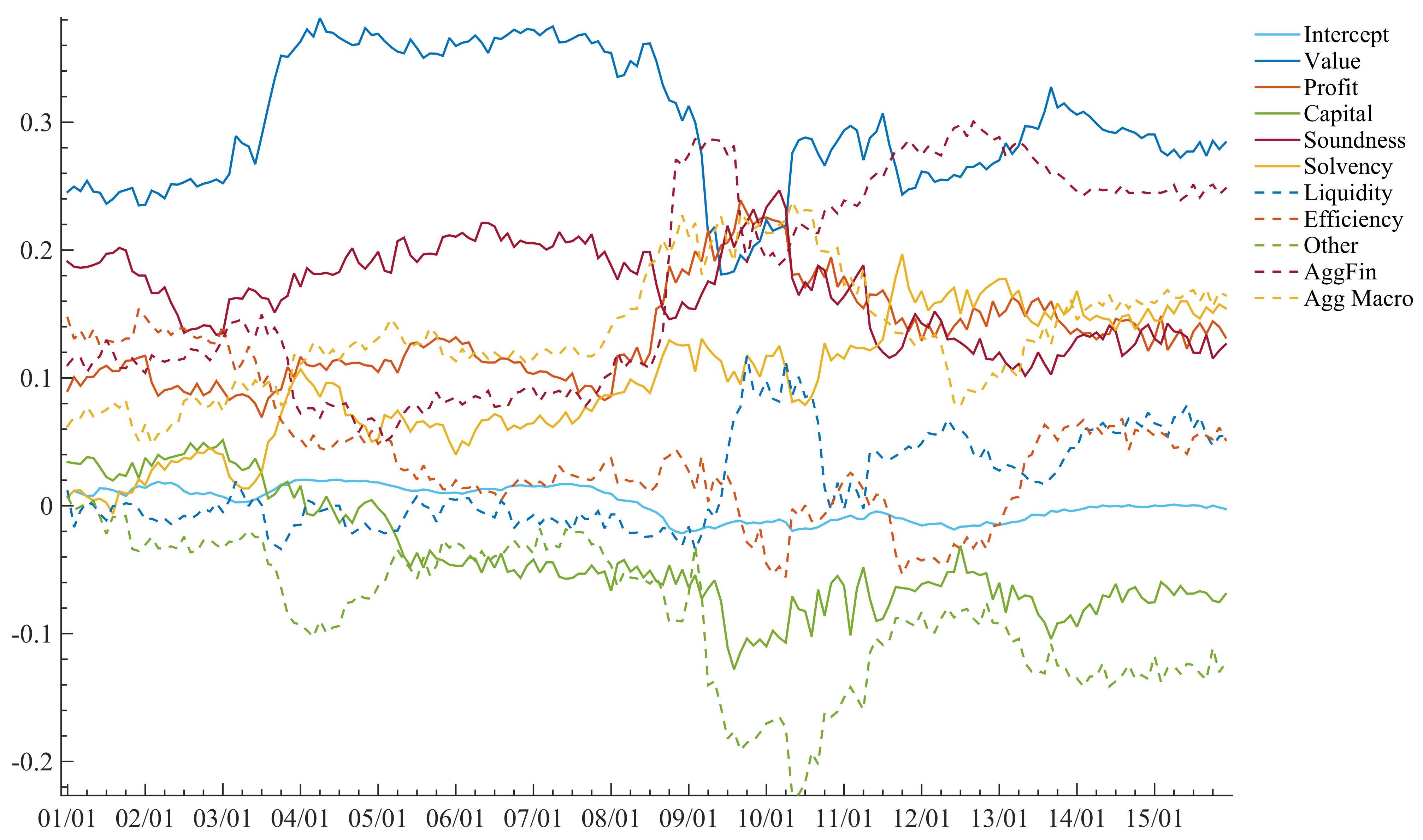}}
\hspace*{-1em}  
\subfigure[Cons. Non-Durable]{\includegraphics[width=0.5\textwidth]{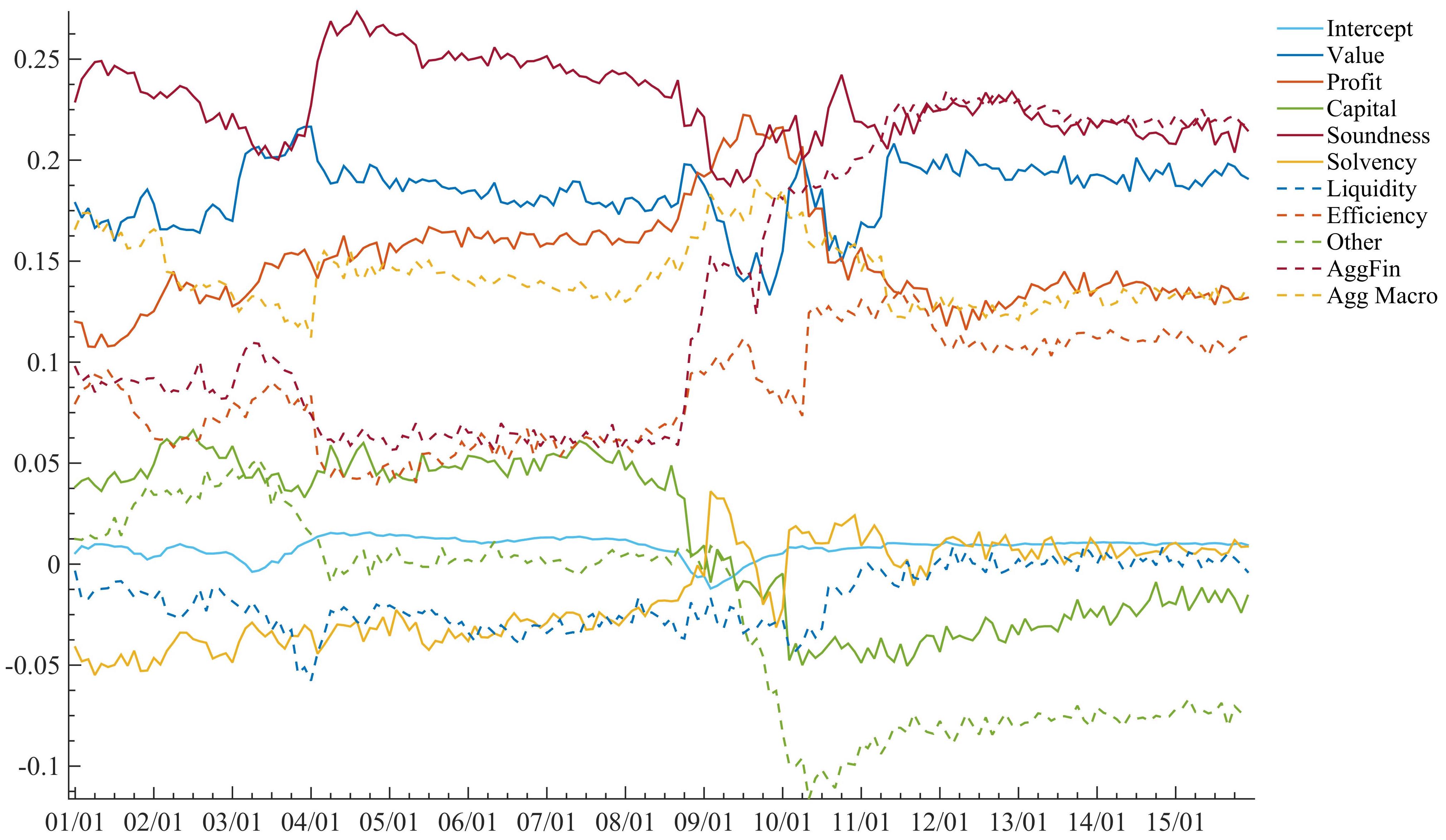}\hspace*{-2.5em}} \\
\hspace*{-4em} 
\subfigure[Manufacturing]{\includegraphics[width=0.5\textwidth]{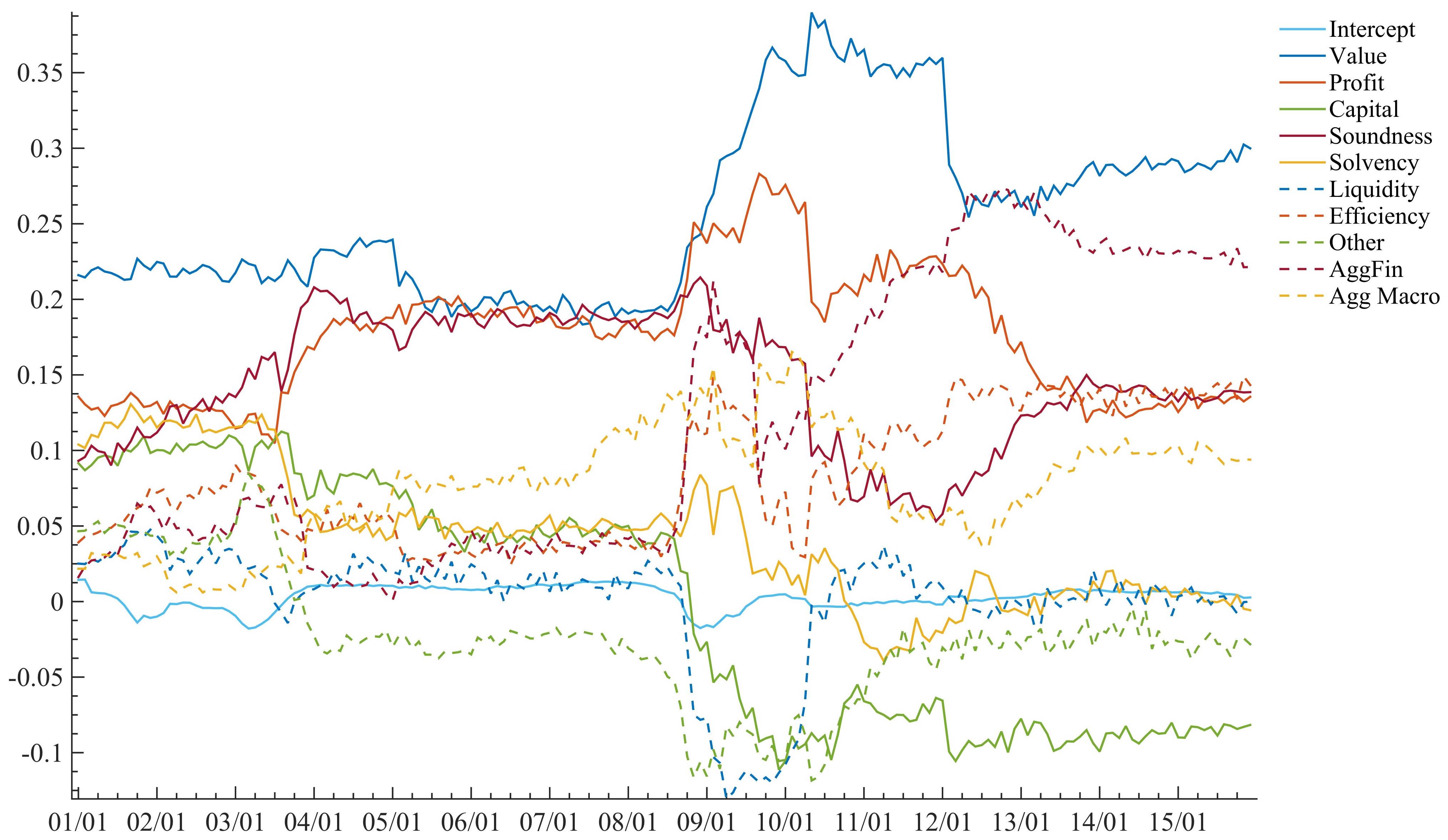}}
\hspace*{-1em}  
\subfigure[Other]{\includegraphics[width=0.5\textwidth]{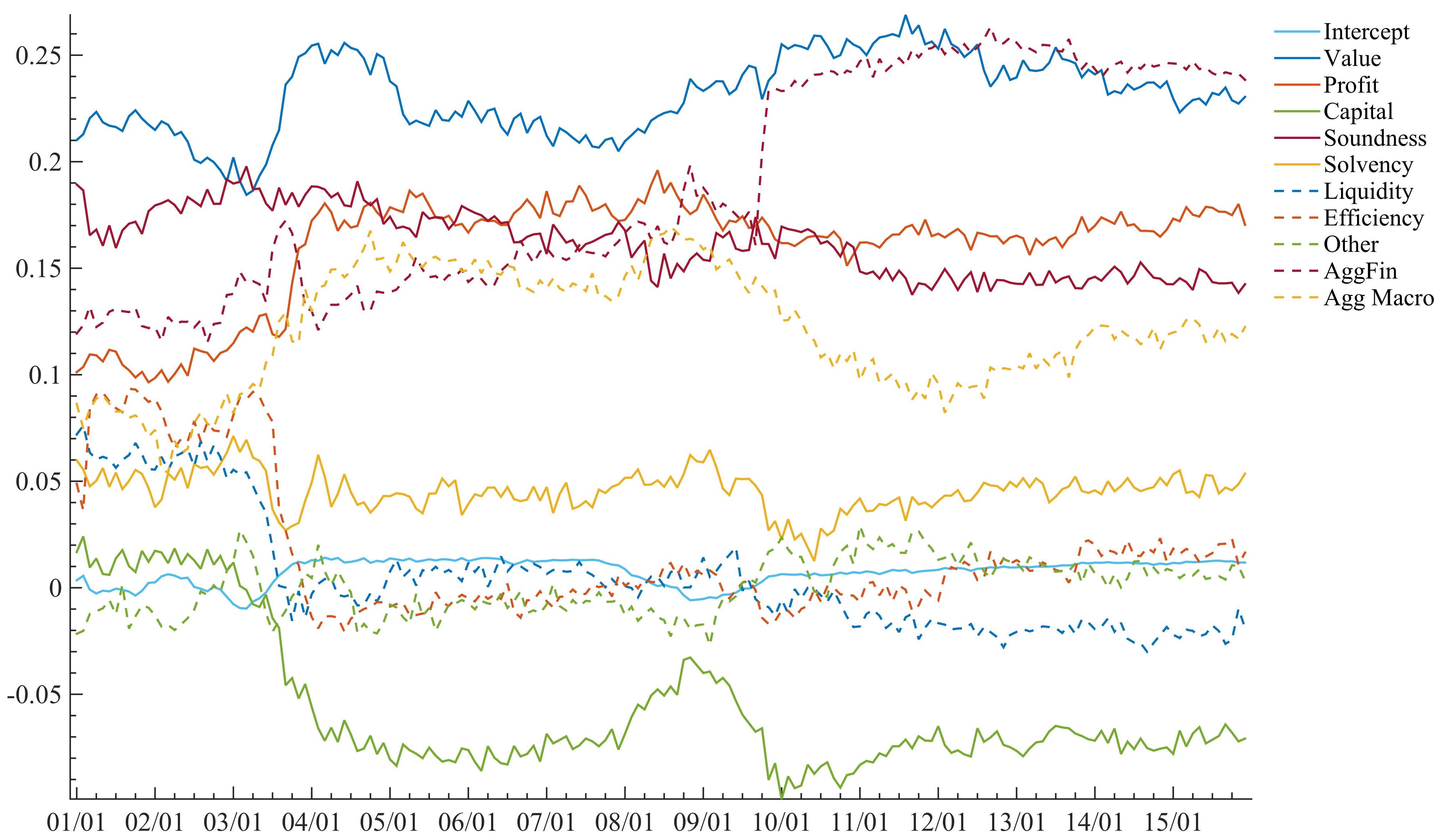}\hspace*{-2.5em}}\\
\hspace*{-4em} 
\subfigure[Utils]{\includegraphics[width=0.5\textwidth]{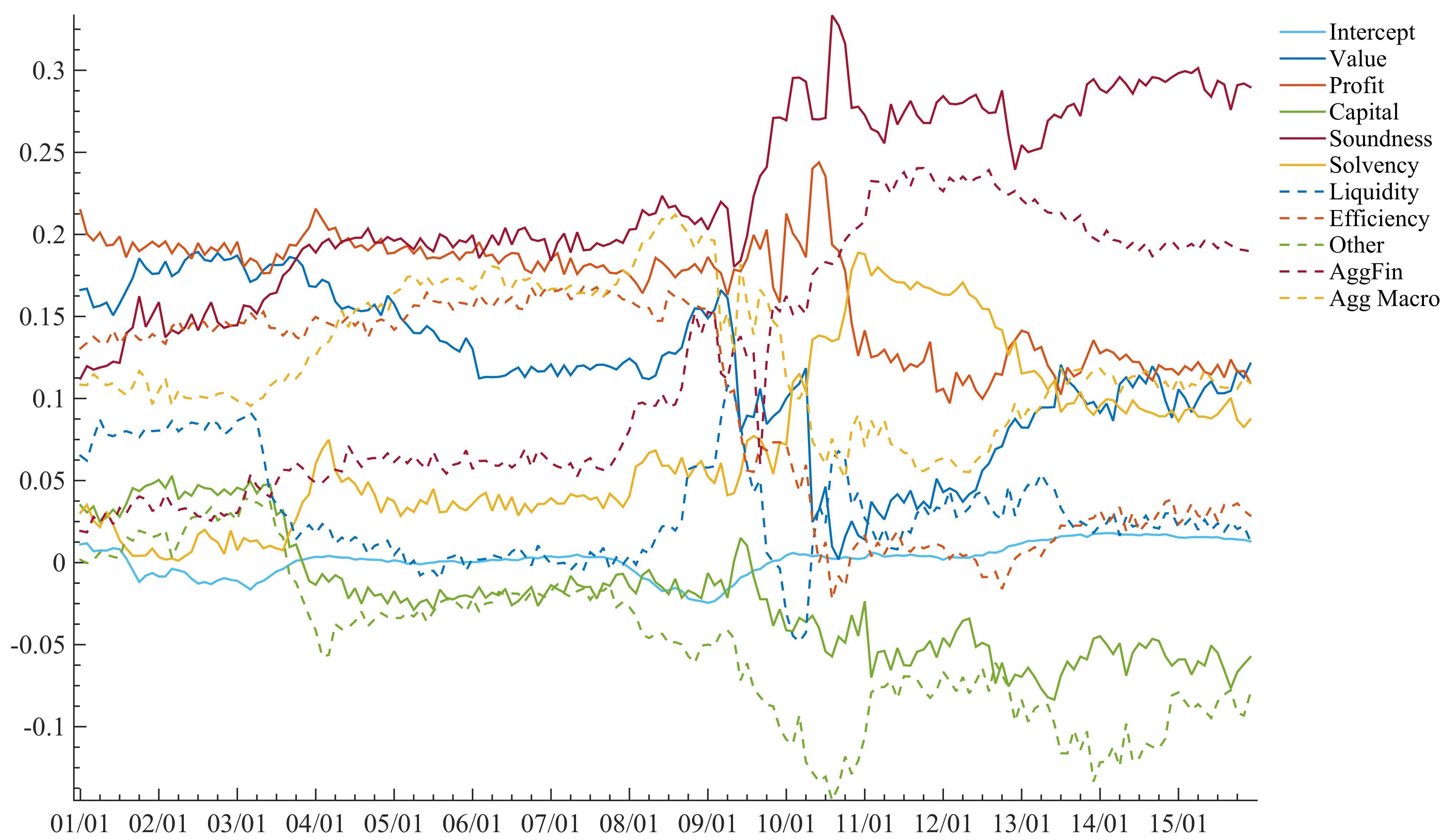}}
\hspace*{-1em}  
\subfigure[Shops]{\includegraphics[width=0.5\textwidth]{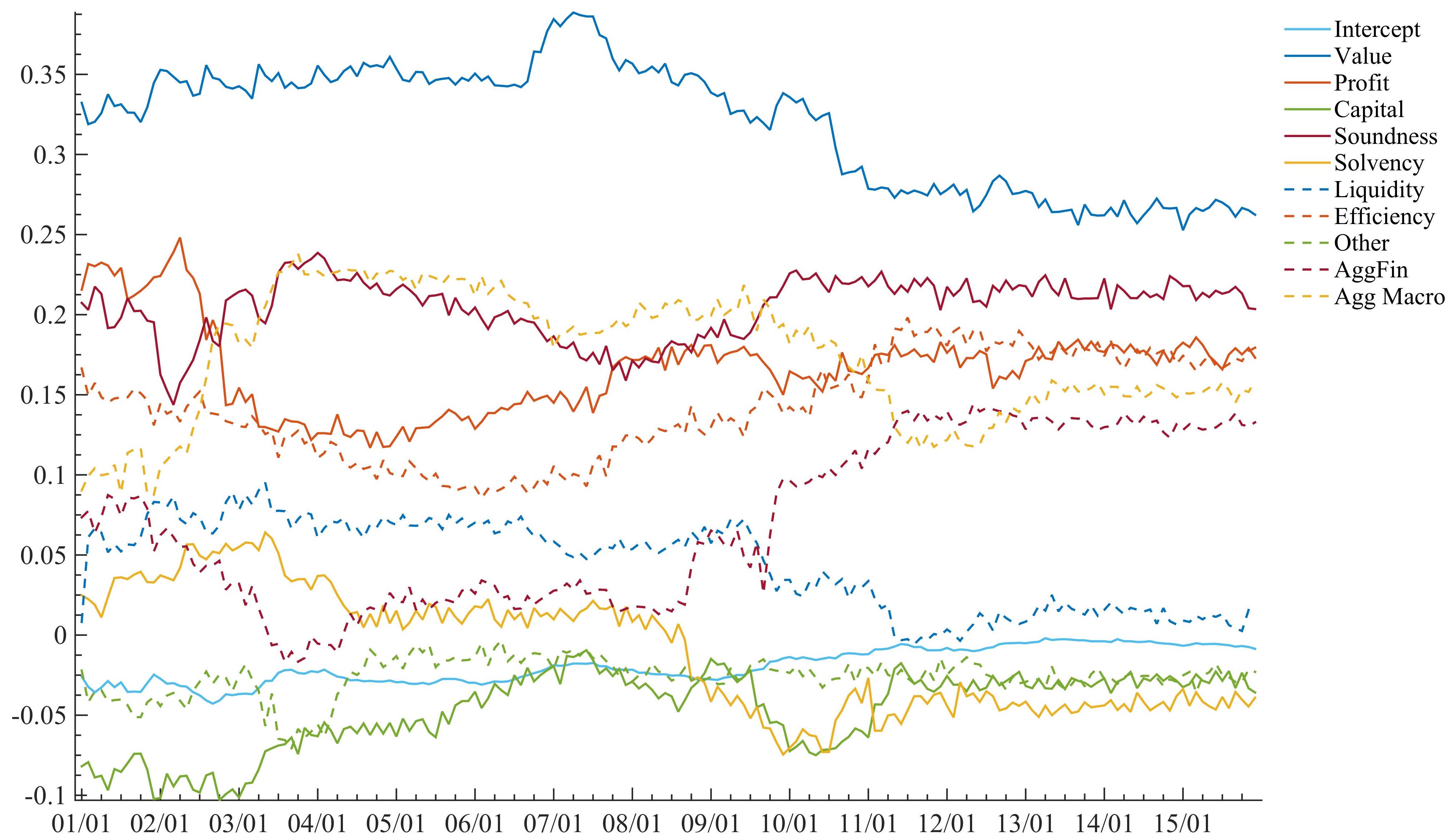}\hspace*{-2.5em}}  
  \label{fig:BPS_Coef_Finance}
 \end{figure}

Figure \ref{fig:BPS_Coef_Finance} shows that, indeed, the flexibility in the DRS coefficients is quite significant, and some interesting aspects related to returns predictability emerge. For instance, the role of Valuation and Financial Soundness highly fluctuates around the financial crisis for both Consumer durables and Consumer non-durables. Financial Soundness indicators involve variables such as cash flow over total debt, short-term debt over total debt, current liabilities over total liabilities, long-term debt over book equity, and long-term debt over total liabilities, among others. These variables assess a company's risk level in the medium-to-long term as evaluated in relation to the company's debt level, and therefore collectively capture the ability of a company to manage its outstanding debt effectively keeping to keep its operating ability. Quite understandably, the interplay between debt (especially medium term debt) and market value increasingly affect risk premia, and therefore the predicted value of future excess returns in a significant manner. 
\[
\left[\text{Insert Figure \ref{fig:BPS_Coef_Finance} about here}\right]
\]

Although there are some similarities in the DRS coefficients across industries, some cross-sectional heterogeneity emerge as well. As a matter of fact, while Valuation and Capitalization tend to dominate for the Other sector, R\&D expenses-- which falls within the Other category in the clusters of predictors-- turns out to be quite relevant for Consumer durables. By looking at the actual composition of predictors and industries these trends turn out to be fully consistent with economic theory. Take the Other sector as an example; in the 10-industry classification we used, the market capitalization of the Other sector is primarily driven by business services, constructions, building materials, financial services, and banking. The capitalization of all these industries, especially the banking and finance sector, has been significantly affected in the aftermath of the great financial crisis. On the one hand, anecdotal evidence and policy making commentaries highlighted how the increasing burden, due to a huge amount of non-performing loans in the banking sectors, generated a liquidity contraction, which ultimately affected those sectors more dependent on bank financing, such as construction and building materials. On the other hand, while the regime of low policy rates might have, in the short term, helped to prevent a disorderly adjustment of balance sheets in distressed banks and provided relief in terms of lower interest payments in those more exposed to mortgages, they also weakened the incentive to repair balance sheets of banks and building societies in the first place. To summarize, the massive amount of non-performing loans and the subsequent liquidity issues, coupled with moral-hazard issues represented significant sources of capital risk for the Other industry and its components. 

The instability in the cross-sectional latent dependencies across group regressors over time is quite evident from Figures \ref{fig:BPS_Coef_Finance}, which shows the covariances across the group-specific predictive densities at different times in the out-of-sample period. However, it should be clear that our goal here is not to over-throw other results from the empirical finance literature with respect to the correlation among predictors, but to deal with the crucial aspect of modeling the dynamic interplay between different, economically motivated, predictive densities in forecasting excess stock returns. Our results show that this is not the case: there is a substantial time-variation and cross-sectional heterogeneity in the marginal predictive power of different groups of covariates. 

So far we have compared the statistical performance of return forecasts generated by economically constrained investors. We next evaluate the economic significance of these return forecasts by considering the optimal portfolio choice of an investor who uses the return forecasts. An advantage of our approach is that it accounts for both parameter and model uncertainty, whose importance has been emphasized in the existing empirical research (see, e.g., \citealt{Barberis:2000}, \citealt{avramov2002stock}, \citealt{Goyal2010}, \citealt{billio2013}, and \citealt{pettenuzzo2014}, among others). Moreover, our approach provides the full predictive density, which means that we are not reduced to considering only mean-variance utility, but can use more general constant relative risk aversion preferences. In particular, we construct a two asset portfolio with a risk-free asset ($r_t^f$) and a risky asset ($y_t$; industry returns) for each $t$, by assuming the existence of a representative investor that needs to solve the optimal asset allocation problem
\begin{align}
\omega_\tau^{\star} & = \arg\underset{w_\tau}{\max}E\left[U\left(\omega_\tau,y_{\tau+1}\right)|\mH_\tau\right],
\end{align}
with $\mH_\tau$ indicating all information available up to time $\tau$, and $\tau=1,...,t$. The investor is assumed to have power utility 
\begin{align}
U\left(\omega_\tau,y_{\tau+1}\right) & = \frac{\left[\left(1-\omega_\tau\right)\exp\left(r_\tau^f\right) + \omega_\tau\exp\left(r_\tau^f + y_{\tau+1}\right)\right]^{1-\gamma}}{1-\gamma},
\end{align}
here, $\gamma$ is the investor's coefficient of relative risk aversion. The time $\tau$ subscript reflects the fact that the investor chooses the optimal portfolio allocation conditional on his available information set at that time. Taking expectations with respect to the predictive density in Eq.~\eqref{eq:theorem}, we can rewrite the optimal portfolio allocation as
\begin{align}
\omega_\tau^{\star} & = \arg\underset{\omega_\tau}{\max}\int U\left(\omega_\tau,y_{\tau+1}\right)p(y_{\tau+1}|\mH_\tau)dy_{\tau+1},
\label{eq:portfolio}
\end{align}
As far as DRS is concerned, the integral in Eq.~\eqref{eq:portfolio} can be approximated using the draws from the predictive density in Eq.~\eqref{eq:theorem}. The sequence of portfolio weights $\omega_\tau^{\star}, \tau=1,...,t$ is used to compute the investor's realized utility for each model-combination scheme. Let $\hat{W}_{\tau+1}$ represent the realized wealth at time $\tau+1$ as a function of the investment decision, 
\begin{align}
\hat{W}_{\tau+1} & = \left[\left(1-\omega_\tau^{\star}\right)\exp\left(r_\tau^f\right) + \omega_\tau^{\star}\exp\left(r_\tau^f + y_{\tau+1}\right)\right],
\label{wealth}
\end{align}
The certainty equivalent return (CER) for a given model is defined as the annualized value that equates the average realized utility. We follow \cite{pettenuzzo2014} and compare the the average realized utility of DRS $\hat{U}_{\tau}$ to the average realized utility of the model based on the alternative predicting scheme $i$, over the forecast evaluation sample:
\begin{align}
CER_i & = \left[\frac{\sum_{\tau=1}^{t}\hat{U}_{\tau,i}}{\sum_{\tau=1}^{t}\hat{U}_{\tau}}\right]^{\frac{1}{1-\gamma}}-1,
\end{align}
with the subscript $i$ indicating a given model combination scheme, $\hat{U}_{\tau,i}=\hat{W}_{\tau,i}^{1-\gamma}/(1-\gamma)$, and $\hat{W}_{\tau,i}$ the wealth generated by the competing model $i$ at time $\tau$ according to Eq.~\eqref{wealth}. Panel A of Table \ref{tab:Table002} shows the results for portfolios with unconstrained weights, i.e. short sales are allowed to maximize the portfolio returns. We assume a risk aversion coefficient equal to 5.   
\[
\left[\text{Insert Table \ref{tab:Table002} about here}\right]
\]

The economic performance of our decouple-recouple strategy is rather stark in contrast to both group-specific forecasts and the competing model combination schemes. The realized CER from DRS is much larger than virtually any of the competing model specifications across different industries. Not surprisingly, given the statistical accuracy of a simple recursive historical mean model is not remarkable, the HA model leads to a very low CER. Interestingly, the equally-weighted linear pooling and a Bayesian model averaging approach in which the forecast densities are mixed with respect to sequentially updated model probabilities turns out to be a strong competitor, although still generates lower CERs. 

Panel B of Table \ref{tab:Table002} shows that the performance gap in favor of DRS is again confirmed under the restriction that the weights have to be positive, i.e., long only strategy with no-short sales allowed. Consistent with existing literature (see, e.g., \citeRFSNP{jagannathan2003}, and \citeRFSNP{demiguel2007}), the economic performance obtained by restricting the portfolio weights tend to improve across different industries, regardless the model combination scheme. Yet, our decouple-recouple model synthesis scheme allows a representative investor to obtain a larger performance than BMA and equal-weight linear pooling. Notably, both the performance of the LASSO and the PCA substantially improve by imposing no-short sales constraints. 

In addition to evaluating the economic values of various predicting models over the full forecast evaluation sample, we also study how the different models perform in real time. Specifically, we first calculate the $CER_{i\tau}$ at each time $\tau$ as
\begin{align}
CER_{i\tau} & = \left[\frac{\hat{U}_{\tau,i}}{\hat{U}_{\tau}}\right]^{\frac{1}{1-\gamma}}-1,
\label{eq:single}
\end{align}
Panel A of Table \ref{tab:Table004} shows the average annualized, single-period CER for the forecasting sample for an unconstrained investor. The results show that the out-of-sample performance is robustly in favor of the DRS model-combination scheme. As for the whole-sample CER reported in Table \ref{tab:Table002}, the equal-weighted linear pooling of predictive densities turns out to be a challenging benchmark to beat. Yet, DRS generates constantly higher average CERs across the forecasting sample. 
\[
\left[\text{Insert Table \ref{tab:Table004} about here}\right]
\]

\begin{figure}[p]
%\captionsetup[subfigure]{position=top}
\caption{US equity return forecasting: Out-of-sample cumulative CER without Constraints}
\centering
  \vspace{0.1in}
\begin{justify} 
\footnotesize{This figure shows the dynamics of the out-of-sample Cumulative Certainty Equivalent Return (CER) for an unconstrained as in Eq.~\eqref{eq:cumulative} obtained for each of the group-specific predictors, by taking the historical average of the stock returns (HA), and the results from a set of competing model combination/shrinkage schemes, e.g., LASSO, Equal Weight, and Bayesian Model Averaging (BMA). For the ease of exposition we report the results for four representative industries, namely, Consumer Durables, Consumer Non-Durables, Telecomm, Health, Shops, and Other. Industry aggregation is based on the four-digit SIC codes of the existing firm at each time $t$ following the industry classification from Kenneth French's website. The sample period is 01:1970-12:2015, monthly.}
\end{justify}
\hspace*{-3em} 
\subfigure[Consumer Durable]{\includegraphics[width=0.5\textwidth]{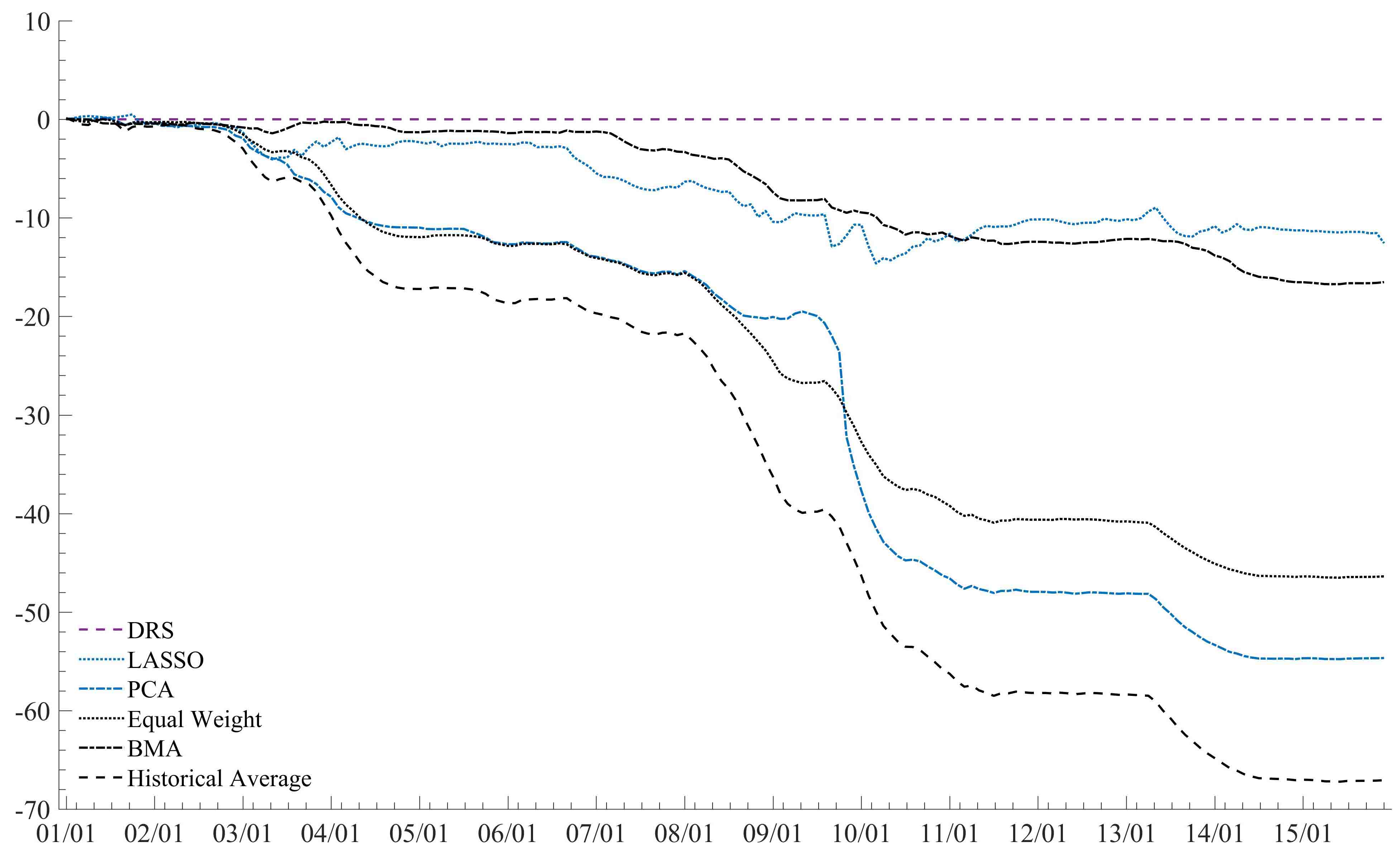}}
\subfigure[Cons. Non-Durable]{\includegraphics[width=0.5\textwidth]{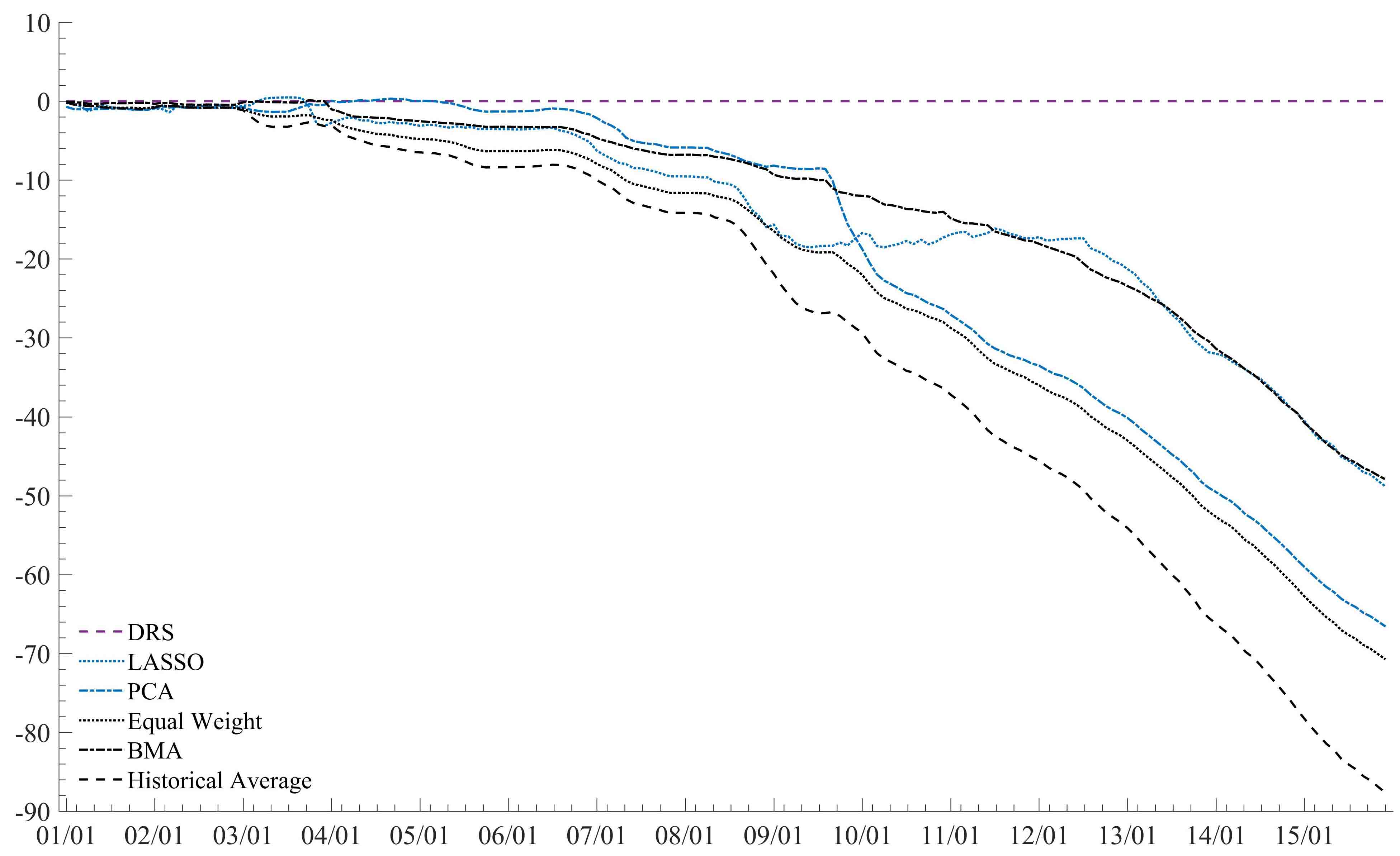}\hspace*{-2.5em}} \\
\hspace*{-3em} 
\subfigure[Telecomm]{\includegraphics[width=0.5\textwidth]{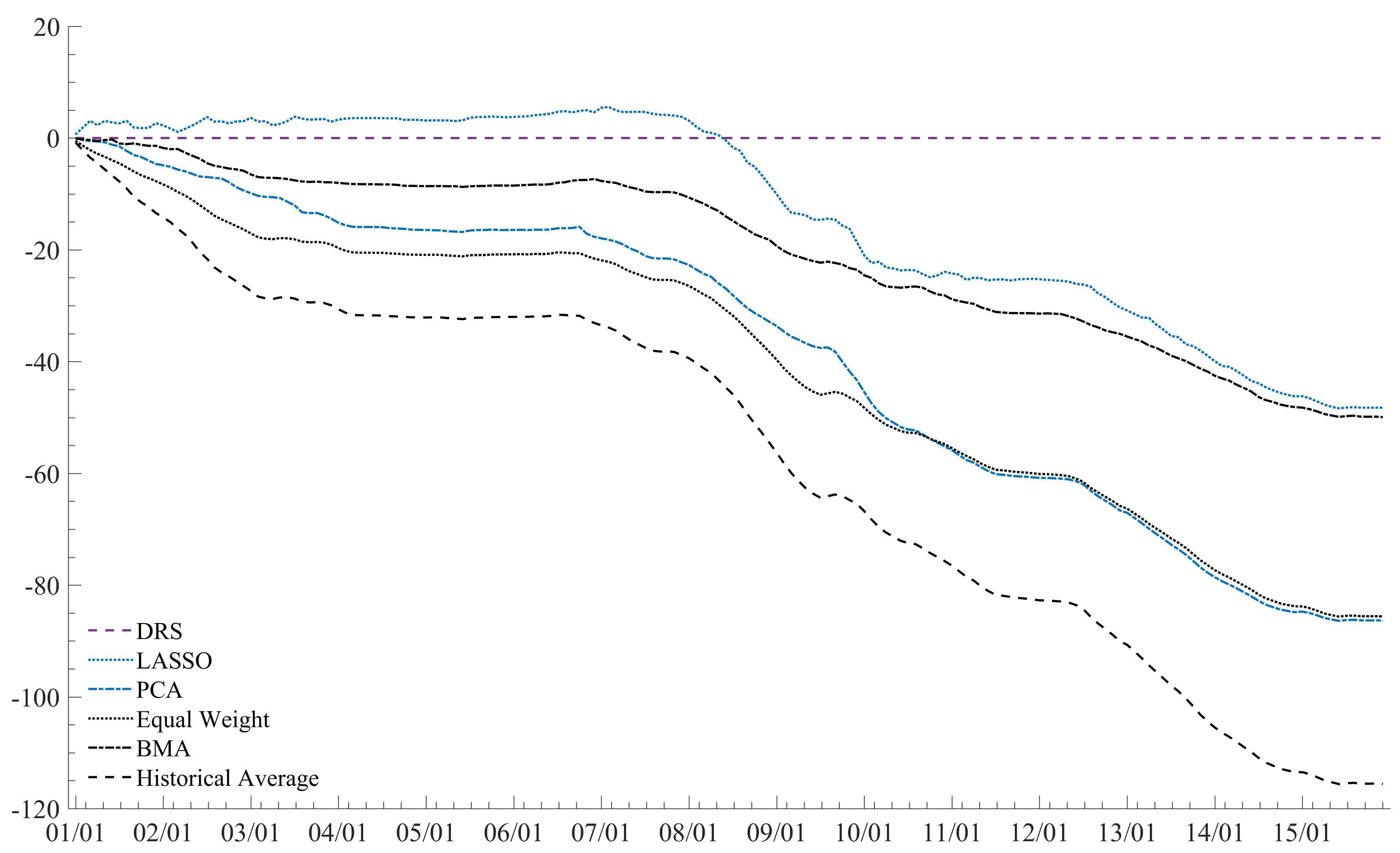}}
\subfigure[Other]{\includegraphics[width=0.5\textwidth]{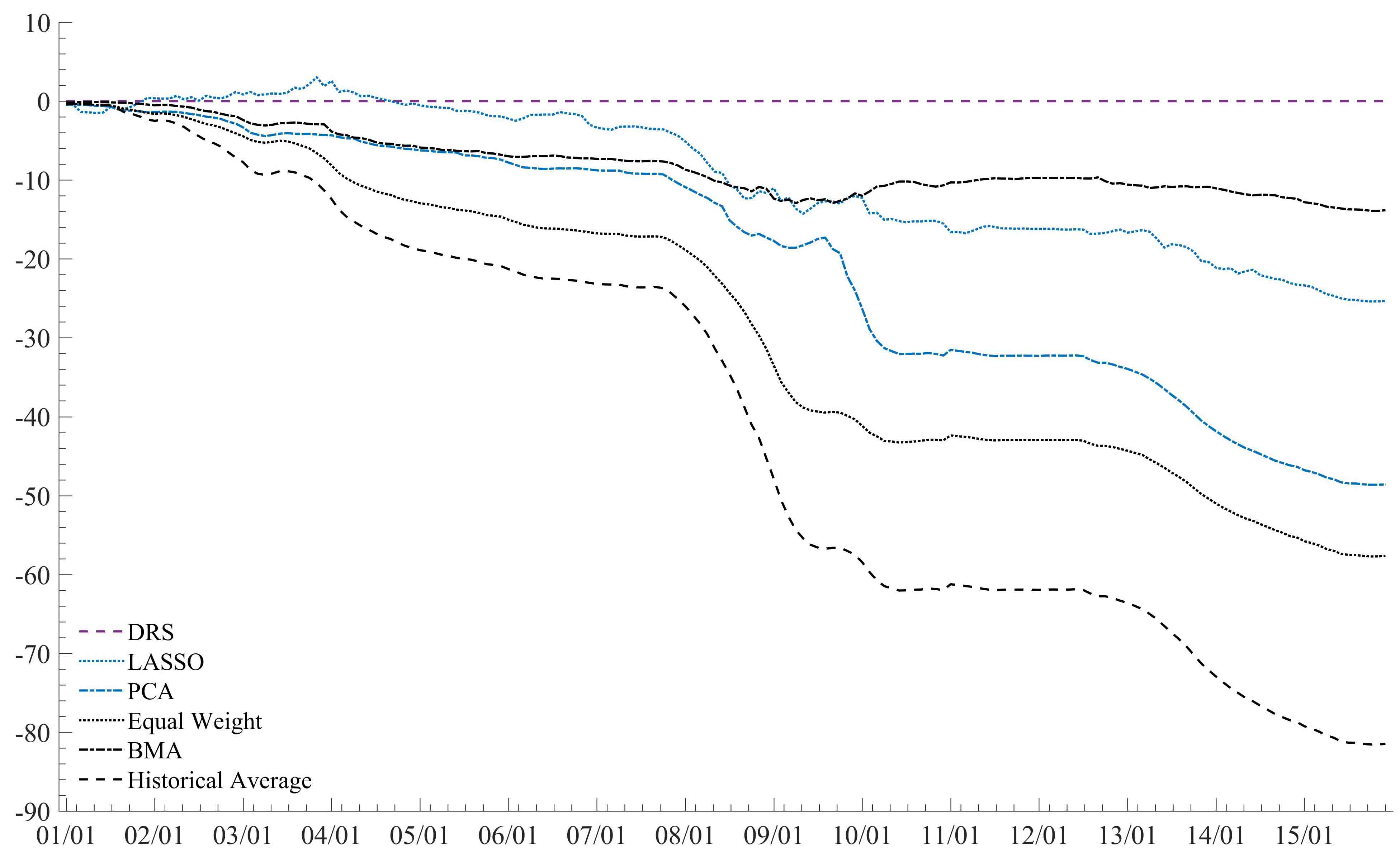}\hspace*{-2.5em}}\\  
\hspace*{-3em} 
\subfigure[Health]{\includegraphics[width=0.5\textwidth]{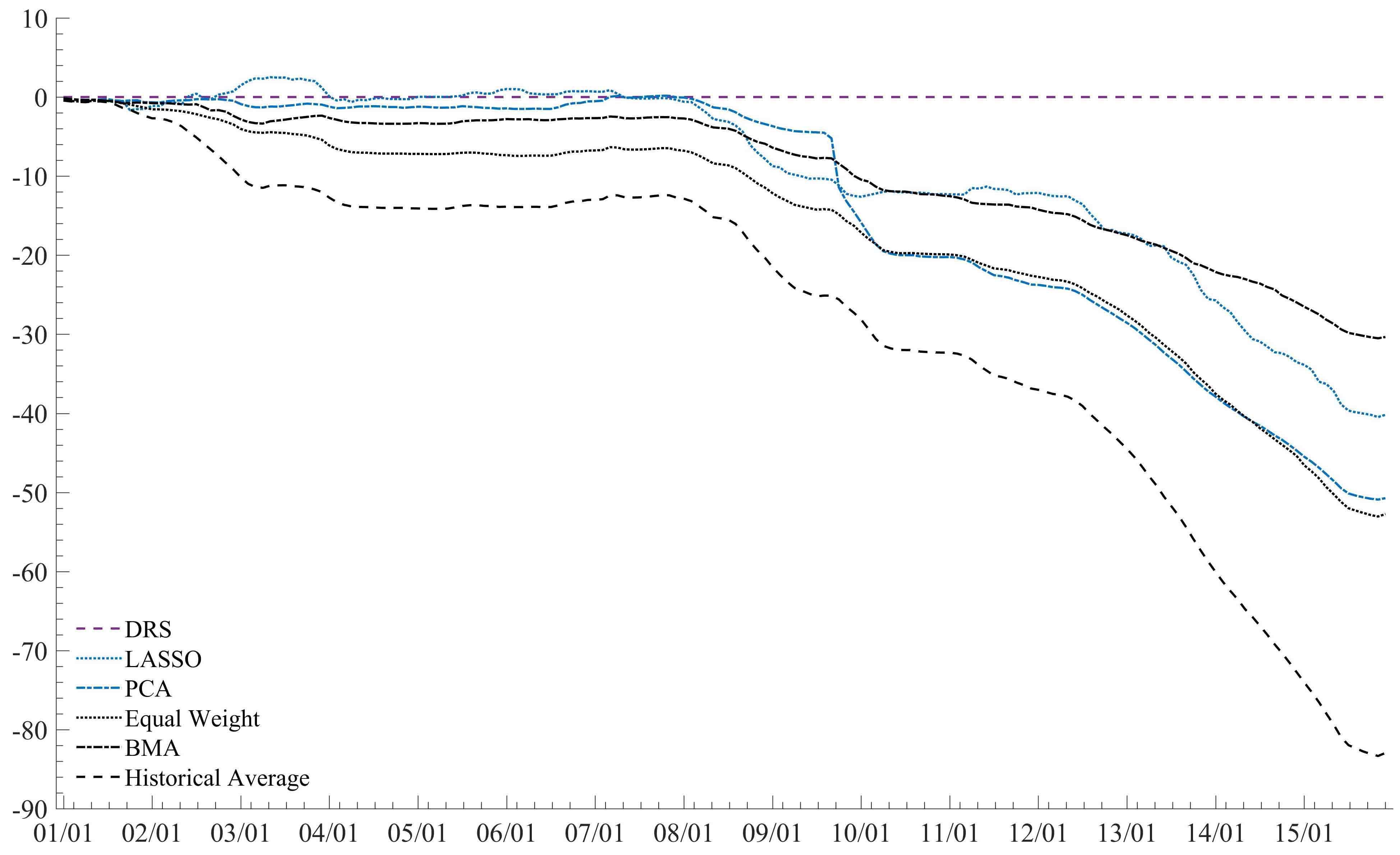}}
\subfigure[Shops]{\includegraphics[width=0.5\textwidth]{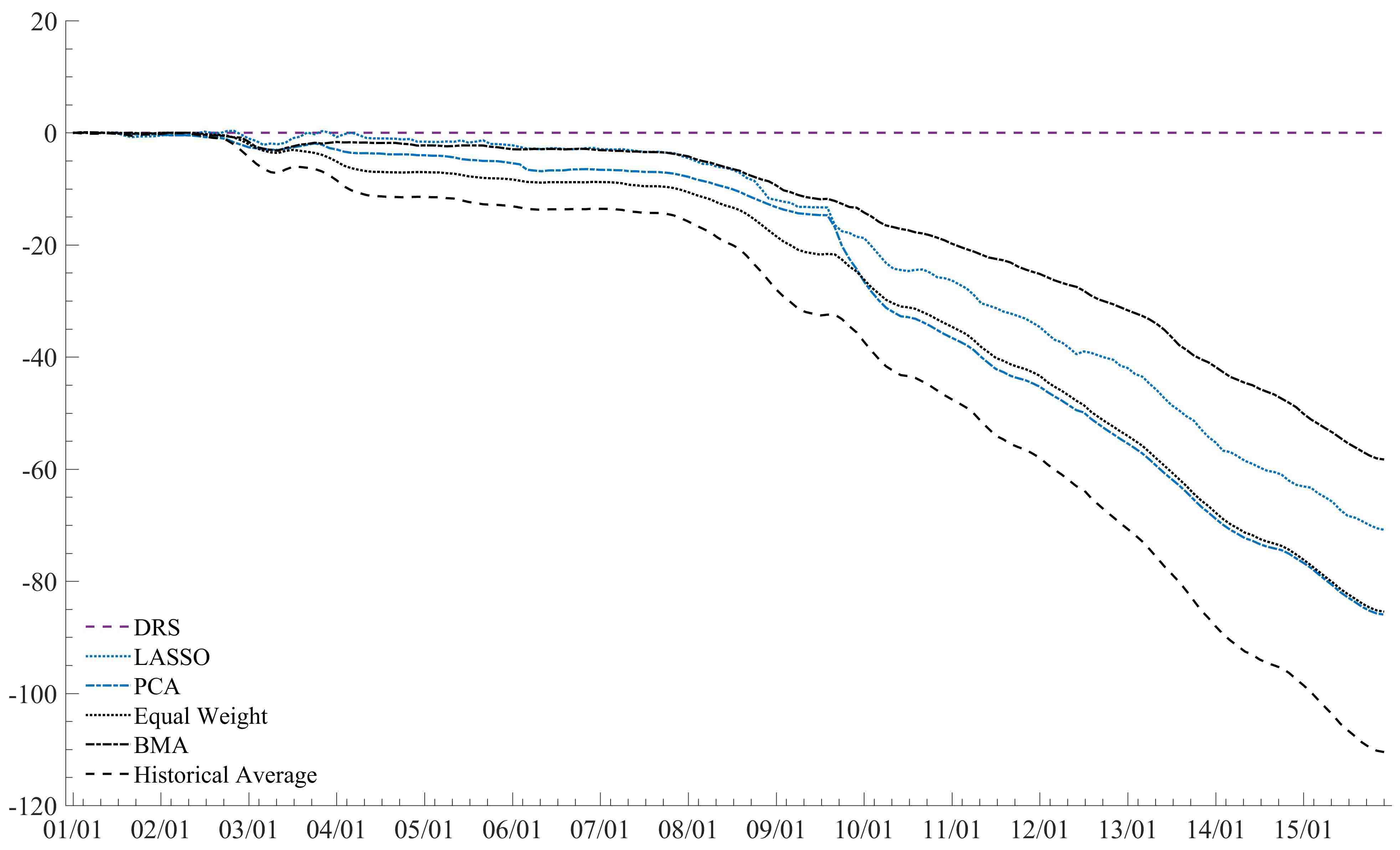}\hspace*{-2.5em}}  
  \label{fig:CCER}
 \end{figure}

Panel B shows the results concerning a short-sale constrained investor. Although the gap between DRS and the competing model combination schemes reduces, the former robustly generates higher performances in the order of 10 to 40 basis points, depending on the industry and the competing strategy. As a whole, Tables \ref{tab:Table002}-\ref{tab:Table004} tell a story whereby by carefully considering the latent dependencies across classes of predictors allows to sensibly improve the out-of-sample economic performance for a power utility investor with moderate risk aversion. To parallel the LPDR in Eq.~\eqref{eq:lpdr}, we also inspect the economic performance of the individual model combination schemes by reporting the cumulative sum of the CERs over time:
\begin{align}
CCER_{it} & = \sum_{\tau=1}^t\log\left(1+CER_{i\tau}\right),
\label{eq:cumulative}
\end{align}
where $CER_{it}$ is calculated as in Eq.~\eqref{eq:single}. Figure \ref{fig:CCER} shows the out-of-sample cumulative CER across the forecasting sample and for the Consumer durable, Consumer non-durable, Telecomm, Health, Shops and Other industrial sectors. Except few nuances, e.g., the pre-crisis period for Telecomm and Other, the DRS combination scheme constantly outperforms the other predictive strategies. Interestingly, although initially generate a good certainty equivalent return, the LASSO failed to adjust to the abrupt underlying changes in the predictability of industry returns around the crisis. As a matter of fact, despite the initial cumulative CER is slightly in favor of the LASSO vis-a-vis DRS, such good performance disappears around the great financial crisis and in the aftermath of the consequent aggregate financial turmoil. As a result, the DRS generates a substantially higher cumulative CER by the end of the forecasting sample, showing much stronger real-time performance. 
\[
\left[\text{Insert Figure \ref{fig:CCER} about here}\right]
\]

\begin{figure}[p]
%\captionsetup[subfigure]{position=top}
\caption{US equity return forecasting: Out-of-sample cumulative CER with short-sale constraints}
\centering
  \vspace{0.1in}
\begin{justify} 
\footnotesize{This figure shows the dynamics of the out-of-sample Cumulative Certainty Equivalent Return (CER) for a short-sale constrained investor as in Eq.~\eqref{eq:cumulative} obtained for each of the group-specific predictors, by taking the historical average of the stock returns (HA), and the results from a set of competing model combination/shrinkage schemes, e.g., LASSO, Equal Weight, and Bayesian Model Averaging (BMA). For the ease of exposition we report the results for four representative industries, namely, Consumer Durables, Consumer Non-Durables, Telecomm, Health, Shops, and Other. Industry aggregation is based on the four-digit SIC codes of the existing firm at each time $t$ following the industry classification from Kenneth French's website. The sample period is 01:1970-12:2015, monthly.}
\end{justify}
\hspace*{-3em} 
\subfigure[Consumer Durable]{\includegraphics[width=0.5\textwidth]{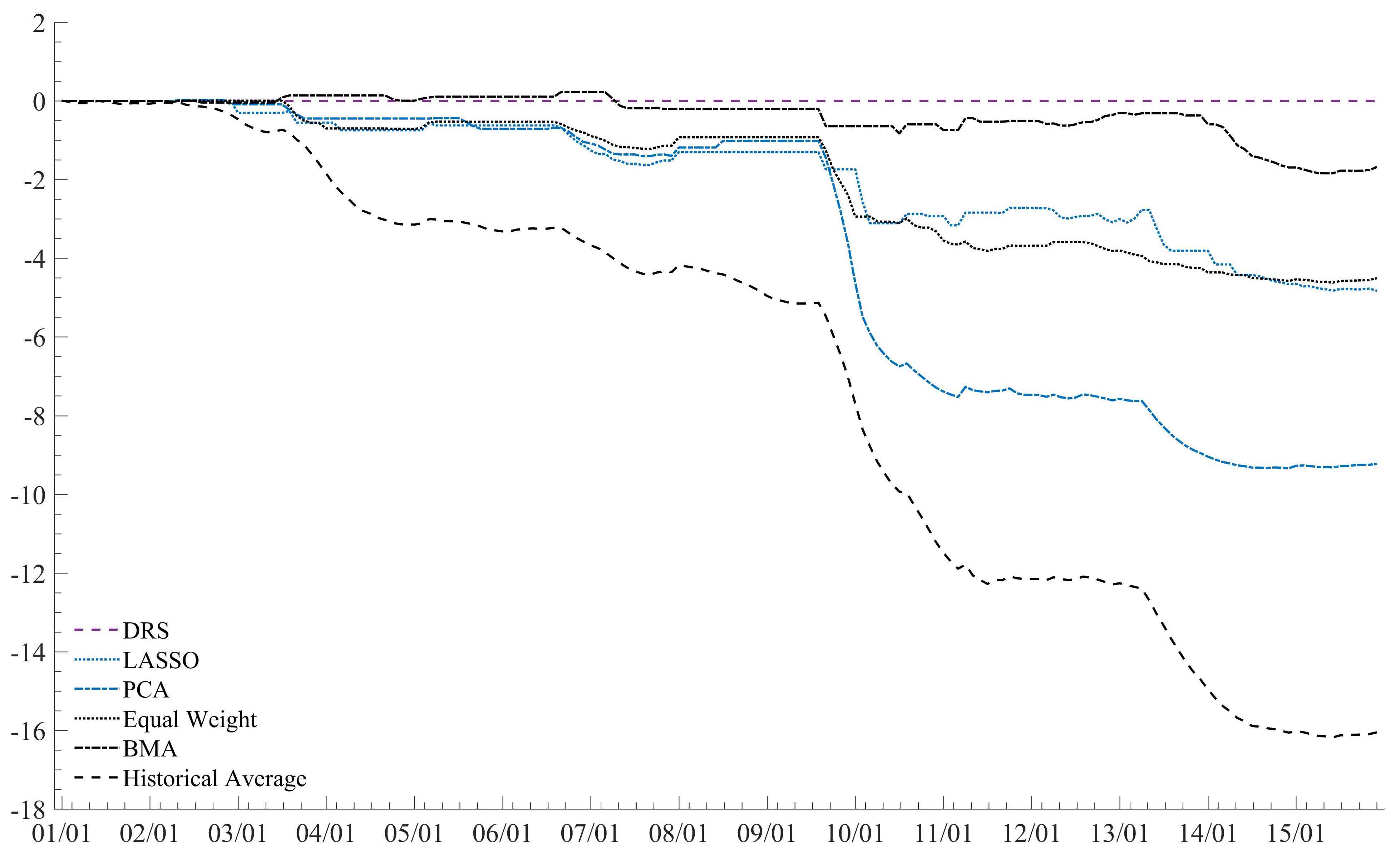}}
\subfigure[Cons. Non-Durable]{\includegraphics[width=0.5\textwidth]{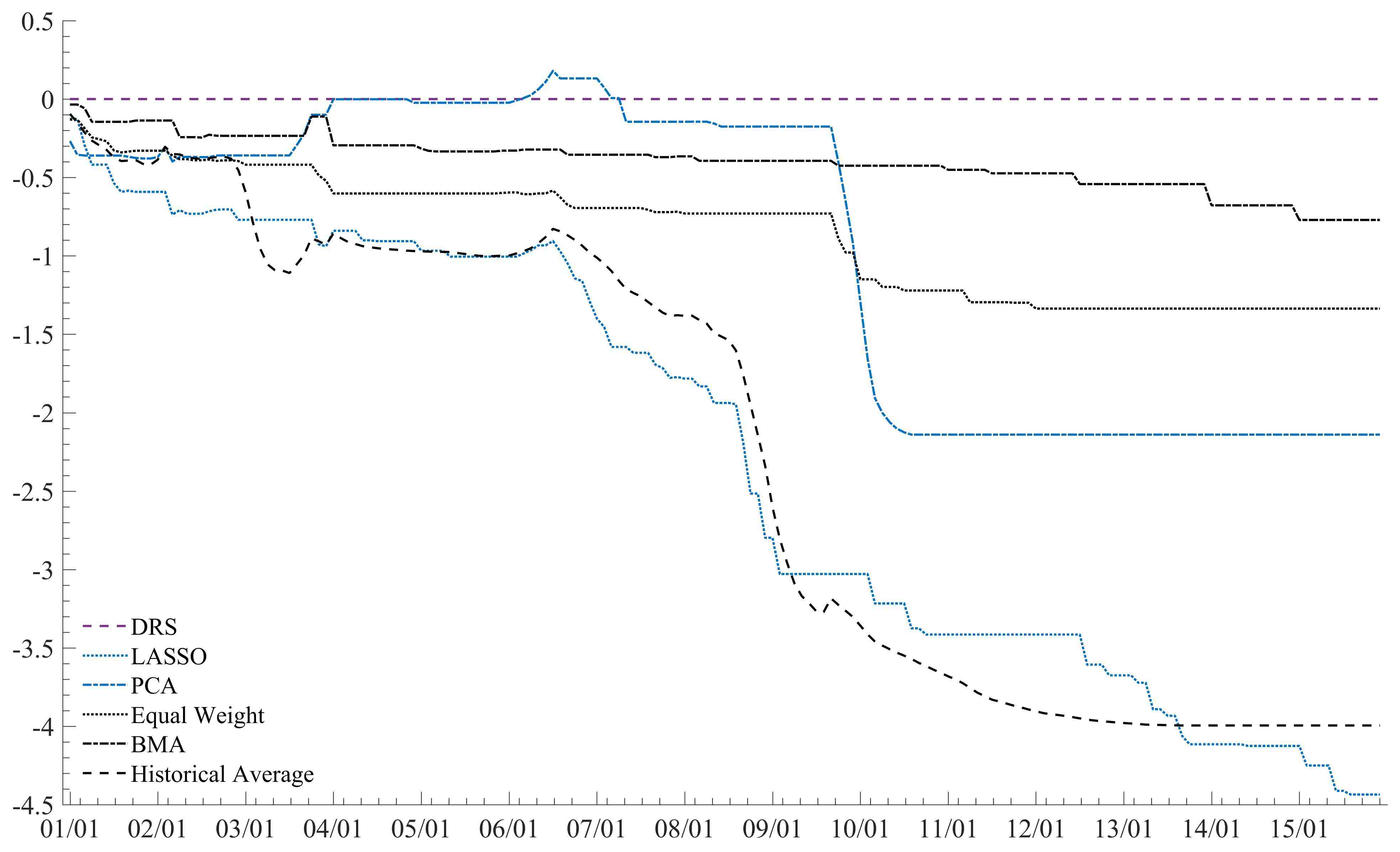}\hspace*{-2.5em}} \\
\hspace*{-3em} 
\subfigure[Telecomm]{\includegraphics[width=0.5\textwidth]{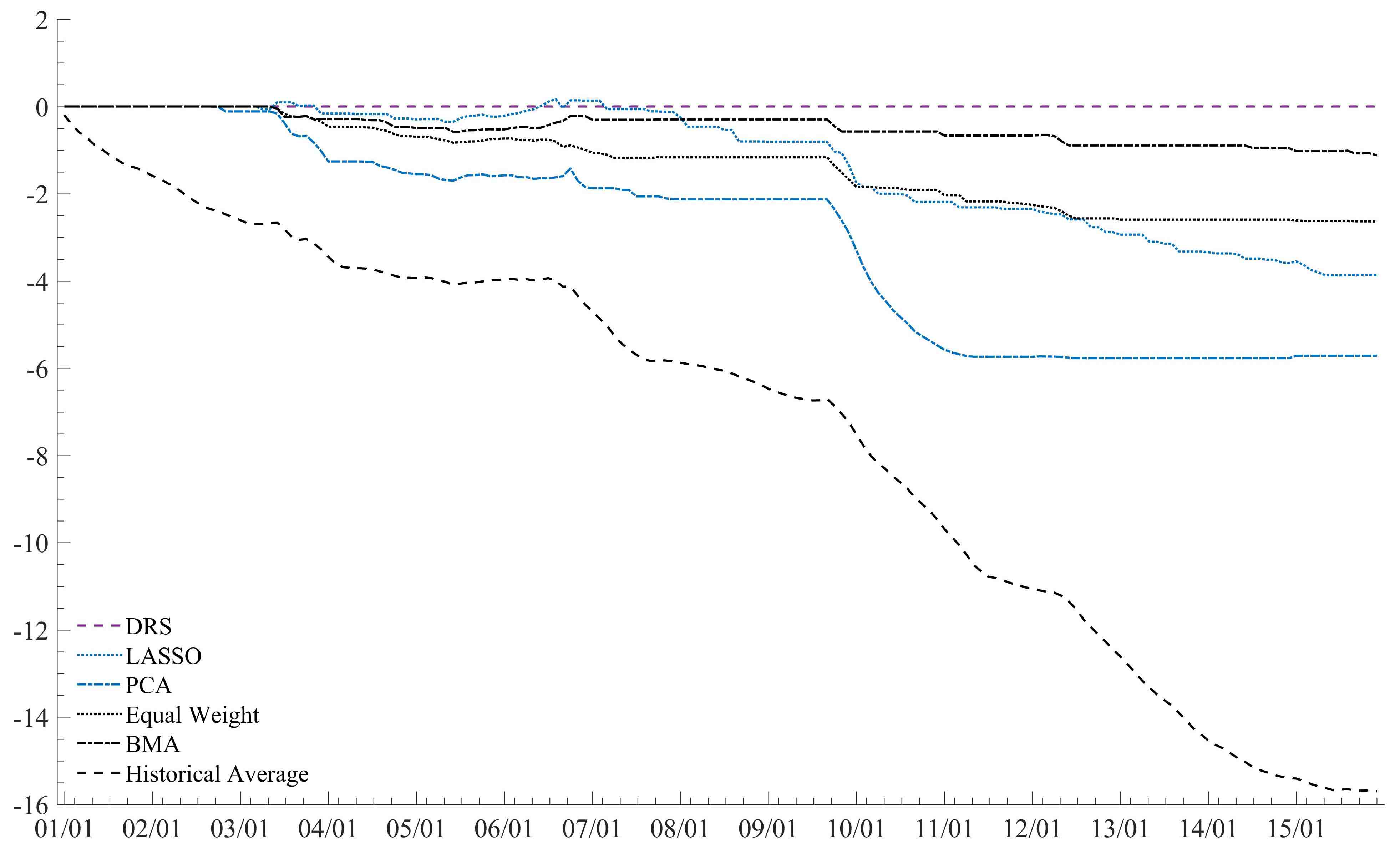}}
\subfigure[Other]{\includegraphics[width=0.5\textwidth]{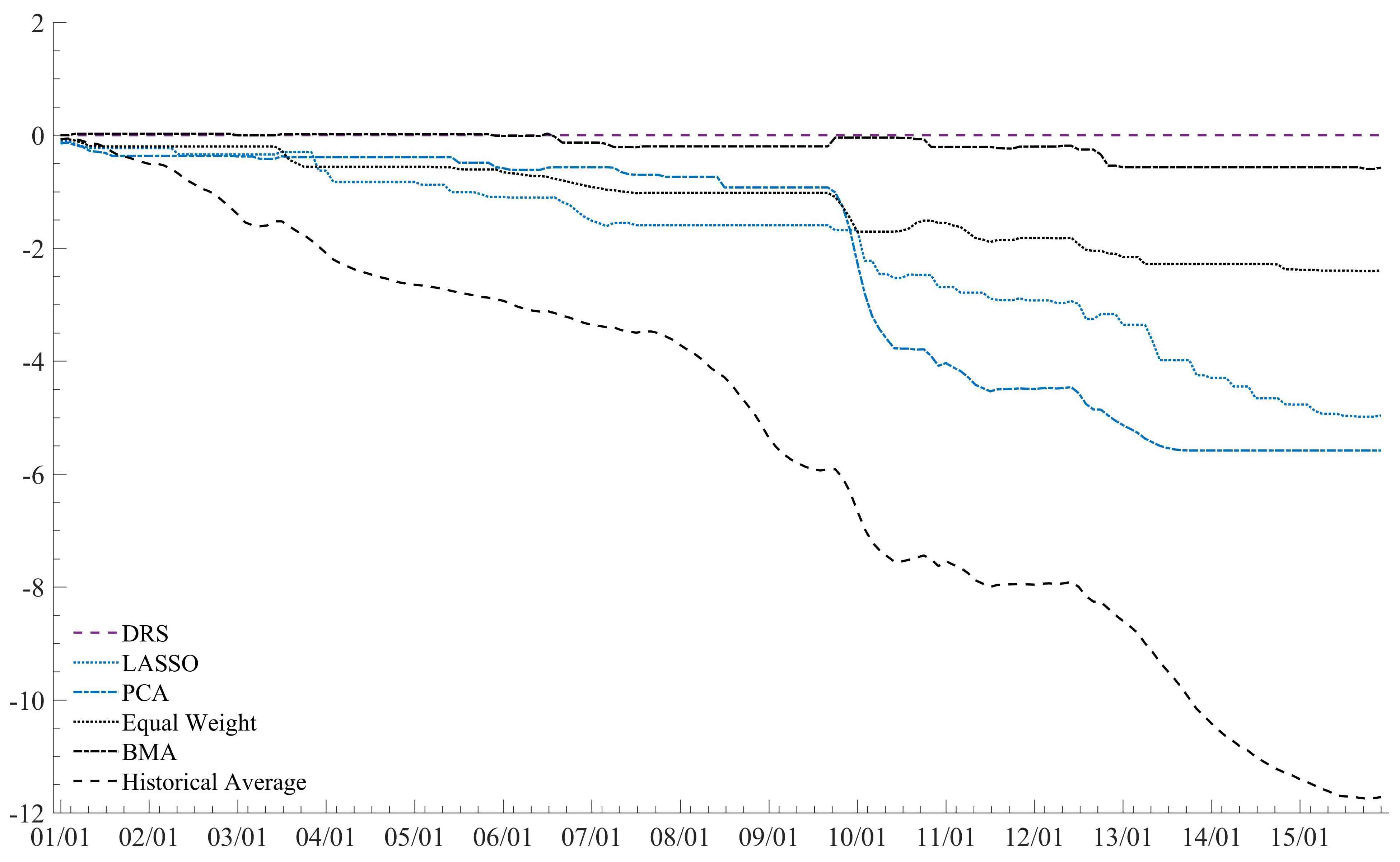}\hspace*{-2.5em}}\\  
\hspace*{-3em} 
\subfigure[Health]{\includegraphics[width=0.5\textwidth]{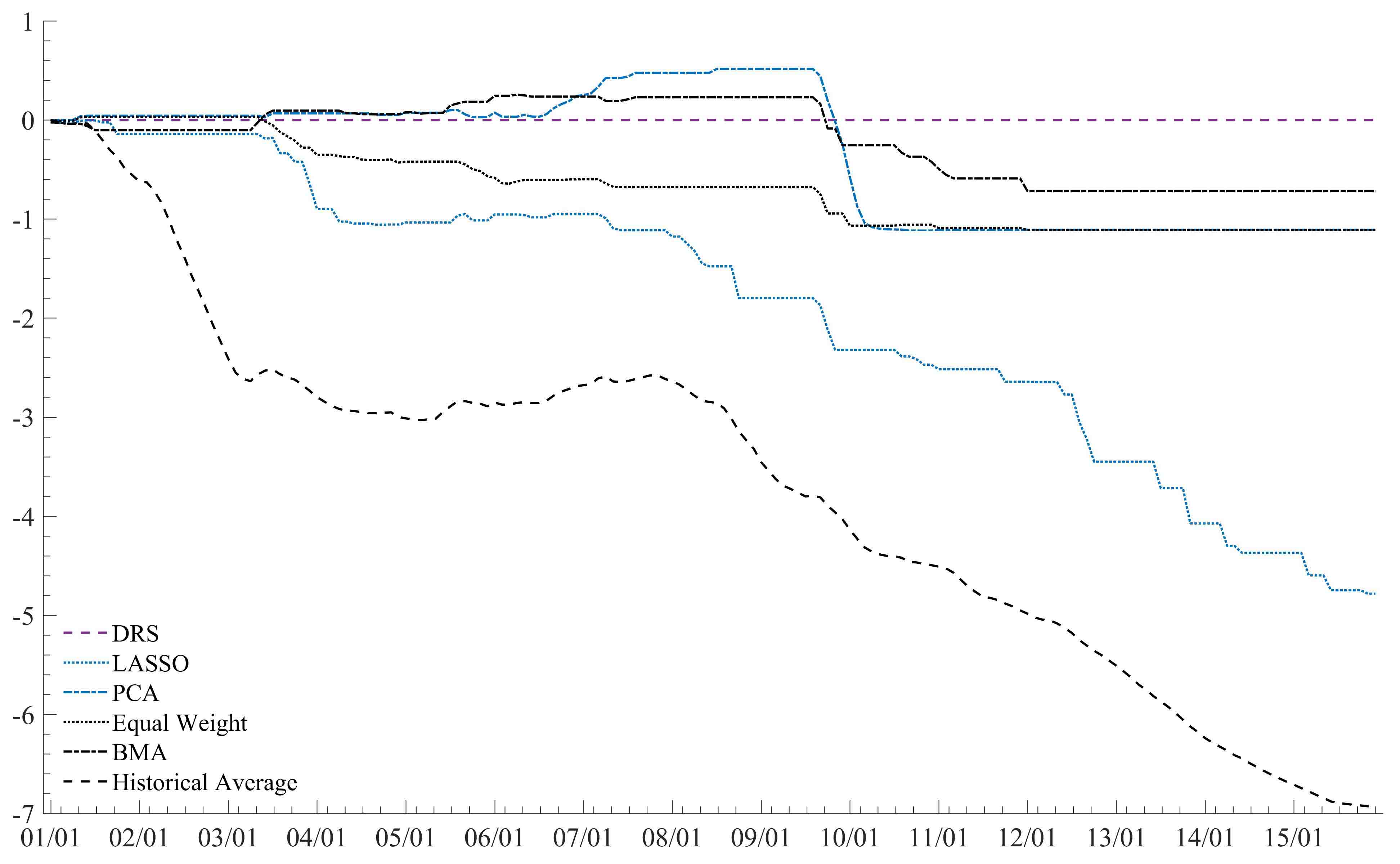}}
\subfigure[Shops]{\includegraphics[width=0.5\textwidth]{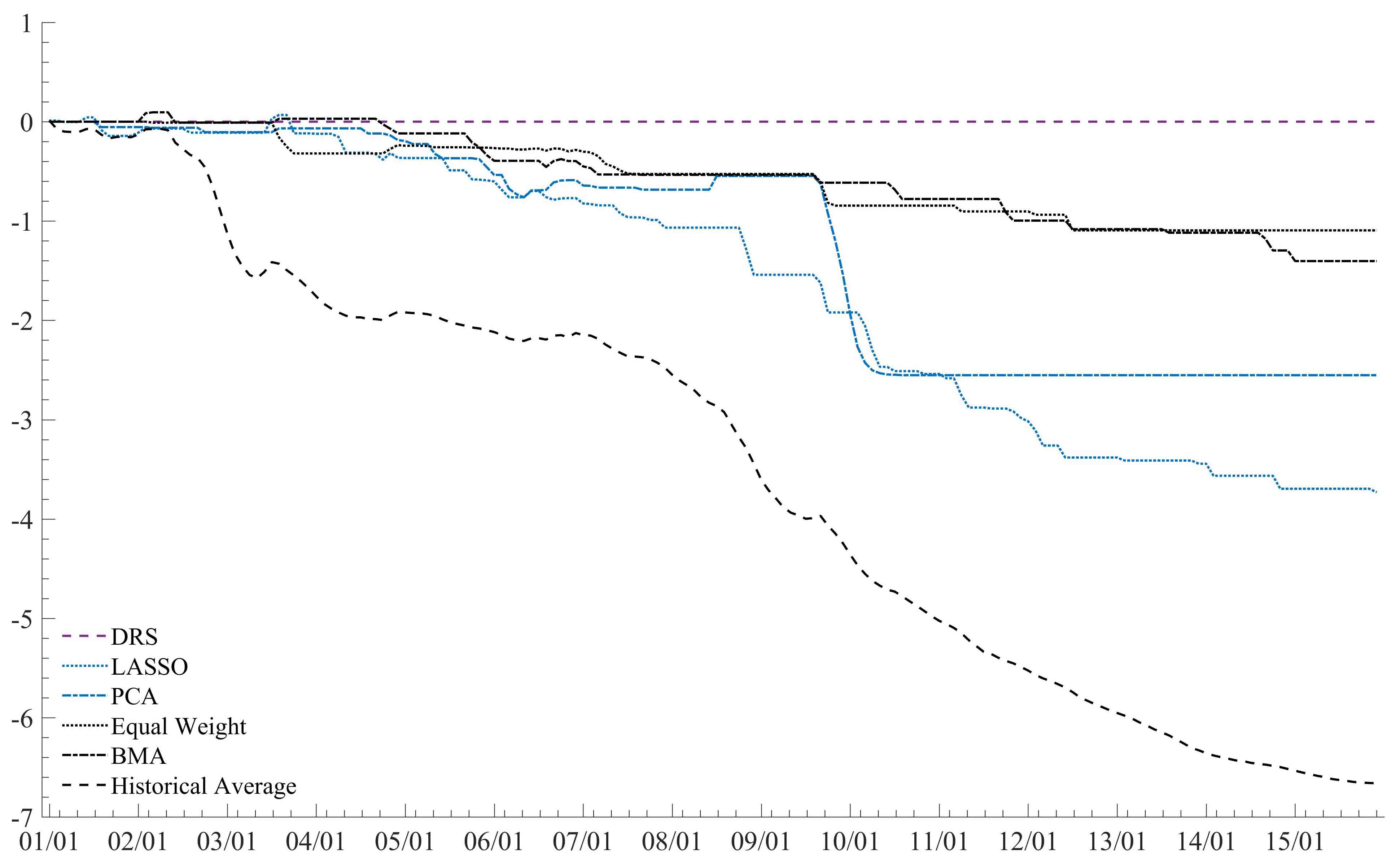}\hspace*{-2.5em}}  
  \label{fig:CCERc}
 \end{figure}

Results are virtually the same by considering an investor with short-sale constraints. Figure \ref{fig:CCERc} shows the out-of-sample cumulative CER across the forecasting sample and for the Consumer durable, Consumer non-durable, Telecomm, Health, Shops and Other industrial sectors, but now imposing that the vector of portfolio weights should be positive and sum to one, i.e. no-short sale constraints. 
\[
\left[\text{Insert Figure \ref{fig:CCERc} about here}\right]
\]

The picture that emerges is the same. Except a transitory period during the great financial crisis for the Health sector, the DRS strategy significantly outperforms the competing specifications. Notice, however, that imposing no-short constraints substantially improves the out-of-sample real-time economic performance of the alternative specifications as well. In this respect, results are consistent with the existing evidence that by restricting portfolio weights we obtain a regularization effect in the model estimations which reduces the effect of the estimation error (see, e.g., \citeRFSNP{jagannathan2003}, and \citeRFSNP{demiguel2007}, and the references therein).

\section{Conclusion}

In this paper, we propose a framework for predictive modeling and decision making when the decision maker is confronted with a large number of predictors. Our new approach retains all of the information available by employing a decouple-recouple strategy by first decoupling a large predictive model into a set of much smaller predictive regressions, which are constructed by similarity among classes of predictors, then recoupling them dynamically using a predictive synthesis approach based on the theoretical foundations of Bayesian predictive synthesis. This is a drastically different approach from the literature where there were mainly three strands of development; shrinking the set of active regressors by imposing regularization via penalized regressions, e.g., LASSO and ridge regression, imposing sparsity through the selection of a sub-set of relevant predictors, e.g., Bayesian variable selection, or assuming a small set of factors can summarize the whole information in an unsupervised manner, e.g., PCA and factor models. Rather than assuming a sparse structure, which might not align with the interest or utility of the decision maker, in our framework, we retain all of the information by treating each of the subgroup of predictors as latent states; latent states, which are learned and calibrated via Bayesian updating, to understand the latent inter-dependencies and biases. These inter-dependencies and biases are then effectively mapped onto a latent dynamic factor model, in order to provide the decision maker with a dynamically updated forecast of the quantity of interest.

We calibrate and implement the proposed methodology on both a macroeconomic and a finance application. We compare forecasts from our framework against sequentially updated Bayesian model averaging (BMA), equal-weighted linear pooling, LASSO-type regularization, as well as a set of simple predictive regressions, one for each class of predictors. Irrespective of the performance evaluation metric, our decouple-recouple model synthesis scheme emerges as the best for forecasting both the annual inflation rate for the U.S. economy as well as the year-on-year, monthly total excess returns across different industries in the U.S market. 

Furthermore, the inference obtained from posterior summaries highlight the benefits of applying our method to these big data problems. The biggest and critical appeal is that it maintains economic interpretability for all the predictors of interest, something that other methods discussed cannot do. Through multiple measurements shown in this paper, we demonstrate how these inter-dependencies can be captured and used to understand the economy/market and how they shift over time. Key economic events highlight these shifts, providing crucial insight that is useful for the decision maker. 

\vskip50pt

%\include{Parts/Appendix}

%\footnotesize
\setlength{\parskip}{.05cm }
\begin{spacing}{0.1}
\bibliography{Biblio/References}
\bibliographystyle{rfs} 
\end{spacing}

\clearpage
\vskip120pt %
%\onehalfspacing 
%% Appendix

\renewcommand{\thesection}{\arabic{section}}
\renewcommand{\thesubsection}{\thesection.\arabic{subsection}}
\setcounter{equation}{0}
\renewcommand{\theequation}{A.\arabic{equation}}
\makeatletter
% we need a period (.) after sectioning numbers, but not in cites thereto.
%\renewcommand{\@seccntformat}[1]{{\csname the#1\endcsname}{\small .}\hspace{0.3em}}
%%%%%%%%%%%%%%%%%%%%%%%%%%%%%%%%%%%%%%%%%%

%\phantomsection%
\addcontentsline{toc}{chapter}{Appendix}
\appendix
%\begin{appendix}
\title{\Huge{\textbf{Appendix}}}

%%%%%%%%%%%%%%%%%%%%%%%%%%%%%%%%%%%%%%%%%%%%%%%%%%%%%%%%%%%%%%%%%%%%%%%%%%%%%%%%%%%%%%%%%%%%%%%%%%%%%%%%%%%%%%%%%%%%%%%%%%%%%%

\section{MCMC Algorithm}
\label{sec:AppendixA}

In this section we provide details of the Markov Chain Monte Carlo (MCMC) algorithm implemented to estimate the BPS recouple step. 
This involves a sequence of standard steps in a customized two-component block Gibbs sampler: the first component learns and simulates from the joint posterior predictive densities of the subgroup models; this the ``learning'' step.
The second step samples the predictive synthesis parameters, that is we ``synthesize'' the models' predictions in the first step to obtain a single predictive density using the information provided by the subgroup models. 
The latter involves the FFBS algorithm central to MCMC in all conditionally normal DLMs (~\citealt{Schnatter1994}; \citealt[][Sect 15.2]{WestHarrison1997book2}; \citealt[][Sect 4.5]{Prado2010}).  

In our sequential learning and forecasting context, the full MCMC analysis is performed in an extending window manner, re-analyzing the data set as time and data accumilates. 
We detail MCMC steps for a specific time $t$ here, based on all data up until that time point. 

\subsection{Initialization:} 

First, initialize by setting $\F_t=(1,x_{t1},...,x_{tJ})'$ for each $t=\seq1T$ at some chosen initial values of the latent states. 
Initial values can be chosen arbitrarily, though following \cite{McAlinn2017} we recommend sampling from the priors, i.e., from the forecast distributions,  $x_{tj} \sim h_{tj}(x_{tj})$ independently for all $t=\seq1T$ and $j=\seq1J$. 

Following initialization, the MCMC iterates repeatedly to resample two coupled sets of conditional posteriors to generate the draws from the target posterior $p(\x_{\seq1T},\bPhi_{\seq1T}|y_{\seq1T}, \mH_{\seq1T}).$ 
These two conditional posteriors and algorithmic details of their simulation are as follows. 

\subsection{Sampling the synthesis parameters $\bPhi_{\seq1T}$ } 
Conditional on any values of the latent agent states, we have a conditionally normal DLM with known predictors. 
The conjugate DLM form, 
\begin{align*}
		y_t&=\F_t'\btheta_t+\nu_t, \quad \nu_t\sim N(0,v_t), \label{eq:DLMa} \\
	\btheta_t&=\btheta_{t-1}+\bomega_t, \quad \bomega_t\sim N(0, v_t\W_t),
\end{align*}
has known elements $\F_t,\W_t$ and specified initial prior at $t=0.$ 
The implied conditional posterior for $\bPhi_{\seq1T}$ then does not depend on  $\mH_{\seq1T}$, reducing to $p(\bPhi_{\seq1T}|\x_{\seq1T},y_{\seq1T})$. 
Standard Forward-Filtering Backward-Sampling algorithm can be applied to efficiently sample these parameters, modified to incorporate the discount stochastic volatility components for $v_t$ (e.g.~\citealt{Schnatter1994}; \citealt[][Sect 15.2]{WestHarrison1997book2}; \citealt[][Sect 4.5]{Prado2010}). 

\subsubsection{Forward filtering:} One step filtering updates are computed, in sequence, as follows:
\begin{itemize} 
	\item[1.]{\em Time $t-1$ posterior:} 
		\begin{align*}
		\btheta_{t-1}|v_{t-1},\x_{\seq1{t-1}},y_{\seq1{t-1}}&\sim N(\m_{t-1}, \C_{t-1}v_{t-1}/s_{t-1}),\\
		v_{t-1}^{-1}|\x_{\seq1{t-1}},y_{\seq1{t-1}}&\sim G(n_{t-1}/2, n_{t-1}s_{t-1}/2),
		\end{align*}
		with point estimates $\m_{t-1}$ of $\btheta_{t-1}$ and $s_{t-1}$ of $v_{t-1}.$ 
	\item[2.]{\em Update to time $t$ prior:} 
		\begin{align*}
		\btheta_{t}|v_t,\x_{\seq1{t-1}},y_{\seq1{t-1}}&\sim N(\m_{t-1}, \R_tv_t/s_{t-1})
		\quad\textrm{with}\quad \R_{t}=\C_{t-1}/\delta, \\
		v_t^{-1}|\x_{\seq1{t-1}},y_{\seq1{t-1}}&\sim G(\beta n_{t-1}/2, \beta n_{t-1}s_{t-1}/2),
		\end{align*}
		with (unchanged) point estimates $\m_{t-1}$ of $\btheta_{t}$ and $s_{t-1}$ of $v_{t},$  but with increased uncertainty relative to the time $t-1$ posteriors, where the level of increased uncertainty is defined by the discount factors.   
	\item[3.]{\em  1-step predictive distribution:} 
		$y_t |\x_{\seq1t},y_{\seq1{t-1}} \sim T_{\beta n_{t-1}}(f_t,q_t)$ where
		$$f_t=\F_t'\m_{t-1}\quad \textrm{and}\quad q_t=\F_t'\R_t\F_t+s_{t-1}.$$
	\item[4.]{\em  Filtering update to time $t$ posterior:}  
		\begin{align*}
		\btheta_{t}|v_{t},\x_{\seq1{t}},y_{\seq1{t}}&\sim N(\m_{t}, \C_{t}v_{t}/s_{t}),\\
		v_{t}^{-1}|\x_{\seq1{t}},y_{\seq1{t}}&\sim G(n_{t}/2, n_{t}s_{t}/2),
		\end{align*}
		with defining parameters as follows:  
		\begin{itemize}
			\item[i.]{For $\btheta_t|v_t:$}  $\m_t=\m_{t-1}+\A_t e_t$ and $ 	\C_{t}=r_t(\R_t-q_t\A_t\A_t'),$
			\item[ii.]{For $v_t:$}  	$n_t=\beta n_{t-1}+1$ and $ s_t=r_ts_{t-1},$ 
		\end{itemize} 
		based on  1-step forecast error  $ e_t=y_t-f_t,$ the state adaptive coefficient vector (a.k.a. \lq\lq Kalman gain'') 
		$\A_t=\R_t\F_t/q_t,$  and volatility estimate ratio $r_t=(\beta n_{t-1}+e_t^2/q_t)/n_t .$ 
\end{itemize} 
\subsubsection{Backward sampling:} Having run the forward filtering analysis up to time $T,$ the backward sampling proceeds as follows. 
	\begin{itemize}
	\item[a.]{\em At time $T$:} Simulate $\bPhi_T=(\btheta_T,v_T)$ from the final normal/inverse gamma posterior 
	  $p(\bPhi_T|\x_{\seq1{T}},y_{\seq1{T}})$ as follows. First, draw $v_T^{-1}$ from $G(n_{T}/2, n_{T}s_{T}/2),$ and then 
	  	draw $\btheta_T$ from $N(\m_T,\C_T v_T/s_T).$
	\item[b.]{\em Recurse back over times $t=T-1, T-2, \ldots, 0:$}  At time $t,$ sample  
		$\bPhi_t=(\btheta_t,v_t)$ as follows:
		\begin{itemize}  
			\item[i.] Simulate the volatility $v_t$ via 
				$v_t^{-1}=\beta v_{t+1}^{-1}+\gamma_t$ where $\gamma_t$ is an independent draw from
					$\gamma_t  \sim G((1-\beta)n_t/2,n_ts_t/2),$
			\item[ii.] Simulate the state $\btheta_t$ from the conditional normal posterior 
				$p(\btheta_{t}|\btheta_{t+1},v_t,\x_{\seq1T},y_{\seq1T})$ with mean 
				vector $\m_{t}+\delta(\btheta_{t+1}-\m_{t})$ and variance matrix 
				$ \C_{t} (1-\delta)(v_t/s_t).$ 				
		\end{itemize} 
	\end{itemize} 

\subsection{Sampling the latent states $\x_{\seq1T}$}

Conditional on the sampled values from the first step,   the MCMC iterate completes with resampling of the posterior joint latent states from $ p( \x_{\seq1t} |  \bPhi_{\seq1t}, y_{\seq1t}, \mH_{\seq1t} ).$   
We note that $\x_t$ are
conditionally independent over time $t$ in this conditional distribution, with time $t$ 
conditionals 
\begin{equation}\label{app:ccforx}
p( \x_t|  \bPhi_t, y_t, \mH_t) \propto N(y_t|\F_t'\btheta_t, v_t) \prod_{j=\seq1J} h_{tj}(x_{tj}) 
	\quad\textrm{where}\quad  \F_t=(1, x_{t1},x_{t2},...,x_{tJ})'. 
\end{equation}
Since $h_{tj}(x_{tj})$ has a density of  $ T_{n_{tj}}(h_{tj},H_{tj})$,
we can express this as a scale mixture of Normal, $ N(h_{tj},H_{tj})$, with  $ \H_t=\textrm{diag}(H_{t1}/\phi_{t1},H_{t2}/\phi_{t2},...,H_{tJ}/\phi_{tJ})$, where $\phi_{tj}$ are 
independent over $t,j$ with gamma distributions, $\phi_{tj} \sim G(n_{tj}/2,n_{tj}/2).$

The posterior distribution for each $\x_t$ is then sampled, given $\phi_{tj}$, from
\begin{equation}\label{app:condx}
	p( \x_t|  \bPhi_t, y_t, \mH_t) = N(\h_{t}+\b_t c_t, \H_t-\b_t\b_t'g_t)
\end{equation}
where $c_t = y_t- \theta_{t0} - \h_t'\btheta_{t,\seq1J}$,    $g_t=v_t+  \btheta_{t,\seq1J}'\q_t\btheta_{t,\seq1J}$, and $\b_t =  \q_t\btheta_{t,\seq1J}/g_t$.
Here, given the previous values of $\phi_{tj}$, we have $\H_t=\textrm{diag}(H_{t1}/\phi_{t1},H_{t2}/\phi_{t2},...,H_{tJ}/\phi_{tJ})$
Then, conditional on these new samples of $\x_t,$  updated samples of the latent scales are drawn
from the implied set of conditional gamma posteriors $\phi_{tj}|x_{tj} \sim G((n_{tj}+1)/2,(n_{tj}+d_{tj})/2)$
where $d_{tj}= (x_{tj}-h_{tj})^2/H_{tj}$, independently for each $t,j$. 
This is easily computed and then sampled independently for each $\seq1T$ to provide resimulated agent states
over $\seq1T.$

\clearpage

%\doublespacing

%\onehalfspacing
%\footnotesize
\normalsize

% Table generated by Excel2LaTeX from sheet 'Regression'

\begin{table}[]
\centering
\caption{Out-of-sample forecast performance: Forecasting inflation.}
\vspace{.1in}
\begin{justify}
\footnotesize{This table reports the out-of-sample comparison of our decouple-recouple framework against each individual model, LASSO, PCA, equal weight average of models, and BMA for inflation forecasting. Performance comparison is based on the Root Mean Squared Error (RMSE), and the Log Predictive Density Ratio (LPDR) as in Eq.~\eqref{eq:lpdr}. The testing period is 2001/1-2015/12, monthly.}
\end{justify}
\renewcommand{\arraystretch}{1.2}
\vspace{.5em}
{\bf Panel A: \normalfont Forecasting 1-Step Ahead Inflation} \vspace{0.2cm}\\ 		
\resizebox{1\textwidth}{!}{\begin{tabular}{lccccccccccccc}
\hline
\multicolumn{9}{c}{Group-Specific Models} & LASSO&PCA & EW & BMA & DRS \\ \cline{2-9}
 &\begin{tabular}[c]{@{}c@{}}Output \&\\ Income\end{tabular} & \begin{tabular}[c]{@{}c@{}}Labor\\ Market\end{tabular} & \begin{tabular}[c]{@{}c@{}}Consump.\end{tabular} & \begin{tabular}[c]{@{}c@{}}Orders \&\\ Invent.\end{tabular} & \begin{tabular}[c]{@{}c@{}}Money\\ \& Credit\end{tabular} & \begin{tabular}[c]{@{}c@{}}Int. Rate \& \\ Ex. Rates\end{tabular} & Prices & \begin{tabular}[c]{@{}c@{}}Stock\\ Market\end{tabular} &        &  &  &    \\ \hline
RMSE&0.2488   & 0.2247   & 0.7339  & 0.2721 & 0.2624 & 0.4258 & 0.2223	 & 0.5027    & 0.3348&0.9329 & 0.2945 & 0.2721                  & 0.2051                  \\
(\%)&-7.35\% & -7.37\%  & -122.06\% & -8.73\%  & -15.75\% & -40.56\% & -6.83\%  & -59.59\%   & -63.24\%&  -354.85\%& -43.59\%  & -32.68\% & -                       \\
LPDR&-40.48  & -42.05   & -233.09   & -59.15   & -56.34    & -134.18   & -20.00 & -171.21     & -3785.15& -285.41   & -88.81  & -60.40                   & -                       \\ \hline
\end{tabular}}

\vspace{1em}
{\bf Panel B: \normalfont Forecasting 3-Step Ahead Inflation} \vspace{0.2cm}\\ 		
\resizebox{1\textwidth}{!}{\begin{tabular}{lccccccccccccc}
\hline
\multicolumn{9}{c}{Group-Specific Models} & LASSO&PCA & EW & BMA & DRS \\ \cline{2-9}
 &\begin{tabular}[c]{@{}c@{}}Output \&\\ Income\end{tabular} & \begin{tabular}[c]{@{}c@{}}Labor\\ Market\end{tabular} & \begin{tabular}[c]{@{}c@{}}Consump.\end{tabular} & \begin{tabular}[c]{@{}c@{}}Orders \&\\ Invent.\end{tabular} & \begin{tabular}[c]{@{}c@{}}Money\\ \& Credit\end{tabular} & \begin{tabular}[c]{@{}c@{}}Int. Rate \& \\ Ex. Rates\end{tabular} & Prices & \begin{tabular}[c]{@{}c@{}}Stock\\ Market\end{tabular} &        &  &  &    \\ \hline
RMSE&0.3594   & 0.3595   & 0.7435  & 0.3640 & 0.3875 & 0.4706 & 0.3577	 & 0.5343    & 0.3991&0.9223 & 0.3777 & 0.3640                  & 0.3348                  \\
(\%)&-21.32\% & -9.57\%  & -257.86\% & -32.68\%  & -27.95\% & -107.66\% & -8.39\%  & -145.14\%   & -19.21\%&175.45\%  & -12.87\%  & -8.73\% & -                       \\
LPDR&-78.65  & -225.75   & -156.59   & -61.96   & -122.27    & -77.76   & -101.55 & -101.82     & -3804.35& -203.12   & -41.00  & -78.54                   & -                       \\ \hline
\end{tabular}}
\label{tab:Table003}

\end{table}%

% Table generated by Excel2LaTeX from sheet 'Regression'
\begin{sidewaystable}[htbp]
\caption{Out-of-sample forecast performance: Forecasting Stock Industry Returns.}
\vspace{.1in}
\begin{justify}
\footnotesize{This table reports the out-of-sample comparison of our predictive framework against standard model combination methodologies, across ten different industries. Performance comparison is based on the Root Mean Squared Error (RMSE), and the Log Predictive Density Ratio (LPDR) as in Eq.~\eqref{eq:lpdr}. We report the results obtained for each of the group-specific predictors, the results obtained by simply taking the historical average of the stock returns (HA), and the results from a set of competing model combination/shrinkage schemes, e.g., LASSO, PCA, Equal Weight, and Bayesian Model Averaging (BMA). The sample period is 01:1970-12:2015, monthly.}
\end{justify}
\renewcommand{\arraystretch}{1.2}
\vspace{1em}
{\bf Panel A: \normalfont Root Mean Squared Error} \vspace{0.2cm}\\ 		
\resizebox{1\textwidth}{!}{\begin{tabular}{lccccccccccccccccc}
\hline
Industry & 				& \multicolumn{10}{c}{Group-Specific Models}                                                              & Full  &LASSO  & EW & BMA & PCA & DRS \\
\cline{3-12}
				 &	HA		& Value	&Profit	& Capital	& Soundness	 &Solvency	&Liquidity	&Efficiency	&Other	&Aggregate Fin	&Macro  &\\
\hline
Durbl    &0.308   &0.193	&0.208	  &0.260	     &0.228	    &0.225	    &0.238	    &0.292	&0.267	        &0.248	&0.205	&0.257	&0.245	&0.169	&0.193	&0.366	&0.154\\
NoDurbl  &0.129	  &0.092	&0.108	  &0.115	     &0.097	    &0.109	    &0.113	    &0.108	&0.111	        &0.100	&0.103	&0.119	&0.119	&0.076	&0.090	&0.121	&0.072\\
Manuf    &0.197	  &0.130	&0.140  	&0.175	     &0.165	    &0.162	    &0.182	    &0.137	&0.180	        &0.152	&0.150	&0.164	&0.140	&0.110	&0.125	&0.174	&0.103\\
Energy   &0.214	  &0.123	&0.151	  &0.176	     &0.155	    &0.142	    &0.166	    &0.181	&0.197	        &0.142	&0.170	&0.206	&0.188	&0.121	&0.123	&0.173	&0.105\\
HiTech   &0.289	  &0.139	&0.231	  &0.206	     &0.184	    &0.212	    &0.185	    &0.192	&0.238	        &0.231	&0.220	&0.196	&0.197	&0.143	&0.139	&0.223	&0.121\\
Health   &0.171	  &0.104	&0.099	  &0.127	     &0.097	    &0.109	    &0.118	    &0.101	&0.111	        &0.108	&0.098	&0.132	&0.140	&0.083	&0.094	&0.115	&0.080\\
Shops    &0.232	  &0.180	&0.159	  &0.182	     &0.131	    &0.157	    &0.163	    &0.157	&0.194	        &0.175	&0.148	&0.155	&0.171	&0.114	&0.131	&0.235	&0.108\\
Telecomm &0.157	  &0.096	&0.101	  &0.128	     &0.107	    &0.119	    &0.118	    &0.120	&0.119	        &0.102	&0.098	&0.119	&0.124	&0.082	&0.096	&0.222	&0.072\\
Utils    &0.261	  &0.142	&0.139	  &0.172	     &0.134	    &0.159	    &0.179	    &0.129	&0.196	        &0.172	&0.144	&0.184	&0.178	&0.106	&0.121	&0.190	&0.098\\
Other    &0.173	  &0.112	&0.110	  &0.132	     &0.117	    &0.143	    &0.136	    &0.114	&0.142	        &0.128	&0.123	&0.158	&0.146	&0.091	&0.097	&0.143	&0.083\\
\hline
    \end{tabular}}%
\vspace{1cm}

{\bf Panel B: \normalfont Log-Predictive Density Ratio} \vspace{0.2cm}\\ 		
\resizebox{1\textwidth}{!}{\begin{tabular}{lccccccccccccccccc}
\hline
Industry & 				& \multicolumn{10}{c}{Group-Specific Models}                                                              & Full  &LASSO  & EW & BMA & PCA & DRS \\
\cline{3-12}
				 &	HA		& Value	&Profit	& Capital	& Soundness	 &Solvency	&Liquidity	&Efficiency	&Other	&Aggregate Fin	&Macro  &\\
\hline
Durbl    &-44.55	&-45.83	&-51.75	&-107.39	&-69.84	&-82.57	&-91.23	&-124.08	&-114.72	&-64.51	&-59.86	&-69.66	     	&-91.93	  &-146.55	&-341.35	&-135.55&-\\
NoDurbl  &-109.50	&-44.08	&-60.24	&-91.38	  &-55.51	&-81.21	&-90.07	&-73.37	&-87.24	    &-45.40	&-67.43	&-94.00	     	&-228.77	&-173.04	&-334.41	&-92.57&-\\
Manuf    &-36.34	&-35.87	&-53.78	&-108.57	&-53.15	&-95.02	&-113.86	&-66.02	&-123.30	&-51.07	&-72.63	&-82.22	     	&-167.35	&-85.53	  &-232.15	&-113.60&-\\
Energy   &-56.17	&-26.65	&-52.81	&-96.36	  &-42.06	&-52.94	&-86.32	&-80.02	&-118.86	  &-45.25	&-84.58	&-59.03	     	&-150.12	&-113.92	&-281.18	&-89.40&-\\
HiTech   &-179.58	&-17.83	&-75.16	&-110.60	&-69.41	&-100.94	&-82.43	&-87.47	&-134.93	&-110.40	&-106.30	&-62.71	 	&-138.79	&-257.24	&-848.69	&-124.59&-\\
Health   &-124.31	&-46.58	&-27.30	&-77.57	  &-26.73	&-53.59	&-65.62	&-41.45	&-63.09	    &-49.76	&-32.44	&-80.54	     	&-200.14	&-275.66	&-437.22	&-60.41&-\\
Shops    &-81.20	&-53.20	&-56.88	&-99.89	  &-27.24	&-75.51	&-90.22	&-75.54	&-117.14	  &-65.67	&-61.57	&-55.11	     	&-161.13	&-166.28	&-468.83	&-108.89&-\\
Telecomm &-112.92	&-42.74	&-60.49	&-99.38	  &-69.44	&-83.03	&-85.85	&-88.10	&-82.07	    &-51.55	&-51.25	&-102.55	   	&-215.60	&-174.92	&-275.45	&-100.66&-\\
Utils    &-165.95	&-64.61	&-49.29	&-98.94	  &-45.90	&-85.61	&-107.23	&-47.20	&-124.94	&-74.01	&-65.39	&-84.44	     	&-162.67	&-203.81	&-410.01	&-123.01&-\\
Other    &-115.31	&-50.11	&-29.38	&-86.38	  &-37.23	&-99.04	&-99.83	&-46.18	&-102.96	  &-71.07	&-73.01	&-66.65	     	&-197.83	&-128.33	&-314.82	&-94.06&-\\
\hline
    \end{tabular}}%
\label{tab:Table001}

\end{sidewaystable}%

% Table generated by Excel2LaTeX from sheet 'Regression'
\begin{sidewaystable}[htbp]
\caption{Out-of-sample economic performance for stock industry returns: certainty equivalent returns}
\vspace{.1in}
\begin{justify}
\footnotesize{This table reports the out-of-sample comparison of our predictive framework against standard model combination methodologies, across ten different industries. Performance comparison is based on the Certainty Equivalent (CER), and its modification whereby short sales are not allowed. We report the results obtained for each of the group-specific predictors, the results obtained by simply taking the historical average of the stock returns (HA), and the results from a set of competing model combination/shrinkage schemes, e.g., LASSO, PCA, Equal Weight, and Bayesian Model Averaging (BMA). The sample period is 01:1970-12:2015, monthly.}
\end{justify}
\renewcommand{\arraystretch}{1.2}
\vspace{1em}
{\bf Panel A: \normalfont Certainty Equivalent} \vspace{0.2cm}\\ 		
\resizebox{1\textwidth}{!}{\begin{tabular}{lcccccccccccccccc}
\hline
Industry & 				& \multicolumn{10}{c}{Group-Specific Models}                                                              & Full  &LASSO  & EW & BMA & PCA  \\
\cline{3-12}
				 &	HA		& Value	&Profit	& Capital	& Soundness	 &Solvency	&Liquidity	&Efficiency	&Other	&Aggregate Fin	&Macro  &\\
\hline

Durbl    &-0.084	&-0.047	&-0.032	&-0.973	&-0.075	&-0.063	&-0.306	&-0.863	&-0.342	&-0.067	&-0.082	&-0.054	&-0.721	&-0.021	&-0.047	&-0.996\\
NoDurbl  &-0.195	&-0.294	&-0.114	&-0.608	&-0.097	&-0.211	&-0.163	&-0.800	&-0.211	&-0.119	&-0.149	&-0.141	&-0.548	&-0.125	&-0.261	&-0.719\\
Manuf    &-0.126	&-0.881	&-0.101	&-0.141	&-0.465	&-0.206	&-0.099	&-0.765	&-0.101	&-0.005	&-0.210	&-0.488	&-0.544	&-0.019	&-0.840	&-0.983\\
Energy   &-0.141	&-0.074	&-0.118	&-0.092	&-0.055	&-0.099	&-0.102	&-0.915	&-0.180	&-0.050	&-0.201	&-0.030	&-0.990	&-0.032	&-0.005	&-0.651\\
HiTech   &-0.149	&-0.029	&-0.079	&-0.165	&-0.077	&-0.135	&-0.119	&-0.082	&-0.257	&-0.152	&-0.113	&-0.076	&-0.203	&-0.070	&-0.029	&-0.361\\
Health   &-0.133	&-0.023	&-0.024	&-0.112	&-0.044	&-0.044	&-0.142	&-0.024	&-0.072	&-0.044	&-0.030	&-0.064	&-0.163	&-0.038	&-0.017	&-0.989\\
Shops    &-0.150	&-0.045	&-0.070	&-0.414	&-0.039	&-0.496	&-0.770	&-0.044	&-0.107	&-0.059	&-0.154	&-0.042	&-0.498	&-0.059	&-0.040	&-0.746\\
Telecomm &-0.174	&-0.067	&-0.097	&-0.208	&-0.104	&-0.106	&-0.104	&-0.426	&-0.077	&-0.078	&-0.050	&-0.113	&-0.864	&-0.081	&-0.067	&-0.749\\
Utils    &-0.216	&-0.104	&-0.078	&-0.462	&-0.072	&-0.122	&-0.185	&-0.160	&-0.160	&-0.103	&-0.098	&-0.095	&-0.474	&-0.109	&-0.075	&-0.439\\
Other    &-0.096	&-0.009	&0.031	&-0.038	&0.022	&-0.105	&-0.089	&0.038	&-0.042	&-0.019	&-0.044	&0.028	&-0.657	&0.016	&0.055	&-0.433\\
%Durbl    &0.044	&0.086	&0.104	&-0.969	&0.054	&0.068	&-0.208	&-0.844	&-0.250	&0.063	&0.047	&0.078	&-0.682	&0.116	&0.086	&-0.996	&0.140\\
%NoDurbl  &0.085	&-0.050	&0.194	&-0.472	&0.217	&0.063	&0.127	&-0.731	&0.063	&0.186	&0.146	&0.157	&-0.391	&0.178	&-0.005	&-0.621	&0.347\\
%Manuf    &0.039	&-0.859	&0.070	&0.022	&-0.364	&-0.056	&0.072	&-0.721	&0.070	&0.184	&-0.060	&-0.391	&-0.458	&0.167	&-0.810	&-0.980	&0.190\\
%Energy   &0.037	&0.119	&0.065	&0.097	&0.142	&0.088	&0.085	&-0.897	&-0.009	&0.147	&-0.035	&0.172	&-0.987	&0.169	&0.202	&-0.579	&0.208\\
%HiTech   &0.032	&0.177	&0.116	&0.012	&0.118	&0.048	&0.068	&0.112	&-0.099	&0.027	&0.074	&0.120	&-0.034	&0.127	&0.177	&-0.226	&0.212\\
%Health   &0.054	&0.187	&0.185	&0.079	&0.161	&0.161	&0.043	&0.185	&0.128	&0.161	&0.179	&0.138	&0.016	&0.169	&0.194	&-0.986	&0.215\\
%Shops    &0.041	&0.170	&0.139	&-0.282	&0.177	&-0.382	&-0.718	&0.171	&0.094	&0.153	&0.036	&0.173	&-0.385	&0.152	&0.175	&-0.689	&0.225\\
%Telecomm &0.070	&0.209	&0.170	&0.025	&0.161	&0.157	&0.161	&-0.257	&0.196	&0.194	&0.230	&0.149	&-0.824	&0.190	&0.209	&-0.675	&0.295\\
%Utils    &0.032	&0.180	&0.215	&-0.292	&0.222	&0.157	&0.073	&0.106	&0.106	&0.181	&0.188	&0.191	&-0.308	&0.173	&0.218	&-0.262	&0.317\\
%Other    &0.057	&0.159	&0.205	&0.124	&0.195	&0.046	&0.065	&0.213	&0.119	&0.147	&0.117	&0.202	&-0.599	&0.187	&0.233	&-0.337	&0.169\\
\hline
    \end{tabular}}%
\vspace{1cm}

{\bf Panel B: \normalfont Certainty Equivalent (no-short sales)} \vspace{0.2cm}\\ 		
\resizebox{1\textwidth}{!}{\begin{tabular}{lcccccccccccccccc}
\hline
Industry & 				& \multicolumn{10}{c}{Group-Specific Models}                                                              & Full  &LASSO  & EW & BMA & PCA  \\
\cline{3-12}
				 &	HA		& Value	&Profit	& Capital	& Soundness	 &Solvency	&Liquidity	&Efficiency	&Other	&Aggregate Fin	&Macro  &\\
\hline

Durbl    &-0.045	&-0.007	&-0.003	&-0.024	&-0.010	&-0.021	&-0.011	&-0.012	&-0.014	&-0.005	&-0.015	&-0.009	&-0.018	&-0.005	&-0.007	&-0.021\\
NoDurbl  &-0.031	&-0.005	&-0.012	&-0.016	&-0.010	&-0.010	&-0.020	&-0.013	&-0.013	&-0.007	&-0.016	&-0.017	&-0.032	&-0.006	&-0.004	&-0.010\\
Manuf    &-0.068	&-0.006	&-0.010	&-0.047	&0.000	&-0.028	&-0.032	&-0.005	&-0.027	&-0.007	&-0.009	&-0.025	&-0.017	&-0.006	&-0.006	&-0.017\\
Energy   &-0.067	&-0.001	&-0.016	&-0.013	&-0.003	&-0.011	&-0.019	&-0.011	&-0.037	&-0.008	&-0.013	&-0.010	&-0.032	&-0.004	&0.003	&-0.018\\
HiTech   &-0.068	&-0.004	&-0.014	&-0.023	&-0.013	&-0.036	&-0.021	&-0.031	&-0.053	&-0.046	&-0.037	&-0.019	&-0.042	&-0.016	&-0.004	&-0.029\\
Health   &-0.041	&-0.003	&-0.005	&-0.021	&-0.003	&-0.009	&-0.007	&-0.006	&-0.010	&-0.008	&-0.005	&-0.013	&-0.027	&-0.005	&-0.004	&-0.005\\
Shops    &-0.058	&-0.010	&-0.013	&-0.064	&-0.005	&-0.014	&-0.023	&-0.007	&-0.021	&-0.015	&-0.009	&-0.013	&-0.024	&-0.009	&-0.004	&-0.025\\
Telecomm &-0.042	&-0.009	&-0.006	&-0.016	&-0.009	&-0.006	&-0.009	&-0.011	&-0.003	&-0.010	&-0.011	&-0.018	&-0.023	&-0.005	&-0.009	&-0.013\\
Utils    &-0.070	&-0.011	&-0.011	&-0.019	&-0.010	&-0.014	&-0.032	&-0.006	&-0.029	&-0.010	&-0.010	&-0.012	&-0.021	&-0.011	&-0.006	&-0.027\\
Other    &-0.055	&-0.009	&-0.006	&-0.017	&-0.009	&-0.018	&-0.017	&-0.005	&-0.023	&-0.015	&-0.007	&-0.015	&-0.037	&-0.007	&-0.005	&-0.015\\

%Durbl    &0.044	&0.085	&0.090	&0.067	&0.082	&0.069	&0.080	&0.079	&0.077	&0.087	&0.076	&0.083	&0.073	&0.087	&0.085	&0.070	&0.092\\
%NoDurbl  &0.084	&0.113	&0.105	&0.100	&0.107	&0.107	&0.096	&0.104	&0.104	&0.111	&0.101	&0.100	&0.083	&0.112	&0.114	&0.108	&0.119\\
%Manuf    &0.039	&0.109	&0.105	&0.064	&0.116	&0.085	&0.080	&0.111	&0.086	&0.108	&0.106	&0.088	&0.096	&0.109	&0.110	&0.097	&0.116\\
%Energy   &0.037	&0.110	&0.094	&0.097	&0.108	&0.100	&0.090	&0.099	&0.070	&0.102	&0.097	&0.101	&0.076	&0.107	&0.115	&0.091	&0.111\\
%HiTech   &0.032	&0.103	&0.091	&0.081	&0.092	&0.067	&0.084	&0.072	&0.049	&0.056	&0.065	&0.085	&0.060	&0.089	&0.103	&0.075	&0.107\\
%Health   &0.054	&0.095	&0.093	&0.075	&0.096	&0.089	&0.091	&0.092	&0.087	&0.089	&0.094	&0.084	&0.069	&0.093	&0.094	&0.093	&0.099\\
%Shops    &0.041	&0.093	&0.090	&0.033	&0.099	&0.089	&0.079	&0.096	&0.081	&0.087	&0.094	&0.089	&0.078	&0.094	&0.100	&0.077	&0.104\\
%Telecomm &0.070	&0.106	&0.109	&0.098	&0.106	&0.109	&0.106	&0.103	&0.113	&0.105	&0.104	&0.097	&0.090	&0.110	&0.106	&0.102	&0.116\\
%Utils    &0.032	&0.098	&0.098	&0.089	&0.099	&0.094	&0.074	&0.103	&0.077	&0.099	&0.099	&0.096	&0.086	&0.098	&0.103	&0.080	&0.110\\
%Other    &0.057	&0.108	&0.111	&0.099	&0.108	&0.097	&0.099	&0.112	&0.092	&0.101	&0.110	&0.101	&0.076	&0.111	&0.113	&0.101	&0.118\\
\hline
    \end{tabular}}%
\label{tab:Table002}

\end{sidewaystable}%

% Table generated by Excel2LaTeX from sheet 'Regression'
\begin{sidewaystable}[htbp]
\caption{Out-of-sample economic performance for stock industry returns: average single-period certainty equivalent returns}
\vspace{.1in}
\begin{justify}
\footnotesize{This table reports the out-of-sample comparison of our predictive framework against standard model combination methodologies, across ten different industries. Performance comparison is based on the single-period Certainty Equivalent (CER) (see Eq.~\eqref{eq:single}), and its modification whereby short sales are not allowed. We report the results obtained for each of the group-specific predictors, the results obtained by simply taking the historical average of the stock returns (HA), and the results from a set of competing model combination/shrinkage schemes, e.g., LASSO, PCA, Equal Weight, and Bayesian Model Averaging (BMA). The sample period is 01:1970-12:2015, monthly.}
\end{justify}
\renewcommand{\arraystretch}{1.2}
\vspace{1em}
{\bf Panel A: \normalfont Average Single-Period Certainty Equivalent} \vspace{0.2cm}\\ 		
\resizebox{1\textwidth}{!}{\begin{tabular}{lcccccccccccccccc}
\hline
Industry & 				& \multicolumn{10}{c}{Group-Specific Models}                                                              & Full  &LASSO  & EW & BMA & PCA  \\
\cline{3-12}
				 &	HA		& Value	&Profit	& Capital	& Soundness	 &Solvency	&Liquidity	&Efficiency	&Other	&Aggregate Fin	&Macro  &\\
\hline
Durbl    &-0.373	&-0.092	&-0.116	&-0.300	  &-0.097	&-0.200	&-0.227	&-0.280	&-0.282	&-0.145	&-0.161	&-0.153	&-0.070	&-0.258	&-0.092	&-0.304\\
NoDurbl  &-0.487	&-0.266	&-0.305	&-0.430	  &-0.265	&-0.367	&-0.415	&-0.365	&-0.416	&-0.288	&-0.323	&-0.364	&-0.271	&-0.393	&-0.266	&-0.370\\
Manuf    &-0.361	&-0.070	&-0.112	&-0.255	  &-0.087	&-0.229	&-0.256	&-0.164	&-0.312	&-0.133	&-0.166	&-0.236	&-0.048	&-0.247	&-0.085	&-0.240\\
Energy   &-0.444	&-0.066	&-0.116	&-0.329	  &-0.067	&-0.160	&-0.241	&-0.229	&-0.380	&-0.190	&-0.216	&-0.112	&-0.216	&-0.283	&-0.073	&-0.205\\
HiTech   &-0.550	&-0.135	&-0.363	&-0.411	  &-0.286	&-0.396	&-0.396	&-0.396	&-0.497	&-0.447	&-0.426	&-0.334	&-0.252	&-0.439	&-0.135	&-0.430\\
Health   &-0.461	&-0.208	&-0.192	&-0.343	  &-0.164	&-0.256	&-0.306	&-0.216	&-0.308	&-0.238	&-0.175	&-0.267	&-0.223	&-0.293	&-0.169	&-0.282\\
Shops    &-0.453	&-0.158	&-0.164	&-0.348	  &-0.068	&-0.222	&-0.297	&-0.192	&-0.381	&-0.256	&-0.206	&-0.168	&-0.141	&-0.320	&-0.077	&-0.270\\
Telecomm &-0.614	&-0.324	&-0.388	&-0.515	  &-0.378	&-0.407	&-0.472	&-0.468	&-0.434	&-0.381	&-0.358	&-0.486	&-0.394	&-0.475	&-0.324	&-0.478\\
Utils    &-0.642	&-0.319	&-0.268	&-0.476	  &-0.262	&-0.415	&-0.501	&-0.304	&-0.543	&-0.426	&-0.361	&-0.312	&-0.268	&-0.476	&-0.278	&-0.480\\
Other    &-0.524	&-0.204	&-0.167	&-0.347	  &-0.180	&-0.397	&-0.387	&-0.222	&-0.424	&-0.349	&-0.275	&-0.158	&-0.168	&-0.372	&-0.157	&-0.275\\
\hline
    \end{tabular}}%
\vspace{1cm}

{\bf Panel B: \normalfont Average Single-Period Certainty Equivalent (no-short sales)} \vspace{0.2cm}\\ 		
\resizebox{1\textwidth}{!}{\begin{tabular}{lcccccccccccccccc}
\hline
Industry & 				& \multicolumn{10}{c}{Group-Specific Models}                                                              & Full  &LASSO  & EW & BMA & PCA  \\
\cline{3-12}
				 &	HA		& Value	&Profit	& Capital	& Soundness	 &Solvency	&Liquidity	&Efficiency	&Other	&Aggregate Fin	&Macro  &\\
\hline

Durbl    &-0.089	&-0.009	&-0.009	&-0.040	&-0.010	&-0.026	&-0.026	&-0.031	&-0.036	&-0.011	&-0.017	&-0.015	&-0.027	&-0.025	&-0.009	&-0.051\\
NoDubl   &-0.022	&-0.005	&-0.013	&-0.017	&-0.008	&-0.011	&-0.021	&-0.012	&-0.014	&-0.007	&-0.016	&-0.018	&-0.025	&-0.007	&-0.004	&-0.012\\
Manuf    &-0.041	&-0.006	&-0.010	&-0.028	&-0.001	&-0.022	&-0.028	&-0.007	&-0.030	&-0.007	&-0.010	&-0.027	&-0.019	&-0.011	&-0.006	&-0.023\\
Energy   &-0.054	&-0.003	&-0.011	&-0.014	&-0.008	&-0.010	&-0.017	&-0.014	&-0.035	&-0.009	&-0.011	&-0.018	&-0.033	&-0.008	&-0.002	&-0.020\\
HiTech   &-0.086	&-0.003	&-0.017	&-0.032	&-0.014	&-0.027	&-0.025	&-0.024	&-0.050	&-0.035	&-0.024	&-0.023	&-0.043	&-0.025	&-0.003	&-0.043\\
Health   &-0.038	&-0.003	&-0.006	&-0.021	&-0.002	&-0.009	&-0.007	&-0.005	&-0.010	&-0.007	&-0.005	&-0.016	&-0.027	&-0.006	&-0.004	&-0.006\\
Shops    &-0.065	&-0.009	&-0.012	&-0.035	&-0.004	&-0.015	&-0.027	&-0.008	&-0.030	&-0.018	&-0.010	&-0.015	&-0.028	&-0.013	&-0.003	&-0.031\\
Telecomm &-0.037	&-0.008	&-0.006	&-0.016	&-0.009	&-0.006	&-0.009	&-0.013	&-0.003	&-0.010	&-0.011	&-0.019	&-0.021	&-0.006	&-0.008	&-0.014\\
Utils    &-0.087	&-0.011	&-0.009	&-0.022	&-0.008	&-0.015	&-0.040	&-0.006	&-0.033	&-0.012	&-0.012	&-0.015	&-0.021	&-0.015	&-0.006	&-0.032\\
Other    &-0.059	&-0.009	&-0.006	&-0.016	&-0.010	&-0.020	&-0.017	&-0.005	&-0.020	&-0.015	&-0.007	&-0.016	&-0.036	&-0.008	&-0.005	&-0.013\\
\hline
    \end{tabular}}%
\label{tab:Table004}

\end{sidewaystable}%

%\setcounter{table}{0}
%\renewcommand{\thetable}{B.\arabic{table}}
%
%\input{Tables/Regression_1_Global_Weekly_Appendix}

%\begin{figure}[p]
%\captionsetup[subfigure]{position=top}
%\caption{US inflation rate forecasting: Root Mean Squared Error}
%\centering
  %\vspace{0.1in}
%\begin{justify} 
%\footnotesize{This figure shows the dynamics of the out-of-sample Root Mean Squared Error (RMSE) obtained for each of the group-specific predictors, by taking the results from a set of competing model combination/shrinkage schemes, e.g., LASSO, Equal Weight, and Bayesian Model Averaging (BMA). The sample period is 01:2001-12:2015, monthly.}
%\end{justify}
%%\vspace*{-2em}
%\includegraphics[width=1\textwidth]{./Figures/rmse_macro.jpg} 
%\label{fig:RMSE_Inflation}

% \end{figure}

 %\begin{figure}[p]
%\captionsetup[subfigure]{position=top}
%\caption{3-Step ahead US inflation forecasting: On-line posterior means of synthesis coefficients}
%\centering
  %\vspace{0.1in}
%\begin{justify} 
%\footnotesize{This figure shows the three-step ahead latent interdependencies across groups of predictive densities-- measured through the predictive coefficients-- used in the recoupling step. These latent components are sequentially computed at each of the $t=\seq1{180}$ months. The objective is to forecast annual inflation three-step ahead.}
%\end{justify}
%%\vspace*{-2em}
%\includegraphics[width=1\textwidth]{./Figures/BPS_coef3.jpg} 
%\label{fig:BPS_Inflation3}
 %\end{figure}
 
%\begin{figure}[p]
%\captionsetup[subfigure]{position=top}
%\caption{US equity return forecasting: Root Mean Squared Error}
%\centering
  %\vspace{0.1in}
%\begin{justify} 
%\footnotesize{This figure shows the dynamics of the out-of-sample Root Mean Squared Error (RMSE) obtained for each of the group-specific predictors, by taking the historical average of the stock returns (HA), and the results from a set of competing model combination/shrinkage schemes, e.g., LASSO, Equal Weight, and Bayesian Model Averaging (BMA). For the ease of exposition we report the results for four representative industries, namely, Manufacturing, Utils, Consumer Durables and Other. Industry aggregation is based on the four-digit SIC codes of the existing firm at each time $t$ following the industry classification from Kenneth French's website. The sample period is 01:1970-12:2015, monthly.}
%\end{justify}
%\hspace*{-3em} 
%\subfigure[Consumer Durable]{\includegraphics[width=0.5\textwidth]{./Figures/rmse_durbl.jpg}}
%\subfigure[Cons. Non-Durable]{\includegraphics[width=0.5\textwidth]{./Figures/rmse_nodurbl.jpg}\hspace*{-2.5em}} \\
%\hspace*{-3em} 
%\subfigure[Telecomm]{\includegraphics[width=0.5\textwidth]{./Figures/rmse_Telcm.jpg}}
%\subfigure[Other]{\includegraphics[width=0.5\textwidth]{./Figures/rmse_Other.jpg}\hspace*{-2.5em}}\\  
%\hspace*{-3em} 
%\subfigure[Hi-Tech]{\includegraphics[width=0.5\textwidth]{./Figures/rmse_HiTech.jpg}}
%\subfigure[Utils]{\includegraphics[width=0.5\textwidth]{./Figures/rmse_Utils.jpg}\hspace*{-2.5em}}  
  %\label{fig:RMSE_Finance}
 %\end{figure}

%\doublespacing

%\onehalfspacing
%\footnotesize

\end{document}